\documentclass{aa}  

\usepackage{graphicx}
\usepackage{txfonts}
\begin{document}

   \title{Gas and dust from metal-rich AGB stars}


   \author{P. Ventura\inst{1}, F. Dell'Agli\inst{1}, M. Lugaro\inst{2,3,4},  
           D. Romano\inst{5}, M. Tailo\inst{6}, A. Yag\"ue\inst{3}
          }

   \institute{INAF, Observatory of Rome,
              Via Frascati 33, 00077 Monte Porzio Catone (RM), Italy\\
              \email{paolo.ventura@inaf.it} \and
             Konkoly Observatory, Research Centre for Astronomy and Earth Sciences, 
             Konkoly Thege Mikl\'{o}s \'{u}t 15-17, H-1121 Budapest, Hungary \and
             ELTE E\"{o}tv"{o}s Lor\'and University, Institute of Physics, Budapest 1117, 
             P\'azm\'any P\'eter s\'et\'any 1/A, Hungary \and
             School of Physics and Astronomy, Monash University, VIC 3800, Australia  \and
             INAF, Astrophysics and Space Science Observatory,
              Via Piero Gobetti 93/3, 40129 Bologna, Italy \and
             Dipartimento di Fisica e Astronomia “Galileo Galilei”, Univ. di Padova, Vicolo 
             dell’Osservatorio 3, Padova, IT-35122
             }

   \date{Received September 15, 1996; accepted March 16, 1997}


 
  \abstract
   {Stars evolving through the asymptotic giant branch (AGB) phase provide significant
   feedback to their host system, in form of both gas enriched in nuclear-burning products and dust formed in their winds, which they eject into the
   interstellar medium. Therefore AGB stars are an
   essential ingredient for the chemical evolution of the
   Milky Way and other galaxies.}
   {We study AGB models with super-solar metallicity, to complete our large
   database, so far extending from metal-poor to solar chemical compositions. We provide
   chemical yields for masses in the range $1-8~M_{\odot}$ and metallicities $Z=0.03$ and $Z=0.04$. We also
   study dust production in this metallicity domain.}
   {We calculated the evolutionary sequences from the pre main sequence through the
    whole AGB phase. We follow the variation of the surface chemical composition to calculate
    the chemical yields of the various species and model dust formation in the winds to
    determine the dust production rate and the total dust mass produced by each star during
    the AGB phase.}
   {The physical and chemical evolution of the star is sensitive to the initial mass: 
    $M> 3 M_{\odot}$ stars experience hot bottom burning, whereas the surface chemistry of 
    the lower mass counterparts is altered only by third dredge-up. The carbon-star phase 
    is reached by $2.5-3.5~M_{\odot}$ stars of metallicity $Z=0.03$, whereas all the 
    $Z=0.04$ stars (except the $2.5 M_{\odot}$) remain O-rich for the whole AGB phase. 
    Most of the dust produced by metal-rich AGBs is in the form of silicates particles. 
    The total mass of dust produced increases with the mass of the star, reaching 
    $\sim 0.012~M_{\odot}$ for $8~M_{\odot}$ stars.}
   {}


   \maketitle
%

\section{Introduction}


The recent years have witnessed a growing interest towards stars evolving through the
asymptotic giant branch (AGB) phase because they have been recognized to play an important
role in several astrophysical contexts, from the interpretation of the chemical
patterns traced by Milky Way stars (e.g. Romano et al. 2010), to the formation of multiple population 
in globular clusters \citep{ventura01, dercole08}, and the contribution to the overall dust 
budget in the Local Universe and at high redshift \citep{valiante09}.
Several groups have modelled this stellar evolutionary phase, characterized
by the occurrence of a series of thermal pulses (TP),
providing an accurate description of the main evolutionary and structural properties
of AGB stars of different mass and metallicity and the chemical yields from these objects, which are 
essential ingredients to understand the feedback from these stars to the host system
\citep{cristallo11, cristallo15, karakas10, karakas14b, karakas18}.

Recent models also couple the modelling of the AGB evolution with the description
of the dust formation process that takes place in the wind expanding from the central
star (e.g., Ventura et al. 2012, 2014b, Nanni et al. 2013, 2014). 
Our research has been so far mostly focused on metal-poor \citep{ventura12,
dicriscienzo13} and sub-solar metallicity \citep{ventura14a} AGB models. The former were
used to explore the dust contribution from AGB stars at high redshift and in Local Group 
galaxies harbouring only metal-poor populations \citep{flavia19}; the $Z=4-8\times 10^{-3}$
models were the starting point to characterize the evolved populations of the Magellanic
Clouds \citep{flavia14b, flavia15a, flavia15b}.
The advent of Gaia pushed the interest towards solar metallicities, which were studied 
by \citet{dicriscienzo16} and \citet{ventura18}. In this paper we make a step forward, 
studying the behaviour of AGB stars of super-solar metallicity, $Z=0.03$ and $Z=0.04$. 

The exploration of the super-solar metallicity regime is particularly
important given the recent results from Galactic surveys \citep{casagrande11, bensby14},
that showed that stars with ages in the range $5-10$ Gyr currently found in the solar 
neighbourhood span the metallicity range from
$0.2$ to $2.5$ solar, which can be interpreted as the effects of migration mechanisms
taking place within the Galaxy \citep{minchev13, kubryk15, spitoni15}. Furthermore, 
recent studies showing evidences for a significant fraction of metal-rich stars in the 
Galactic centre \citep{schultheis19} suggest that this region is characterized by a 
very high average metallicity, of the order of $1.5$ solar. \citet{thorsbro} analysed
a sample of giants in the Galactic centre and found a metallicity distribution 
extended to $[$Fe$/$H$]=+0.5$. Lastly, we recall recent 
results selected from LAMOST, combined with Gaia DR2 data, that outline the presence of 
super metal-rich stars, both with thin disk and thick disk kinematics, which again invoke 
radial migration. Out of the Milky Way, super metal-rich stars were suggested as possible 
major component in the nuclear regions of massive elliptical galaxies \citep{bertola95}.

While understanding the gas pollution from metal-rich stars is crucial for the studies 
of the Milky Way and other galaxies, 
most current grids of stellar yields in literature cover up only to solar metallicity. When the 
yields are implemented in galactic chemical evolution models, some interpolation is needed 
between adiacent metallicity grids. Unless the yields vary monotonically with metallicity, this is 
clearly an unsafe procedure. When the metallicity exceeds solar, the yields are either 
kept the same as their solar values or extrapolated somehow, with the first choice being 
the most common one. While this does not affect the results of chemical evolution studies 
focusing on dwarf galaxies or the external regions of the Milky Way, it might severely 
affect the predictions about the inner Galaxy evolution, as well as the interpretation of 
data for massive ellipticals at both low and high redshifts. In fact, when large fractions 
of super-solar metallicity stars are formed, the adoption of solar-metallicity yields may 
result in spurious results. As far as AGB stars are concerned, this is especially true for 
elements such as He, $^{13}$C, $^{14}$N, and $^{17}$O that are produced in significant 
amounts by intermediate-mass stars.

Models of nucleosynthesis and dust formation for AGB stars of super-solar metallicity are 
also needed to address the origin of meteoritic stardust grains, the vast majority of 
which originated from AGB stars \citep[see, e.g., the review by][]{zinner14}. Based on 
their isotopic anomalies, it has been proposed that a significant fraction of silicon 
carbide (SiC) grains extracted from meteorites originated from AGB stars of super-solar 
metallicity \citep{lugaro14,lugaro18} and that the efficiency of the formation of SiC dust 
around AGB stars as function of the metallicity may be responsible for the higher than 
expected number of grains from super-solar metallicity AGB stars \citep{lewis13}. 
Furthermore, SiC grains from such AGB stars of super-solar metallicity are the best 
candidates to be the mineral carriers responsible for the anomalies in the elements 
heavier than iron predominant in different Solar System bodies, and provide us information 
on the evolution of dust in the protosolar disc \citep{ek19}. Also to address the problems 
related to meteoritic stardust, the models of metallicities up to above solar, including 
both nucleosynthesis and dust formation, represent one of the most promising current tools.

The paper is organised as follows: in Section \ref{uncert} we discuss the main uncertainties
affecting the modelling of the AGB phase; Section \ref{modinput} presents the numerical
and physical ingredients used to model the 
evolution of the stars and of the dust formation process; the evolution of stars
before and during the AGB phase are described in Section \ref{preAGB} and \ref{evol}, 
respectively; Section \ref{yields} regards the chemical yields of the stars. In 
Section \ref{comp} our results are compared to models of similar metallicity 
available in the literature; in Section \ref{dust} the properties of the dust formed 
in the wind of AGB stars, in terms of grain size distribution, the dust production 
rate, and the mass of dust formed, are reported; in Section \ref{end} the conclusions 
are given.

\section{The uncertainties affecting AGB modelling}
\label{uncert}
Before describing the physical and numerical ingredients used in the present
investigation, we believe important to stress that the results regarding the AGB
modelling are rendered uncertain by the scarse knowledge of two physical mechanisms,
relevant for the description of the evolution of these stars, still poorly known from first
principles: convection and mass loss \citep{karakas14b}.

The importance of convection on the AGB evolution is twofold, as the description of 
convective regions concerns both the efficiency of the convective modality of
transport of energy and the location of the border of the instability regions,
within which mixing of chemicals takes place.

The efficiency of convection reflects into the temperature gradient and is particularly
relevant to determine the temperature at the base of the external mantle, which, in turn,
is connected with the possible ignition of hot bottom burning (hereinafter HBB), 
which consists in the activation of proton captures in the most internal regions of the 
convective envelope, once the temperatures exceed $\sim 30$ MK \citep{renzini81,
blocker91, sackmann91}. \citet{ventura05a} showed 
that the strength of HBB is extremely sensitive to convection modelling and,
within the classic mixing length theory schematization, to the choice of the free 
parameter $\alpha$, giving the mixing length in terms of the local value of the pressure
scale height. The AGB phase is the only case, within the stellar astrophysical context, where
the choice of the convective model affects not only the external temperature and the
colours, but also the physical evolution of the star, namely the luminosity reached
and the duration of this peculiar evolutionary phase \citep{ventura05a}.

An important point related to convection is the location of the borders of the
instability regions. Within the context of AGB modelling, it is of paramount 
importance to determine the extent of the third dredge-up (hereinafter TDU). The TDU consists 
in the inwards penetration of the convective envelope, taking place after each thermal 
pulse of the He burning shell, down to layers previously affected by nucleosynthesis 
via the triple-alpha reactions, thus enriched in carbon \citep{iben74}. When the plain
Schwartzschild criterion is adopted, with no assumptions regarding possible extra-mixing from
the bottom of the convective envelope, the extent of TDU is too small to reproduce the
observational evidence, particularly the luminosity function of carbon stars in the
Magellanic Clouds (but see Straniero et al. 1997 on the possibility that models without 
extra-mixing are able to produce a deep TDU). An early algorithm aimed at calculating the
extent of the extra-mixed zone was proposed by \citet{lattanzio86} and is still used by
the team using the MONASH code \citep{karakas14a, karakas14b, karakas18}. Several 
research groups adopt a velocity profile decaying from the base of the envelope towards
the stellar interior, with the e-folding distance of the decay being treated as a 
free parameter \citep{herwig00, herwig05, cristallo09, weiss09}. We reiterate here that all these
treatments require some ad hoc assumptions, because the details of the mixing mechanism,
and more generally of the convective phenomenon, close to the borders, is substantially
unknown from first principles: the extent of TDU can be empirically determined by 
fitting the observational scenario, but cannot be found on the basis of solid
physical arguments.

The description of mass loss has an extreme importance in the modelling of the AGB phase,
comparable to the role played by convection. The rate at which the mass of the envelope of
AGB stars is expelled into the interstellar medium affects the duration of the AGB phase
and has relevant effects onto the largest luminosity reached, the degree of nucleosynthesis
associated to HBB and the amount of carbon that is gradually accumulated in the surface
regions via TDU \citep{ventura05b}.

Mass loss is commonly described by analytical relations, that allow the computation of
$\dot M$ as a function of the main stellar parameters, in particular the radius and the
surface gravity. 

Several evolutionary codes adopt the classic period - mass loss rate relation by
\citet{vw93}, which was calibrated on the basis of Galactic Mira variables and
pulsation OH/IR stars in the Galaxy and the Large Magellanic Cloud. The \citet{vw93}
recipe is used both for O-rich and carbon stars \citep{cristallo09, karakas14b}.

Some groups \citep{weiss09} model mass loss during the O-rich AGB phases via the empirical 
period - mass loss relation by \citet{jacco05}, which is based on the observations 
of dust-enshrouded M-stars in the Magellanic Clouds. One of the most widely used prescriptions 
for mass loss by M-rich AGBs is the treatment by \citet{blocker95}, based on hydrodynamical 
models of pulsating M-type stars \citep{karakas18, pignatari16}.

Regarding the C-rich phase, alternatively to \citet{vw93}, several research teams 
have based their computations on the theoretical radiation-hydrodynamical models published 
by the Berlin group \citep{wachter02, wachter08}, which consider dust production in
C-rich winds, and the effects of radiation pressure on the carbonaceous dust particles
formed in the circumstellar envelope \citep{weiss09, ventura18}.

The main shortcoming of the use of the above prescriptions is their application
to a wide range of mass and metallicities, despite these relationships have been derived
on the basis of limited sample of stars, mostly homogeneous in the chemical composition.
Furthermore, some of these prescriptions contain free parameters (e.g. the Reimers
parameter entering the \citet{blocker95} recipe) than needs further calibration.

The situation is even more complex for C-stars, if we consider that the formulae by 
\citet{wachter02, wachter08} do not include any dependence on the actual carbon excess 
with respect to oxygen, which intuitively should affect the amount of dust formed in the 
wind of the stars. \citet{wachter02} showed that the dependence of the mass loss rate on 
this quantity is weak enough to be ignored in comparison with all other uncertainties, but this 
statement was seriously argued by \citet{lars08}, who stressed the decisive role of the
carbon excess in the dust formation process and in the determination of the mass loss rate.

\section{Physical and numerical input}
\label{modinput}
\subsection{Stellar evolution modelling}
\label{agbinput}
The models presented in this paper were calculated with the ATON stellar evolution code 
\citep{ventura98}.
An exhaustive description of the numerical details of the code and 
the most recent updates can be found in \citet{ventura13}. 
The models span the mass interval $1M_{\odot} \leq M \leq 8~M_{\odot}$. The 
metallicities used are $Z=0.03$ and $Z=0.04$ and the scaled-solar mixture adopted is taken from 
\citet{gs98}. The initial helium is $Y=0.30$.
The models not undergoing the 
helium flash were evolved from the pre-main sequence until the almost total consumption of 
the envelope. Low-mass models ($M \leq 2~M_{\odot}$) experiencing the helium flash 
were evolved from the horizontal branch, starting from the total mass, core mass and surface 
chemical composition calculated until the tip of the red giant branch.

The temperature gradient within regions unstable to convection is calculated via the full 
spectrum of turbulence (FST) model \citep{cm91}. Overshoot of convective eddies within
radiatively stable regions is modeled by assuming that the velocity of convective elements 
decay exponentially beyond the neutrality point, fixed via the Schwartzschild criterion. 
The e-folding distance of the velocity decays during the core (hydrogen and helium) burning 
phases and during the AGB phase is taken as $0.02H_{\rm P}$ and $0.002H_{\rm P}$, respectively. 
The former was calibrated on the basis of the observed width of the main sequences of
open clusters \citep{ventura98}, while the latter was found by reproducing the
luminosity function of carbon stars in the Magellanic Clouds \citep{ventura14c}.

The mass loss rate for oxygen-rich models is determined via the \citet{blocker95} 
treatment with the parameter entering the \citet{blocker95}'s formula set to $\eta=0.02$, 
following the calibration given in \citet{ventura00}. For carbon stars 
we implemented the description of mass loss from the Berlin group \citep{wachter02, wachter08}. 

The radiative opacities are calculated according to the OPAL release, in the version documented 
by \citet{opal}. The molecular opacities in the low-temperature regime ($T < 10^4$ K) are 
calculated with the AESOPUS tool \citep{marigo09}. The opacities are 
constructed self-consistently, by following the changes in the chemical composition of the 
envelope, particularly of the individual abundances of carbon, nitrogen, and oxygen.

\begin{figure*}
\begin{minipage}{0.48\textwidth}
\resizebox{1.\hsize}{!}{\includegraphics{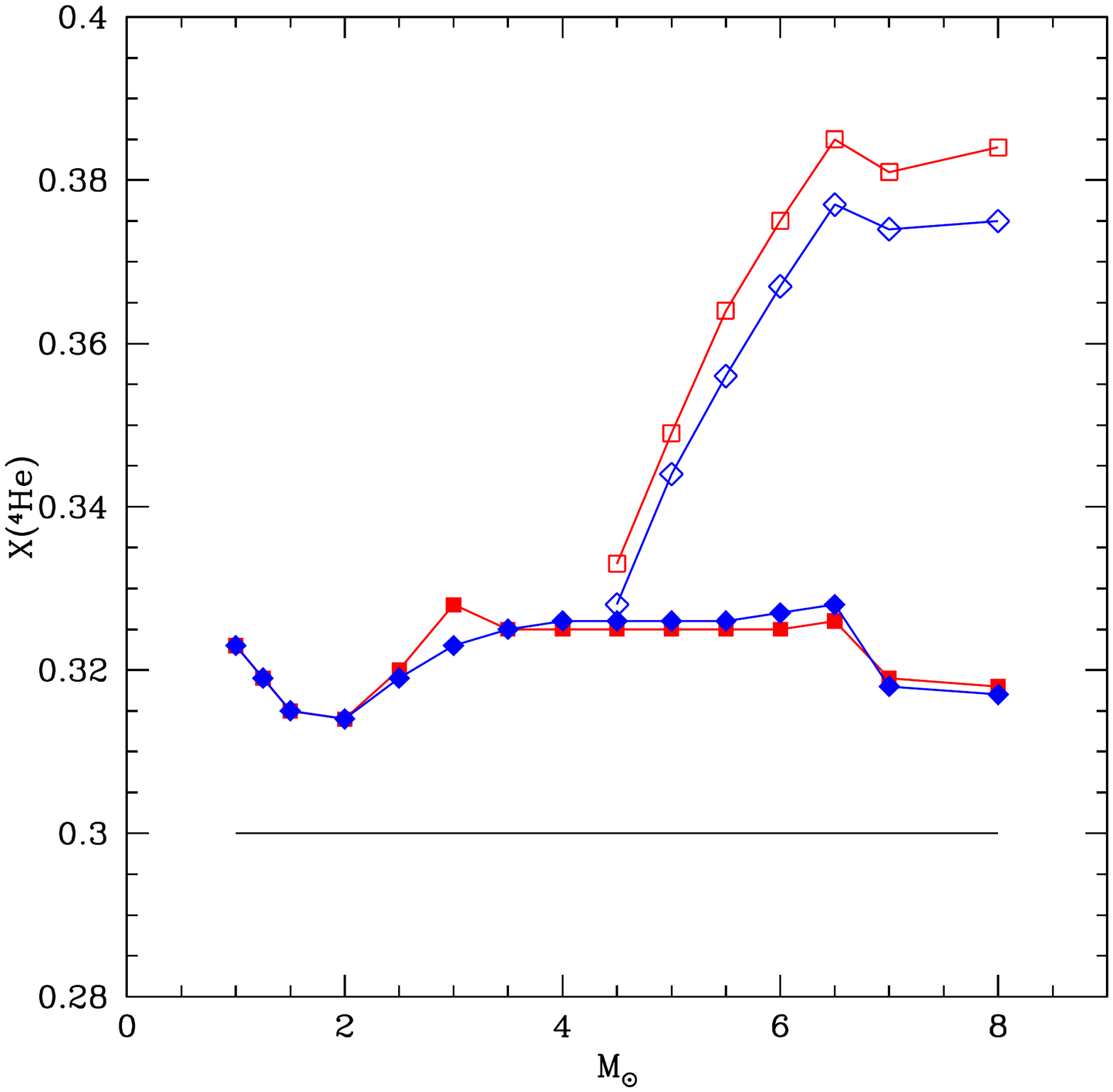}}
\end{minipage}
\begin{minipage}{0.48\textwidth}
\resizebox{1.\hsize}{!}{\includegraphics{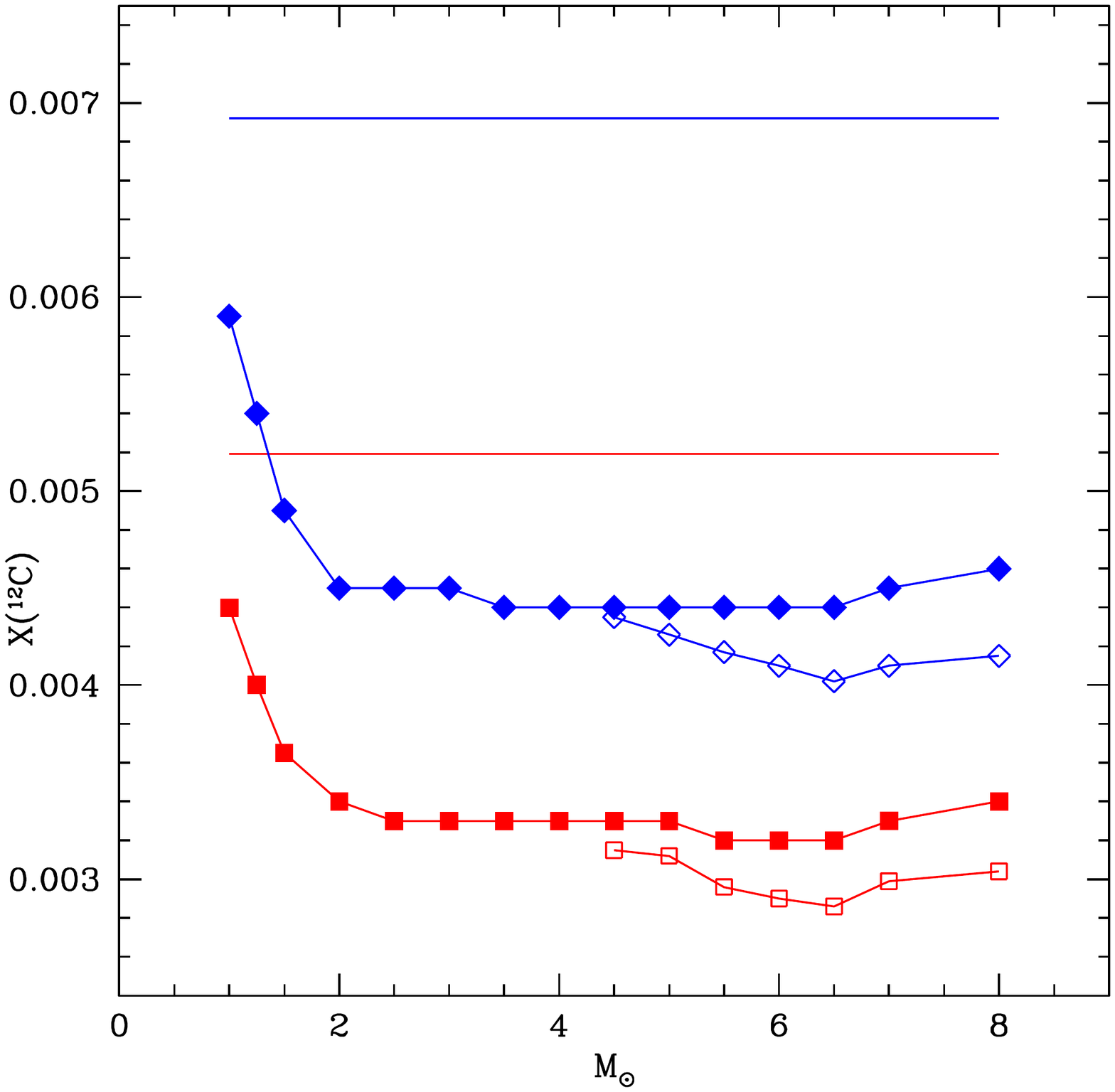}}
\end{minipage}
\vskip-80pt
\begin{minipage}{0.48\textwidth}
\resizebox{1.\hsize}{!}{\includegraphics{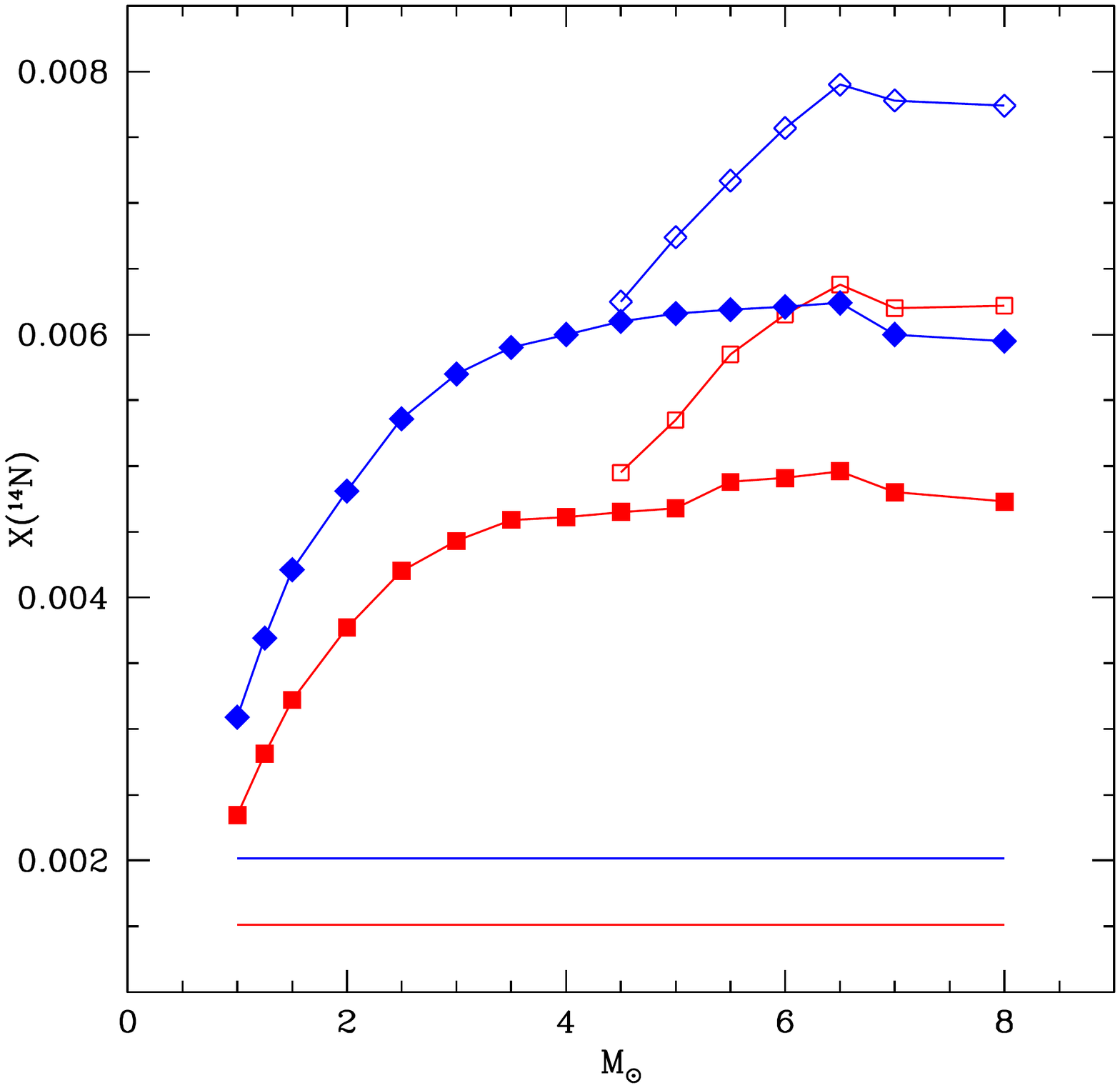}}
\end{minipage}
\begin{minipage}{0.48\textwidth}
\resizebox{1.\hsize}{!}{\includegraphics{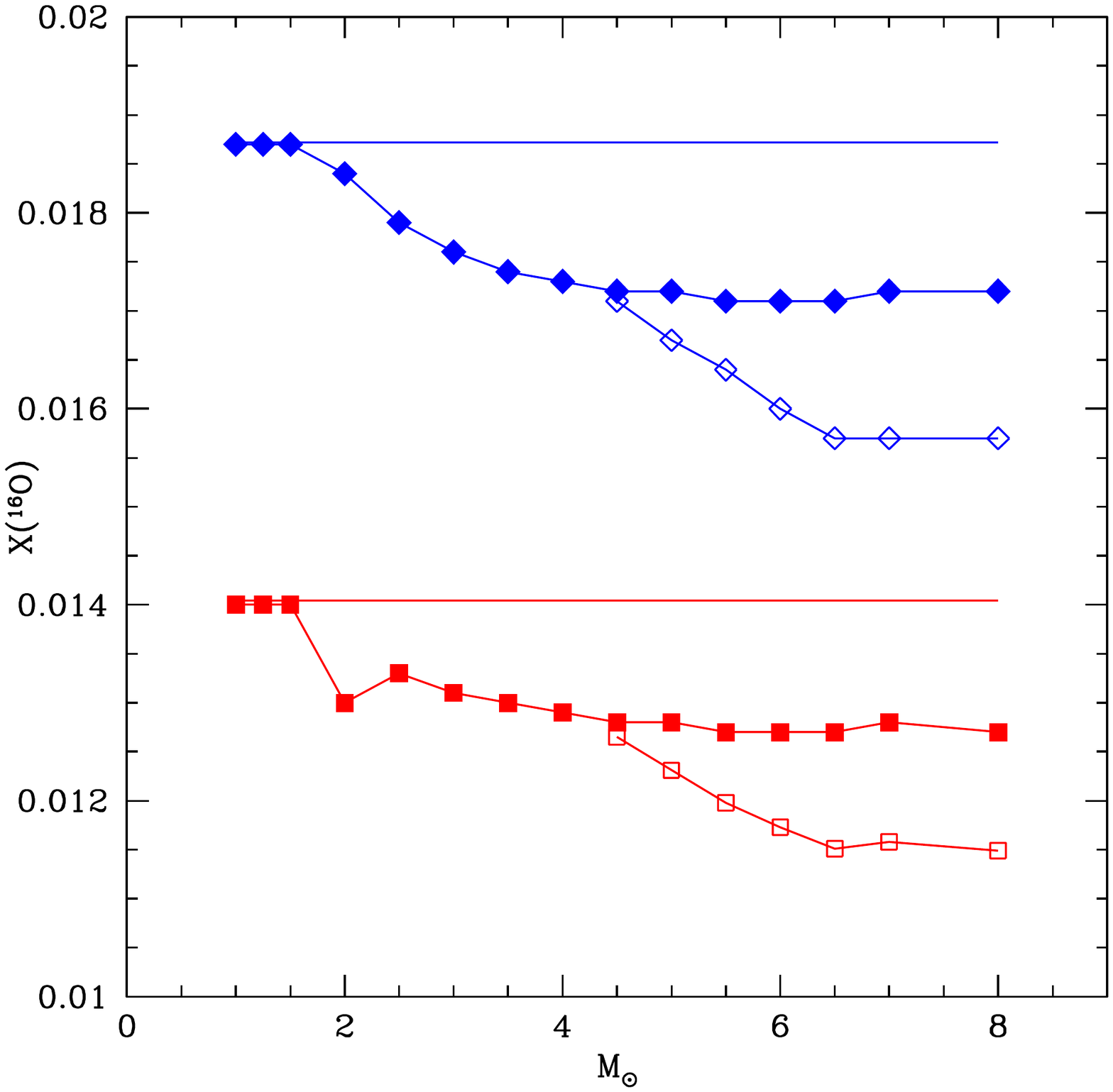}}
\end{minipage}
\vskip-60pt
\caption{The changes in the surface mass fraction of helium (top, left panel), 
$^{12}$C (top, right), $^{14}$N and $^{16}$O after the first (full points)
and the second (open points) dredge-up episode of the AGB stars of metallicity 
$Z=0.03$ (red squares) and $Z=0.04$ (blue diamonds). The horizontal lines in each 
panel indicate the initial mass fractions of the same species.}
\label{fdup}
\end{figure*}

\subsection{Dust production}
The formation and growth of dust particles in the wind of AGB stars is described according to 
the schematization proposed by the Heidelberg group \citep{fg06}, previously used by 
our team \citep{ventura12, ventura14b, ventura15, ventura16} and in a series of papers by 
the Padua group \citep{nanni13, nanni14, nanni16, nanni18, nanni19, nanni20}. All the
relevant equations can be found in \citet{ventura12}. Here we only provide a brief description
of the methodology.

\begin{figure}
 \resizebox{\hsize}{!}{\includegraphics{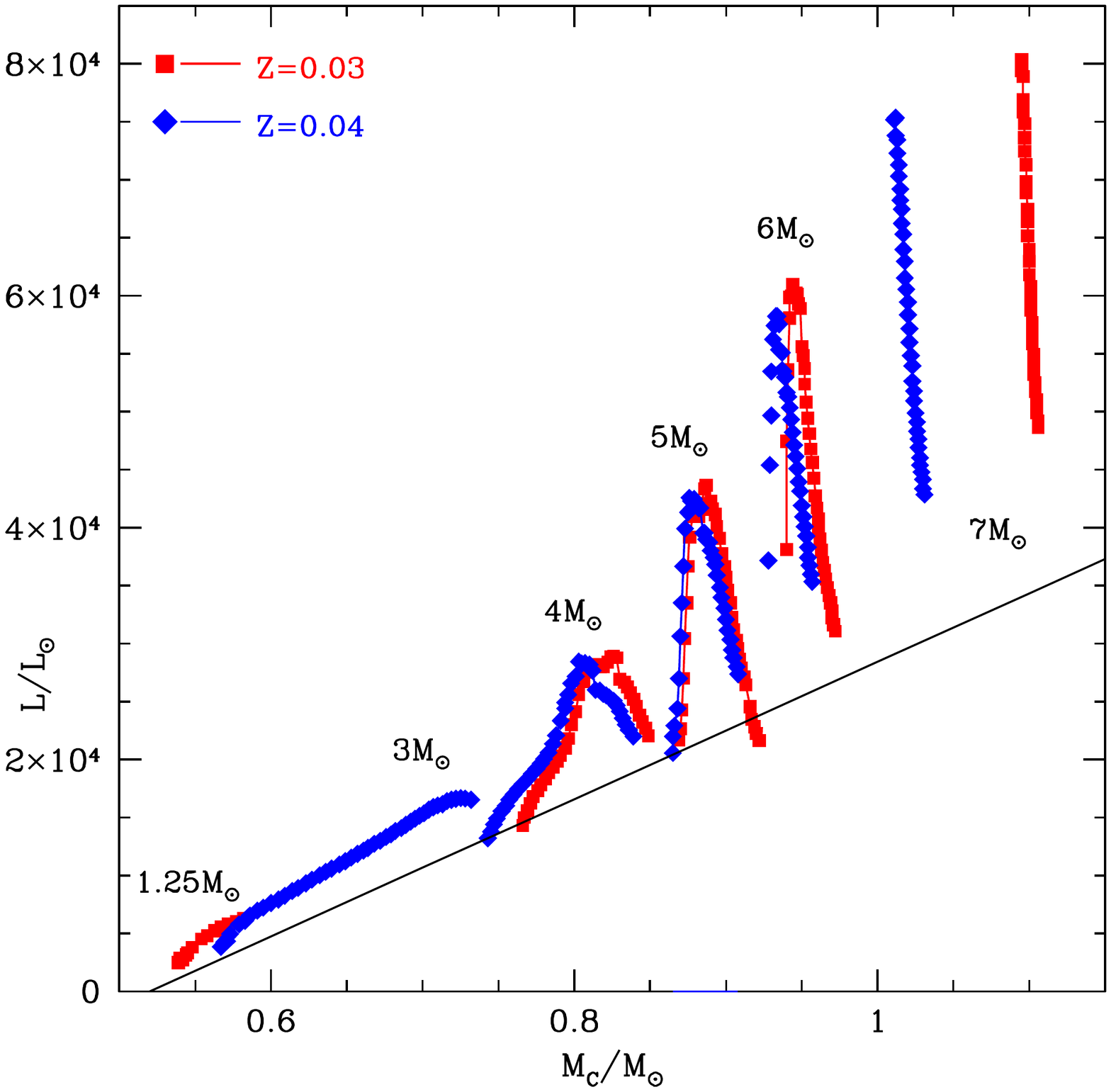}}
\vskip-70pt
 \caption{The evolution of high-metallicity AGB models in the core mass - luminosity
 plane. The black line shows the theoretical relationship derived by \citet{pac70}.
}
 \label{fmcl}
\end{figure}

Dust particles are assumed to form and grow in the wind, which expands isotropically from 
the central star. The dynamics of the wind is described by the momentum equation,
where the acceleration is determined by the balance between gravity and radiation pressure
acting on the newly formed dust grains. The coupling between grain growth and wind
dynamics is given by the extinction coefficients, describing absorption and scattering of
the radiation by dust particles.
The evolution with time of dust grains is determined by the balance between the growth
and the vaporization rate. The former is given by the gas molecules hitting the already
formed grains and the latter is related to the vapour pressure of gaseous molecules 
over the solid compounds.
Regarding the dust species formed, in oxygen-rich environments the most stable compounds
are considered: silicates, alumina dust (Al$_2$O$_3$) and solid iron. In the winds of carbon
stars the species considered are silicon carbide (SiC), solid carbon and solid iron.
This description allows the determination of the surface fraction of 
gaseous silicon, aluminium, iron, and carbon condensed into dust particles (see eq. 20-23 and 
34-35 in Ferrarotti \& Gail 2006) and the dust production rate for each dust species, which 
depends on the gas rate of mass loss, the surface mass fractions of the afore mentioned
chemical elements, and the fraction of the latter species condensed into dust
(see Section 5.2 in Ferrarotti \& Gail 2006).

\begin{table*}
\caption{Main structural properties of our stellar models: initial mass in units of $M_{\odot}$;
duration of core H burning ($\tau_{H}$), core He burning ($\tau_{He}$), early AGB ($\tau_{EAGB}$) and 
thermally pulsating ($\tau_{AGB}$) AGB phases; the core mass at the first TP $M_{\rm C}^{\rm 1TP}$; the final mass of the star ($M_{\rm f}$); the maximum
luminosity ($L^{\rm max}$) and temperature at the bottom of the convective envelope ($T_{\rm b}^{\rm max}$) reached
during the AGB phase; and the final C$/$O surface ratio. Time-scales, luminosities and temperatures
are given as $x(y)$, where $x(y)=x\times 10^y$.}             
\label{tabmod}      
\centering          
\begin{tabular}{c c c c c c c c c c }     
\hline       
M$/$M$_{\odot}$ & $\tau_{H}$ & $\tau_{He}$ & $\tau_{EAGB}$ & $\tau_{AGB}$ &
M$_{\rm C}^{\rm 1TP}/$M$_{\odot}$ & M$_{\rm f}/$M$_{\odot}$ & $L^{\rm max}/L_{\odot}$ & 
$T_{\rm b}^{\rm max}(K)$ & $($C$/$O$)_{\rm f}$\\ 
\hline                    
 & & & & $Z=0.03$ & & & & & \\
\hline
1.00 & 1.14(10) & 8.65(7) & 2.76(7) & 7.57(5) & 0.538 & 0.560 & 4.86(3) & 2.24(6) & 0.38 \\
1.25 & 4.91(9) & 1.12(8) & 1.73(7) & 1.14(6) & 0.539 & 0.582 & 6.30(3) & 2.98(6) & 0.38 \\
1.50 & 2.77(9) & 1.09(8) & 2.90(7) & 1.16(6) & 0.551 & 0.615 & 8.17(3) & 3.68(6) & 0.34 \\
2.00 & 1.20(9) & 1.51(8) & 1.93(7) & 2.34(6) & 0.530 & 0.653 & 1.12(4) & 6.73(6) & 1.03 \\
2.50 & 6.35(8) & 2.11(8) & 2.16(7) & 2.61(6) & 0.527 & 0.688 & 1.45(4) & 1.35(7) & 1.08 \\
3.00 & 3.83(8) & 1.11(8) & 1.20(7) & 1.71(6) & 0.575 & 0.699 & 1.56(4) & 2.15(7) & 1.27 \\
3.50 & 2.52(8) & 6.38(7) & 6.41(6) & 9.77(5) & 0.662 & 0.786 & 2.20(4) & 5.45(7) & 1.09 \\
4.00 & 1.76(8) & 3.92(7) & 3.73(6) & 4.12(5) & 0.766 & 0.849 & 2.89(4) & 7.55(7) & 0.018 \\
4.50 & 1.30(8) & 2.67(7) & 2.22(6) & 2.35(5) & 0.837 & 0.893 & 3.72(4) & 8.09(7) & 0.075 \\
5.00 & 9.94(7) & 1.89(7) & 1.53(6) & 2.02(5) & 0.869 & 0.922 & 4.37(4) & 8.28(7) & 0.065 \\
5.50 & 7.89(7) & 1.45(7) & 1.01(6) & 1.65(5) & 0.905 & 0.951 & 5.13(4) & 8.51(7) & 0.031 \\
6.00 & 6.42(7) & 1.09(7) & 7.38(5) & 1.35(5) & 0.940 & 0.972 & 6.10(4) & 8.75(7) & 0.032 \\
6.50 & 5.36(7) & 8.74(6) & 5.11(5) & 1.01(5) & 0.990 & 1.022 & 7.06(4) & 9.06(7) & 0.030 \\
7.00 & 4.53(7) & 7.59(6) & 3.11(5) & 8.39(4) & 1.095 & 1.052 & 8.04(4) & 9.38(7) & 0.033 \\
8.00 & 3.42(7) & 5.34(6) & 1.99(5) & 4.32(4) & 0.000 & 1.210 & 9.57(4) & 9.91(7) & 0.048 \\
\hline
 & & & & $Z=0.04$ & & & & & \\
\hline       
1.00 & 1.35(10) & 1.31(8) & 1.82(7) & 6.92(5) & 0.540 & 0.560 & 4.74(3) & 2.26(6) & 0.38 \\
1.25 & 5.73(9) & 1.21(8) & 1.91(7) & 1.03(6) & 0.551 & 0.578 & 5.98(3) & 2.92(6) & 0.38 \\
1.50 & 3.19(9) & 1.28(8) & 1.56(7) & 1.37(6) & 0.550 & 0.608 & 7.87(3) & 3.73(6) & 0.35 \\
2.00 & 1.38(9) & 1.52(8) & 2.30(7) & 2.12(6) & 0.538 & 0.666 & 1.14(4) & 6.08(6) & 0.59 \\
2.50 & 7.07(8) & 2.32(8) & 2.39(7) & 2.58(6) & 0.543 & 0.699 & 1.40(4) & 1.13(7) & 1.08 \\
3.00 & 4.21(8) & 1.19(8) & 1.41(7) & 2.02(6) & 0.574 & 0.732 & 1.67(4) & 2.06(7) & 0.83 \\
3.50 & 2.73(8) & 6.89(7) & 7.73(6) & 1.18(6) & 0.648 & 0.769 & 2.00(4) & 4.79(7) & 0.88 \\
4.00 & 1.90(8) & 4.31(7) & 4.47(6) & 5.02(5) & 0.744 & 0.839 & 2.84(4) & 7.10(7) & 0.017 \\
4.50 & 1.39(8) & 2.93(7) & 2.81(6) & 2.54(5) & 0.834 & 0.882 & 3.65(4) & 7.80(7) & 0.019 \\
5.00 & 1.05(8) & 2.12(7) & 1.86(6) & 2.18(5) & 0.866 & 0.908 & 4.26(4) & 8.05(7) & 0.022 \\
5.50 & 8.27(7) & 1.48(7) & 1.29(6) & 1.71(5) & 0.900 & 0.946 & 5.04(4) & 8.17(7) & 0.021 \\
6.00 & 6.71(7) & 1.18(7) & 9.43(5) & 1.53(5) & 0.928 & 0.957 & 5.82(4) & 8.45(7) & 0.022 \\
6.50 & 5.56(7) & 9.30(6) & 6.72(5) & 1.33(5) & 0.968 & 1.023 & 6.76(4) & 8.63(7) & 0.027 \\
7.00 & 4.67(7) & 7.43(6) & 5.09(5) & 9.88(4) & 1.010 & 1.031 & 7.53(4) & 8.73(7) & 0.028 \\
8.00 & 3.49(7) & 5.06(6) & 2.87(5) & 4.31(4) & 0.000 & 1.180 & 8.53(4) & 8.93(7) & 0.031 \\
\hline                  
\end{tabular}
\end{table*}

\begin{figure*}
\begin{minipage}{0.48\textwidth}
\resizebox{1.\hsize}{!}{\includegraphics{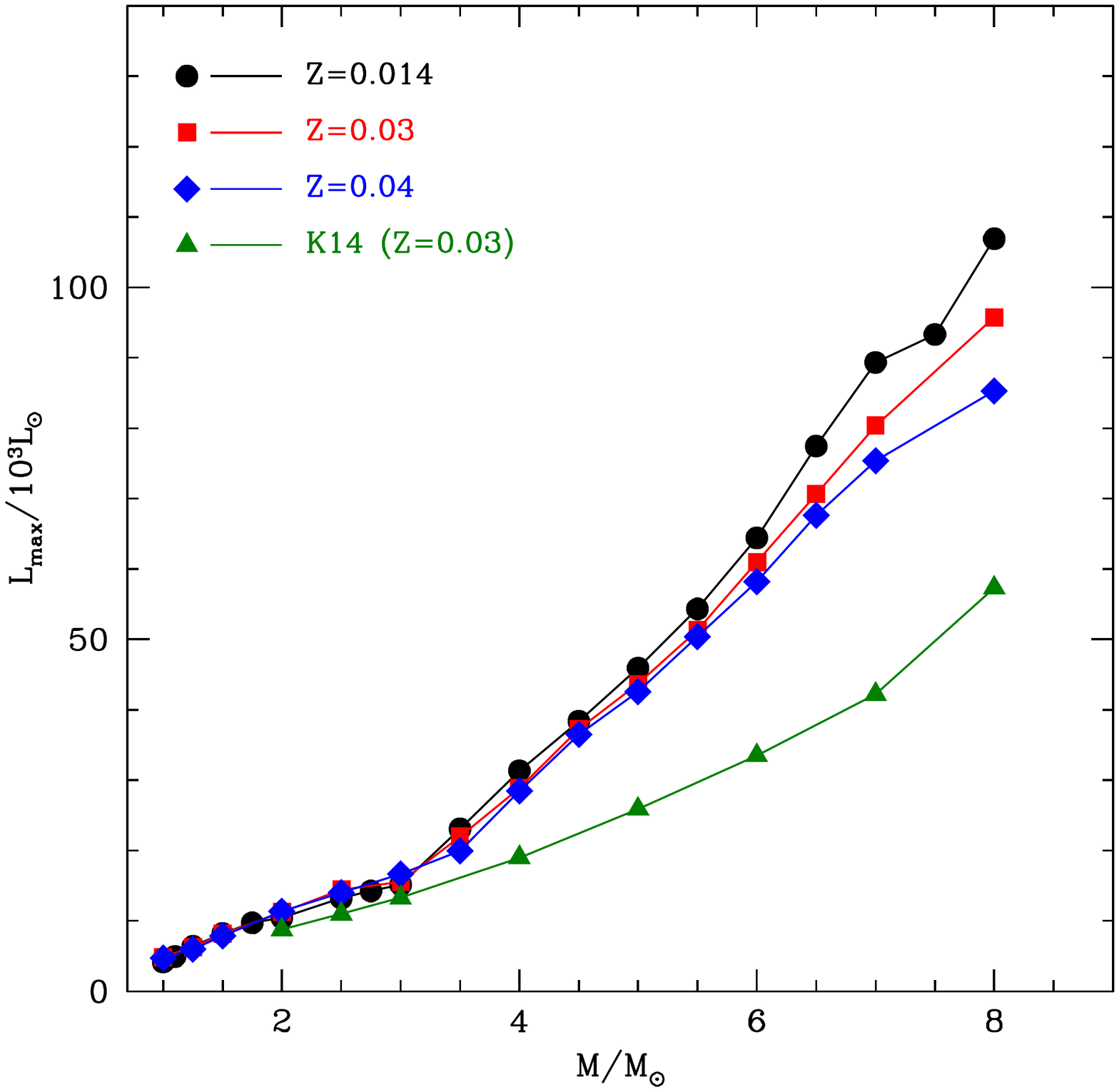}}
\end{minipage}
\begin{minipage}{0.48\textwidth}
\resizebox{1.\hsize}{!}{\includegraphics{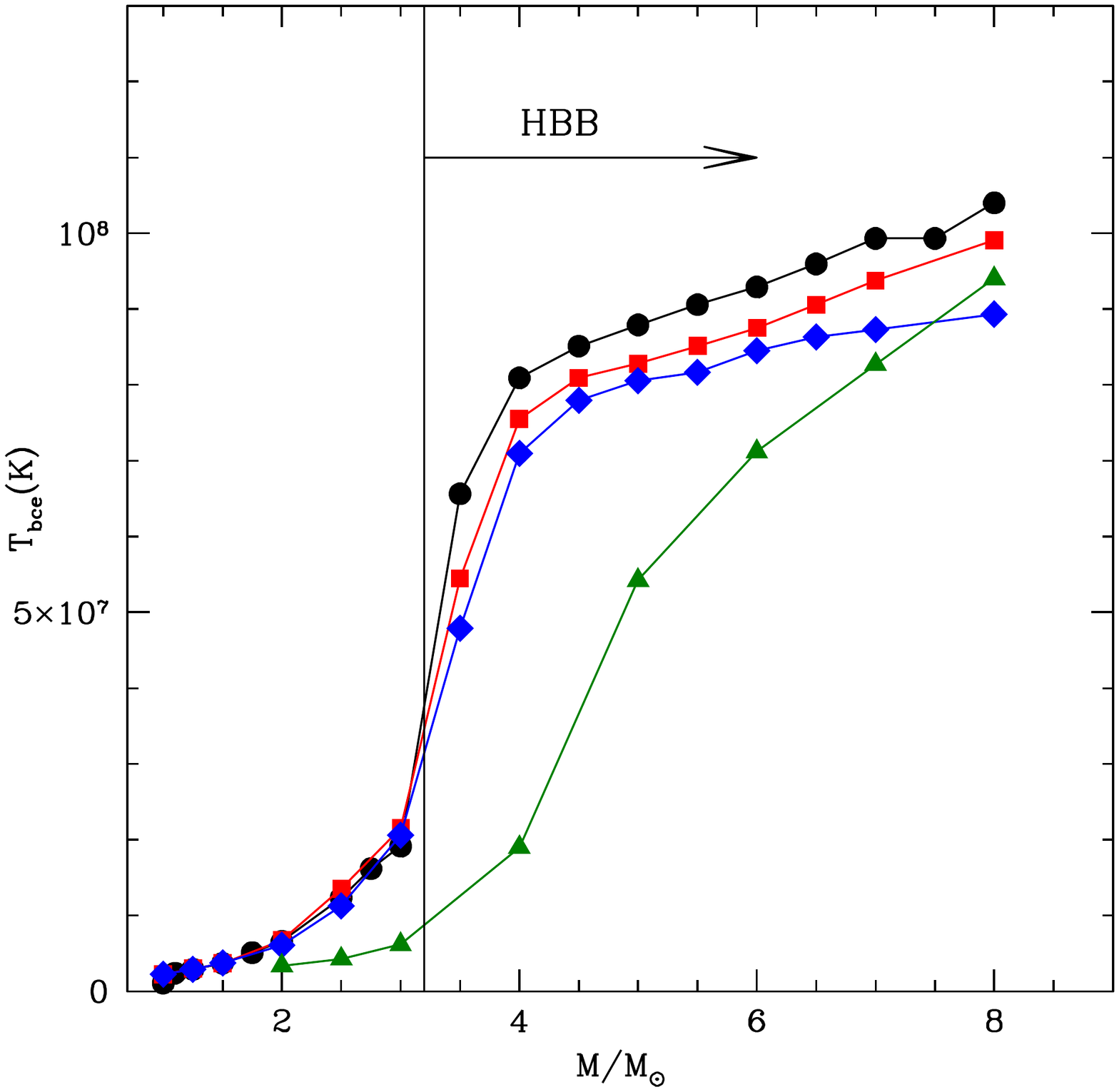}}
\end{minipage}
\vskip-80pt
\begin{minipage}{0.48\textwidth}
\resizebox{1.\hsize}{!}{\includegraphics{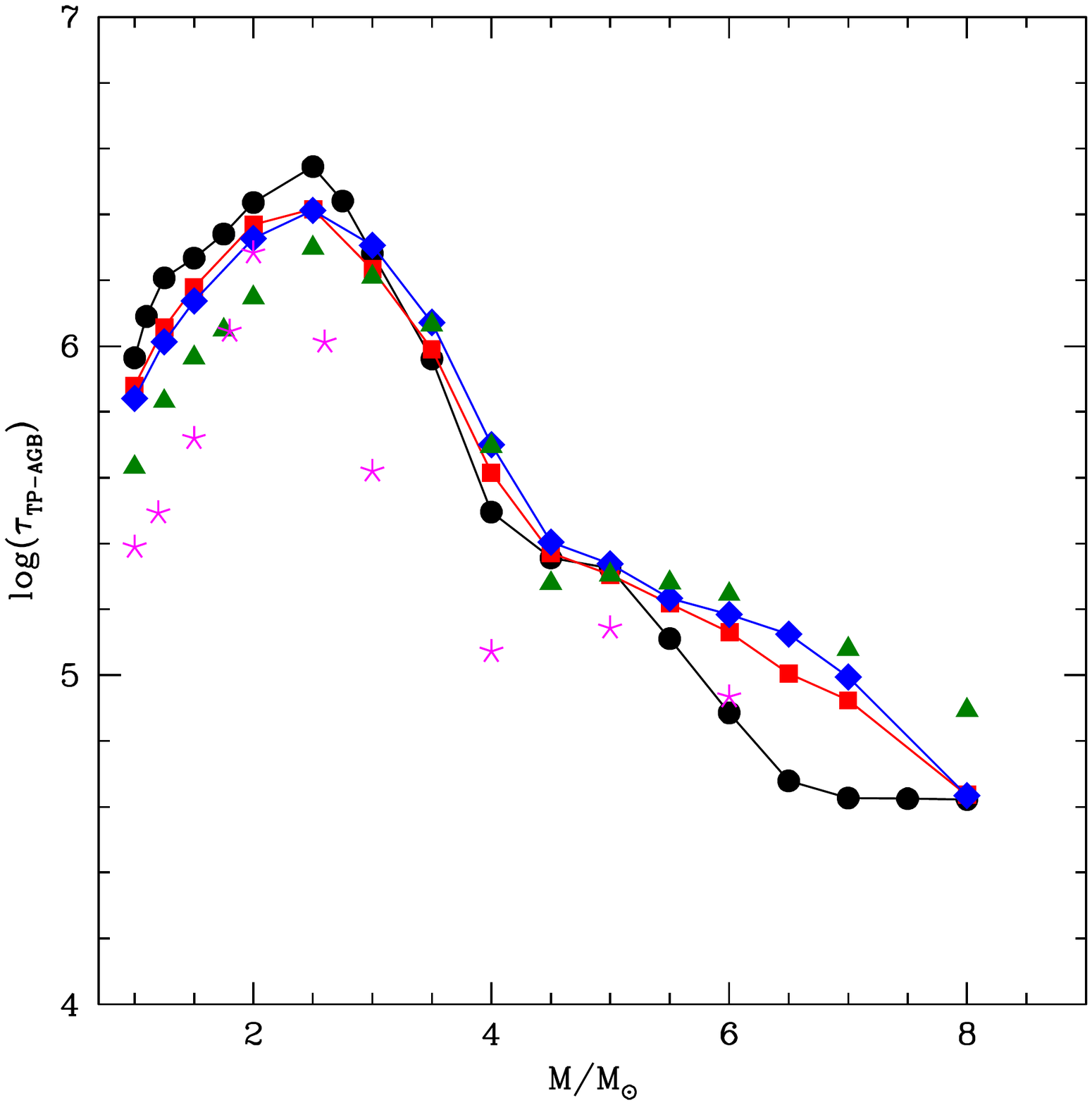}}
\end{minipage}
\begin{minipage}{0.48\textwidth}
\resizebox{1.\hsize}{!}{\includegraphics{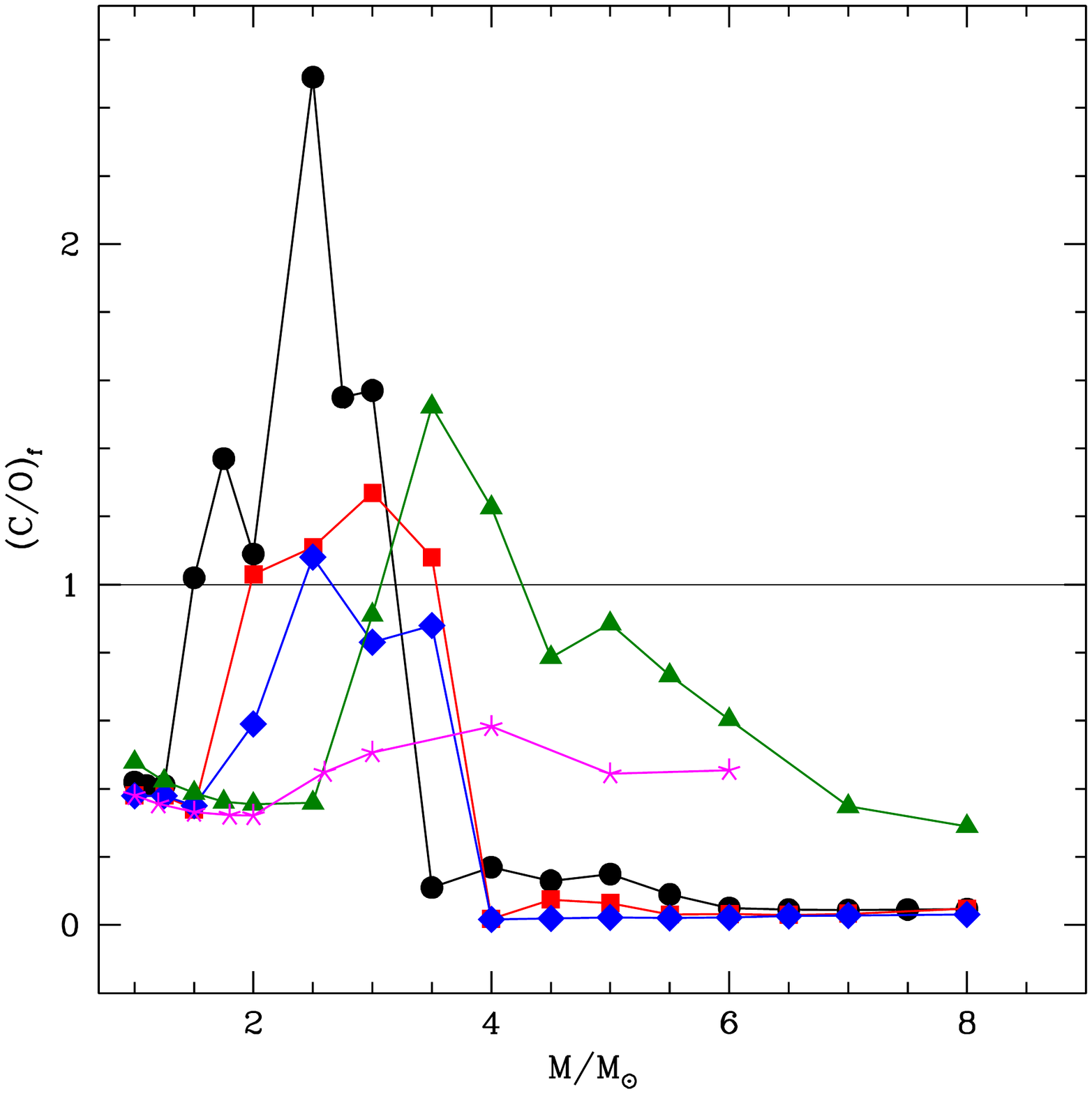}}
\end{minipage}
\vskip-60pt
\caption{The maximum luminosity (top, left panel) and temperature at the base of the envelope
(top, right) experienced by our AGB models of metallicity Z=0.03 (red squares) and Z=0.04
(blue diamonds) as a function of the initial mass. The bottom panels show the duration of
the TP-AGB phase (left) and the final C$/$O (right). For comparison we also show
solar metallicity models by \citet{ventura18} (black points), Z=0.03 models by 
\citet{karakas14a} (green triangles) and Z=0.04 models by \citet{weiss09} (magenta
asterisks).}
\label{fagb}
\end{figure*}

\section{The evolutionary phases before the AGB}
\label{preAGB}
The $Z=0.03$ and $Z=0.04$ evolutionary sequences discussed in the present investigation 
were calculated from the pre-MS phase until the almost complete ejection of the external 
mantle. A summary of the main properties of the models presented here, regarding the core 
hydrogen and helium burning phases and the AGB evolution, are reported in Table \ref{tabmod}.

Before the beginning of the AGB phase the stars experience two episodes during which the
surface convection penetrates inwards, until reaching zones of the star previously 
contaminated by nuclear activity: the first dredge-up episode takes place during the
ascending of the red giant branch, while the second dredge-up occurs after the core helium burning
phase, before the ignition of the first TP. These mixing events are extremely important,
as they determine the surface chemical composition of the star at 
the beginning of the AGB phase. While the first dredge-up changes the surface chemical
composition of all stars, the second dredge-up, in the metallicity domain investigated
here, is experienced only by $M\geq 4.5M_{\odot}$ stars. This second dredge-up event is
relevant for the subsequent evolution of the stars during the AGB phase, as is causes a
significant decrease in the core mass of the star, which will affect the main properties
of the AGB evolution.

Fig.~\ref{fdup} shows the changes in the surface mass fractions of helium and of the
most relevant CNO isotopes determined by the two dredge-up events. 

The first dredge-up favours the rise in the surface helium content, which on the average
increases by $\delta Y \sim 0.02$, although in the stars of mass close to $\sim 2~M_{\odot}$
$\delta Y$ is limited to $\sim 0.01$. A well known consequence of the first dredge-up is 
the drop in the surface mass fraction of $^{12}$C, which we find to be between
$-0.002 < \delta X(^{12}C) < -0.001$, and the rise in the nitrogen content; the latter
quantity is sensitive to the mass of the star, the percentage increase ranging from 
$\sim 50\%$, for solar mass stars, to a factor $\sim 3$, in $M\geq 3~M_{\odot}$ stars. 
The surface content of $^{16}$O decreases by less than $10\%$, the largest depletion 
taking place in the interior of $M>4~M_{\odot}$ stars. In stars of initial mass above
$3~M_{\odot}$ the base of the envelope reaches regions where sodium was produced by
$^{22}$Ne$($p,$\gamma)^{23}$Na reactions: in this mass domain the surface sodium increases
by $40-50\%$.

The extent of the changes of the mass fractions of the various species given above must
be considered as lower limits, because the present results have bee obtained by
considering only convective mixing, without considering additional effects, such
as thermohaline or rotation-induced mixing, that might trigger further changes
in the chemical composition of the surface regions \citep{marc97, charbonnel10,lagarde12}. 

In the discussion of the effects of the second dredge-up we will focus on stars of mass
above $4~M_{\odot}$, as in lower mass objects the inwards penetration of the convective
envelope following the end of the core helium burning phase is not sufficiently deep
to reach regions previously site of nuclear activity.

As discussed previously, one of the most important effect of the second dredge-up is
the decrease in the core mass, which is highly sensitive to the initial mass of the star: 
in $4.5~M_{\odot}$ stars $\delta M_C \sim -0.03~M_{\odot}$, while in 
$8~M_{\odot}$ stars $\delta M_C \sim -0.6~M_{\odot}$.

The decrease in core mass is related to the capability of the external 
mantle to penetrate inwards, past the H-He discontinuity; this process, which is made 
easier by the temporary extinction of the H-burning shell, favours the transportation of
helium-rich material towards the surface layers, which become more and more enriched in 
helium \citep{boothroyd99}. This effect is stronger the higher the mass of the 
star \citep{ventura10}:
the helium enrichment spans the range from $\delta Y=0.01$, for $M=4.5~M_{\odot}$ stars,
to $\delta Y=0.07$, for $M>6M_{\odot}$ stars.

Additional consequences of the occurrence of the second dredge-up are a further
reduction of the surface mass fractions of $^{12}$C and $^{16}$O and the increase in the
surface $^{14}$N. For all these chemical species, similar to helium, the extent of the
change in the surface content increases with the mass of the star, being almost null
in stars of mass around $4.5~M_{\odot}$ and reaching a maximum extent in $M>6M_{\odot}$ 
stars. These results are shown in Fig.~\ref{fdup}.

\begin{table*}
\caption{The properties of the stars related to the occurrence of TDU events and
the achievement (if any) of the C-star stage. The various columns report the initial
mass of the star, the core mass when the first TDU episode takes place, the core mass
and luminosity of the star when becoming a C-star, the percentage duration of the
C-star phase, relative to the whole TP life, the final carbon excess with respect to oxygen 
(defined as $12+\log((n(C)-n(O))/n(H))$, where $n(i)$ is the surface number density of
the i-th species) and the maximum $\lambda$ (see text for definition) experienced.}             
\label{tabTDU}      
\centering          
\begin{tabular}{c c c c c c c}     
\hline       
M$/$M$_{\odot}$ & M$_{\rm C}^{\rm 1TDU}/$M$_{\odot}$ & M$_{\rm C-star}/$M$_{\odot}$ & 
$L_{\rm C-star}/L_{\odot}$ & $\tau_{\rm C-star}$ & (C-O)$_f$ & $\lambda_{\rm max}$\\ 
\hline                    
 & &  & $Z=0.03$ & & & \\
\hline
1.00 &   - & - & - & - & - & - \\
1.25 & 0.57 & - & - & - & - & 0.03   \\
1.50 & 0.60 & - & - & - & - & 0.05   \\
2.00 & 0.58 & 0.65 & 1.08(4) & 1.1 & 7.63 & 0.50   \\
2.50 & 0.59 & 0.68 & 1.36(4) & 0.8 & 8.16 & 0.58   \\
3.00 & 0.60 & 0.69 & 1.44(4) & 7   & 8.56 & 0.60   \\
3.50 & 0.68 & 0.74 & 2.00(4) & 10  & 8.08 & 0.57   \\
4.00 & 0.83 & - & - & - & - & 0.25  \\
4.50 & 0.87 & - & - & - & - & 0.20  \\
5.00 & 0.89 & - & - & - & - & 0.15  \\
5.50 & 0.92 & - & - & - & - & 0.10  \\
6.00 & 0.95 & - & - & - & - & 0.05  \\
6.50 & 0.99 & - & - & - &  -  & -   \\
7.00 & -    & - & - & - &  -  & -   \\
8.00 & -    & - & - & - &  -  & -   \\
\hline
 & & & $Z=0.04$ & & & \\
\hline       
1.00 &   - & - & - & - & - & - \\
1.25 & 0.56 & - & - & - & - & 0.02    \\
1.50 & 0.59 & - & - & - & - & 0.05  \\
2.00 & 0.60 & - & - & - & - & 0.15  \\
2.50 & 0.60 & 0.69 & 1.39e4 & - & 8.23 & 0.40  \\
3.00 & 0.61 & - & - & - & - & 0.45 \\
3.50 & 0.66 & - & - & - & - & 0.48  \\
4.00 & 0.81 & - & - & - & - & 0.15  \\
4.50 & 0.86 & - & - & - & - & 0.10   \\
5.00 & 0.89 & - & - & - & - & 0.07  \\
5.50 & 0.91 & - & - & - & - & 0.05  \\
6.00 & 0.93 & - & - & - & - & 0.05  \\
6.50 & 0.94 & - & - & - & - & 0.04  \\
7.00 & -    & - & - & - &  -  & -  \\
8.00 & -    & - & - & - &  -  & -  \\
\hline                  
\end{tabular}
\end{table*}

\section{The AGB phase of metal-rich stars}
\label{evol}
After the end of the core helium burning phase the stars of mass mass $M\leq 8~M_{\odot}$ 
develop a core composed of carbon and oxygen, which evolves under conditions of electron 
degeneracy. The only exceptions to this general behaviour are the stars of mass 
$M\geq 7~M_{\odot}$, that experience off-center carbon burning and develop a convective 
flame that moves inwards, forming a zone enriched in oxygen and neon close to the stellar 
centre. The subsequent evolution is commonly referred to as "super-AGB" 
(e.g., Garcia-Berro \& Iben 1994, Siess 2007).

The energy supply is provided for $\sim 95\%$ of the time by a CNO burning shell;
periodically a helium-rich region just above the core is ignited in conditions of
thermal instability \citep{schw65}, which explains the use of the terminology 
"thermal pulse" to refer to these episodes. The inter-shell zone between the He-rich 
buffer where the TPs
develop and the CNO burning shell is composed essentially by helium, carbon, oxygen and
neon, with mass fractions (He,C,O,Ne)$\sim (55\%,38\%,4\%,3\%)$. This distribution,
which is relevant for the s-process nucleosynthesis and the relative surface enrichment,
is extremely sensitive to the treatment of convective borders and to the assumption
of extra-mixing from the base of the pulse driven convective shell. 

The core mass is the key factor of the AGB evolution: the higher the core mass
the higher the degree of degeneracy in the internal regions, which reflects into 
larger pressures and temperatures within the CNO burning shell, which supports these
structures on the energetic side for most of the time. This is the reason of the tight 
relationship between core mass and luminosity of AGB stars, initially predicted in a 
classic paper by \citet{pac70}. The core mass -luminosity trend for some of the stars
considered here is shown in Fig.~\ref{fmcl}; the evolutionary sequences in this plane
are similar to those published in Karakas \& Lattanzio (2014, see their figure 18).

The top panels of Fig.~\ref{fagb} shows the highest luminosity and temperature at the base 
of the envelope attained during the AGB phase. To understand the trend with metallicity we 
also show the solar metallicity models published in \citet{ventura18}.
These plots highlight the difference between the
evolution of the stars with initial mass $M > 3~M_{\odot}$ and their lower-mass
counterparts, which evolve to much lower luminosities and temperatures.
The sudden drop of T$_{\rm bce}$ occurring around $3~M_{\odot}$ is clearly visible. 
This behaviour is related to the ignition of HBB,
which requires core masses of the order of $\sim 0.8~M_{\odot}$ \citep{ventura13} and 
strongly affects the surface chemistry. The results 
reported in the bottom, right panel of Fig.~\ref{fagb} show the significant effect of HBB 
on the final C$/$O surface ratio, defined as the ratio of the sum of the surface 
number densities of the carbon isotopes, divided by the sum of the number densities of 
oxygen isotopes: at the base of the convective envelope $^{12}$C nuclei are
exposed to p-capture reactions, which decrease the surface carbon content
and the C$/$O ratio.

The ignition of HBB has also profound effects on the 
physical evolution of AGB stars: it results in a fast rise of the luminosity
\citep{blocker91}, with significant deviations (see Fig.~\ref{fmcl}) from the core mass - 
luminosity relationship predicted by \citet{pac70}. This is accompanied by the
increase in the rate of mass loss, which 
shortens the duration of the AGB phase. This is shown in the bottom, left panel of 
Fig.~\ref{fagb}, where the time scale of the AGB phase of 
$M > 3~M_{\odot}$ stars is on average more than one order of magnitude shorter than for
the lower masses. Specifically, in the low-mass domain the behaviour of the duration of 
the TP-AGB phase, $\tau_{\rm AGB}$, is not monotonically related to the initial mass of 
the star. In the $1-2.5~M_{\odot}$ range $\tau_{\rm AGB}$ increases with the initial mass, 
because the higher the mass, the larger the number of TPs experienced before the entire 
envelope is lost, which makes the AGB phase longer (see bottom, left panel of Fig.~\ref{fagb}). 
For stars of higher mass the most relevant point to $\tau_{\rm AGB}$ is that the core mass, 
hence the luminosities, are larger, which leads to shorter time scales. 
In the low-mass domain the stars evolving faster along the AGB phase, with 
$\tau_{\rm AGB} \sim 1$ Myr, are the $1~M_{\odot}$ and $3.5~M_{\odot}$ stars; the stars 
with the longest AGB life, of the order of $2.5$ Myr, are those with initial mass 
$2.5~M_{\odot}$ (see Column 5 of Table \ref{tabmod}).

The comparison between the present models and those by \citet{ventura13, ventura14a, 
ventura18} indicates that the threshold mass at which the ignition of HBB occurs is in the 
range $3-3.5~M_{\odot}$ for all the metallicities $Z\geq 4\times 10^{-3}$.
The trends with metallicity illustrated in Fig.~\ref{fagb} reflect a well established 
property of AGB stars: the lower the metallicity, the larger is the temperatures at the 
base of the envelope ($T_{\rm bce}$) and, as a consequence, the luminosity \citep{ventura13}. 
This is particularly important in the $M > 3~M_{\odot}$ domain, where the degree of the 
p-capture nucleosynthesis due to HBB is extremely sensitive to 
$T_{\rm bce}$ \citep{flavia18}.
The metallicity effect on the evolutionary time scales for the stars
experiencing HBB is straightforward: the higher the metallicity, the lower is the
luminosity, and the longer the duration of the AGB phase. In the low-mass domain
the behaviour is the opposite: stars of higher metallicity 
evolve at lower effective temperatures and larger radii, thus experience higher rates
of mass loss, which shorten the AGB phase.

\begin{figure*}
\begin{minipage}{0.48\textwidth}
\resizebox{1.\hsize}{!}{\includegraphics{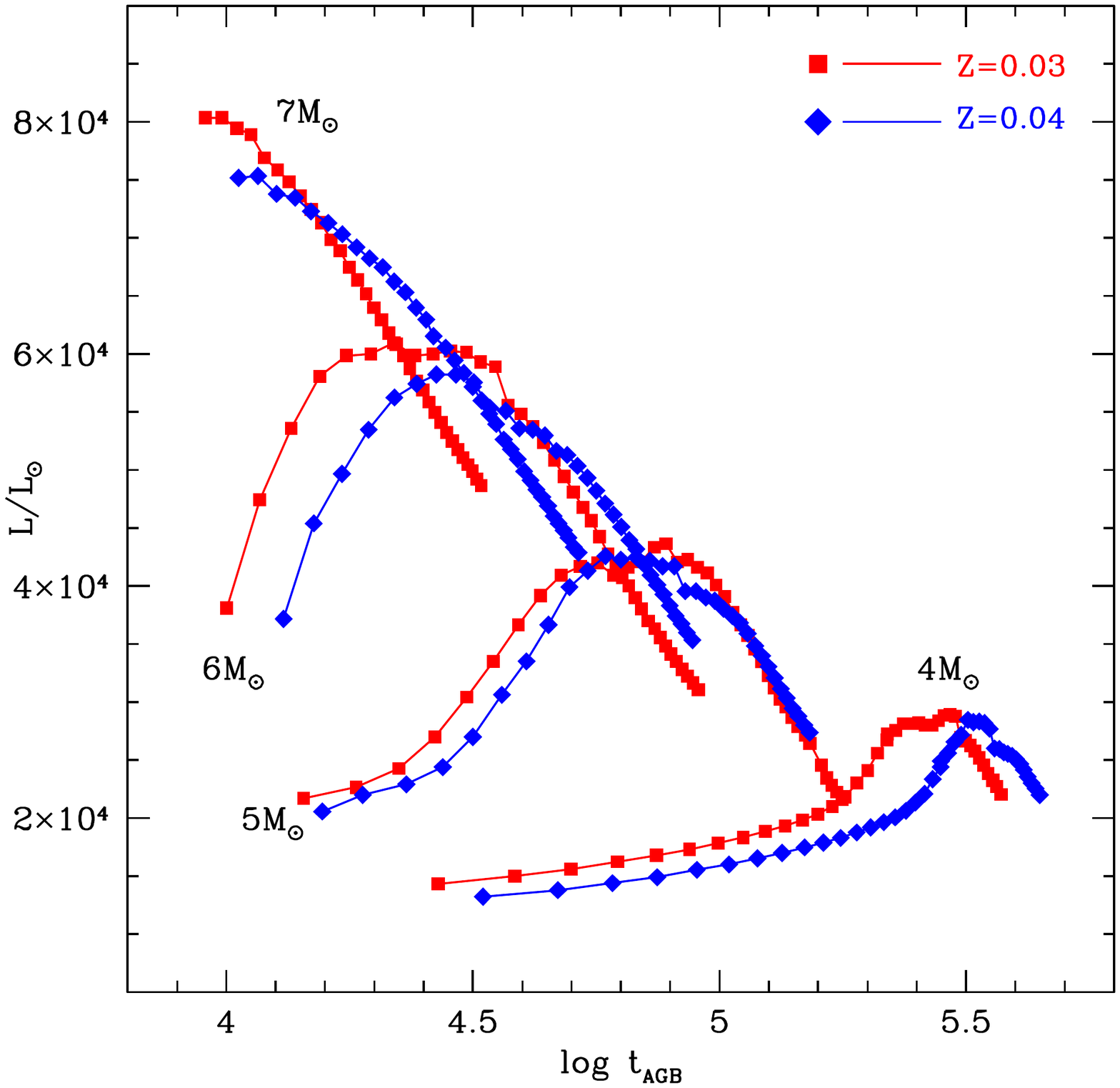}}
\end{minipage}
\begin{minipage}{0.48\textwidth}
\resizebox{1.\hsize}{!}{\includegraphics{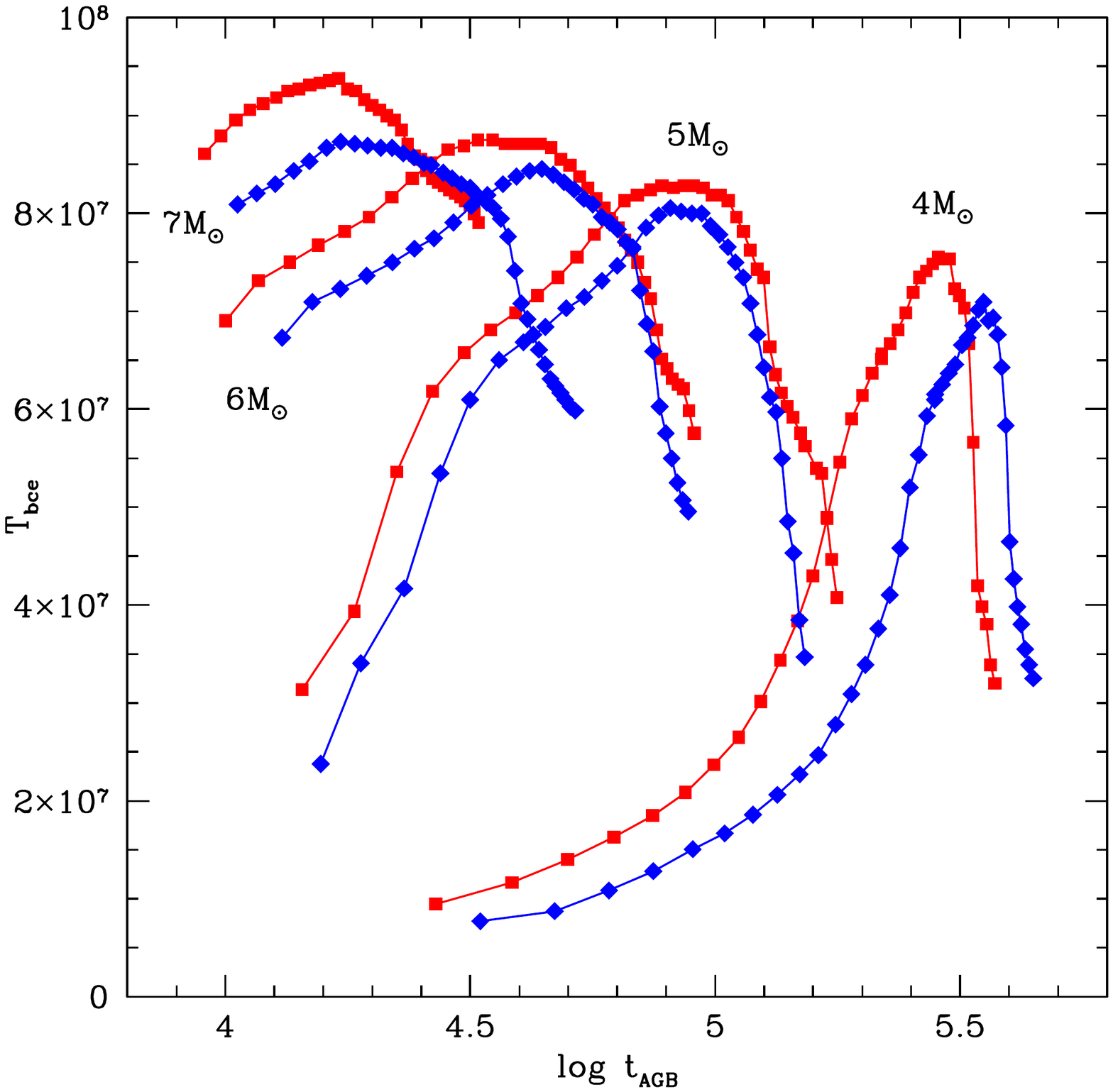}}
\end{minipage}
\vskip-80pt
\begin{minipage}{0.48\textwidth}
\resizebox{1.\hsize}{!}{\includegraphics{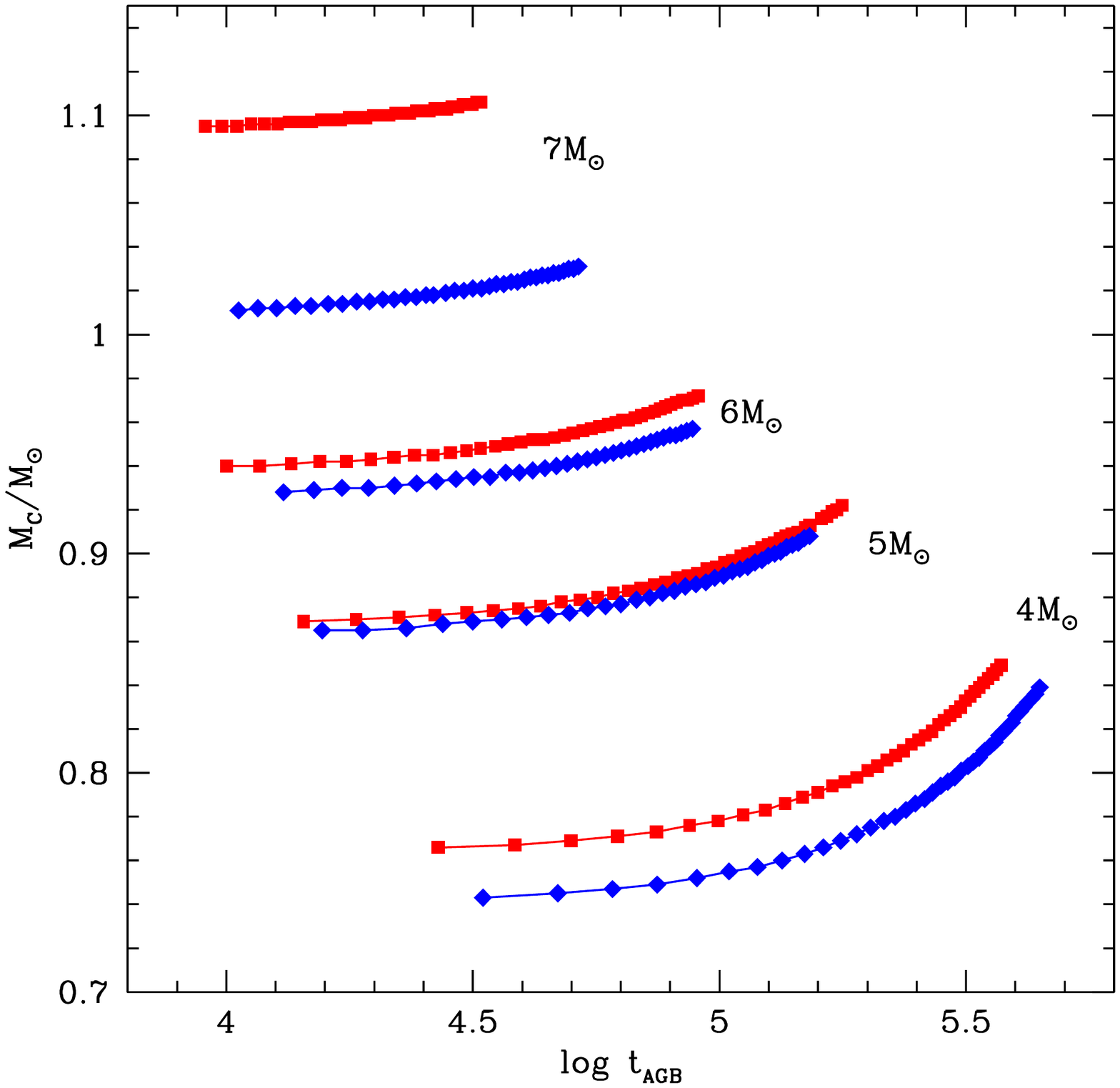}}
\end{minipage}
\begin{minipage}{0.48\textwidth}
\resizebox{1.\hsize}{!}{\includegraphics{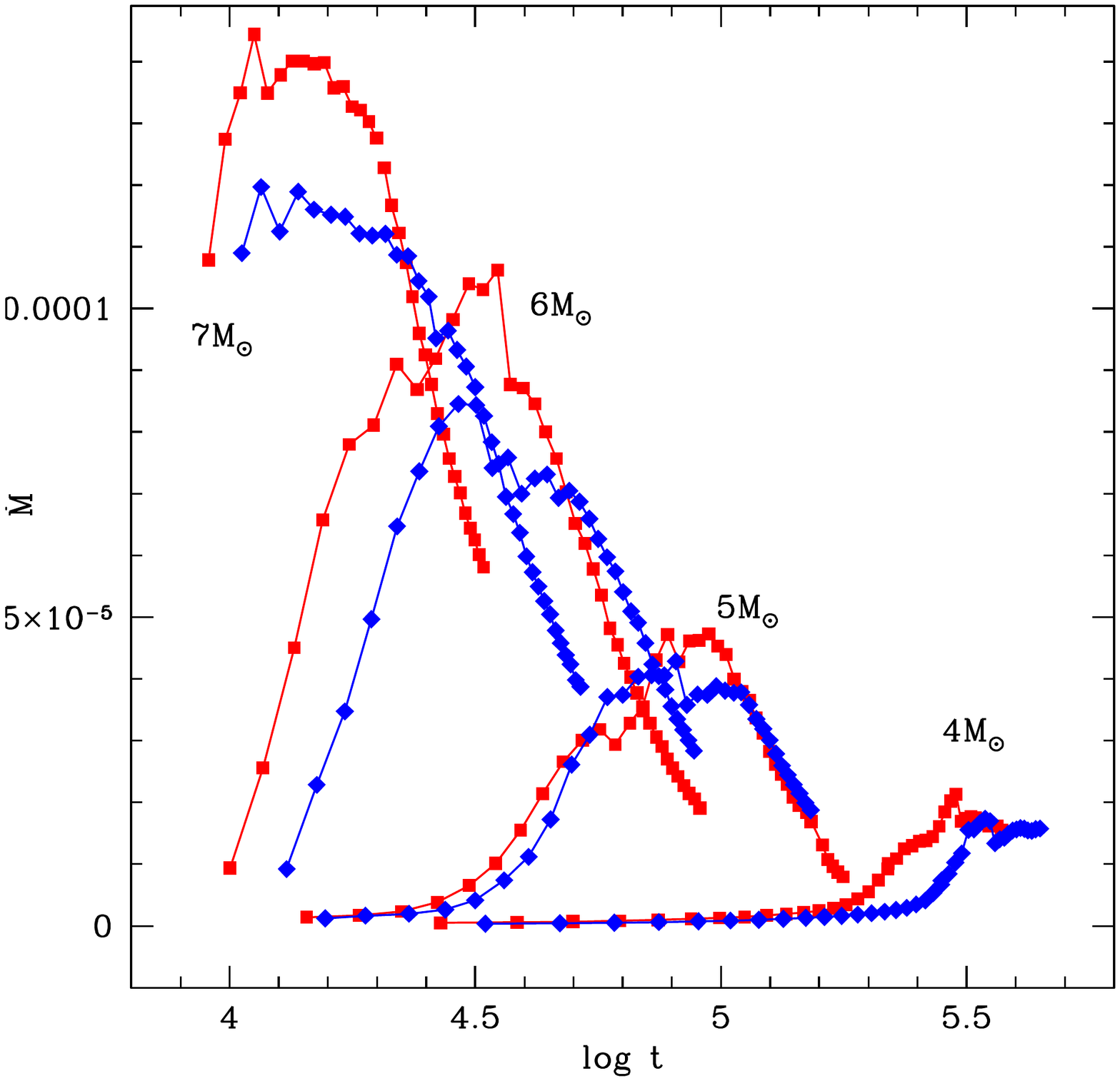}}
\end{minipage}
\vskip-60pt
\caption{The variation with time (starting from the beginning of the TP-AGB phase)
of the luminosity (top, left panel), temperature at the base of the envelope (top, 
right panel), core mass (bottom, left) and mass loss rate (bottom, right) during the 
AGB phase of stars of metallicity $Z=0.03$ (red squares) and $Z=0.04$ (blue diamonds) 
of initial mass $4, 5, 6, 7~M_{\odot}$. The points along the different tracks 
refer to the middle of the inter-pulse phase.}
\label{f1phys}
\end{figure*}

Besides HBB the surface chemical composition of AGB stars can be modified by the
occurence of third dredge-up (hereinafter TDU), which consists in the inwards
penetration of the convective envelope, taking place after each thermal pulse (TP) of the 
He burning shell, down to layers previously affected by nucleosynthesis via the triple-alpha 
reactions, thus enriched in carbon \citep{iben74}. The alteration determined by TDU is
significantly different than HBB: in this case the main effect is the gradual increase in
the surface carbon, which can lead to the formation of a carbon star, once the number
density of carbon atoms exceeds the number density of oxygen. The efficiency of each TDU 
event is commonly described by $\lambda$, defined as the ratio between the decrease in the
core mass favoured by TDU and the increase in the core mass occurred since the previous
TP, during the quiescent CNO burning phase.

In Table \ref{tabTDU} we report for each stellar mass and for the two metallicities
discussed here, the relevant information describing the occurrence and the effects of
TDU, i.e. the core masses of the star when the first TDU takes place, the core mass and
luminosity when the C-star stage is reached, the duration of the C-star phase and the 
maximum $\lambda$ experienced.

Given the significant difference in their evolution, we describe separately the stars in 
the low-mass domain ($M \leq 3.5~M_{\odot}$) and the higher-mass stars that experience HBB.
 
\subsection{The evolution of massive AGB stars}
\label{highm}
The variation with time of the main physical properties of metal-rich, massive AGB 
stars are shown in Fig.~\ref{f1phys}. These stars experience an initial phase during 
which the luminosity and T$_{\rm bce}$ increase (see top panels of Fig.~\ref{f1phys}), 
owing to the gradual rise in the mass of the core (bottom, left panel of Fig.~\ref{f1phys}); 
both quantities decline in the final AGB phases, after a significant fraction 
of the envelope is lost. The mass loss rate, shown in the bottom, right panel of
Fig.~\ref{f1phys}, follows a similar trend, considering the
link between mass loss and luminosity, which is expected on general grounds, and is
particularly tight according to the \citet{blocker95} treatment of mass loss adopted in
the present investigation. The effective temperature $T_{\rm eff}$ of the $Z=0.03$ stars
decreases trend during the AGB life, with $T_{\rm eff} \sim 3500$ K during
the initial TPs, and $T_{\rm eff} \sim 2300$ K during the final AGB phases; this behaviour
is substantially independent of the initial mass of the star. $Z=0.04$ stars follow
a similar decreasing trend, the difference with respect to the $Z=0.03$ case being that
the effective temperatures are $\sim 100$ K cooler.

\begin{figure*}
\begin{minipage}{0.48\textwidth}
\resizebox{1.\hsize}{!}{\includegraphics{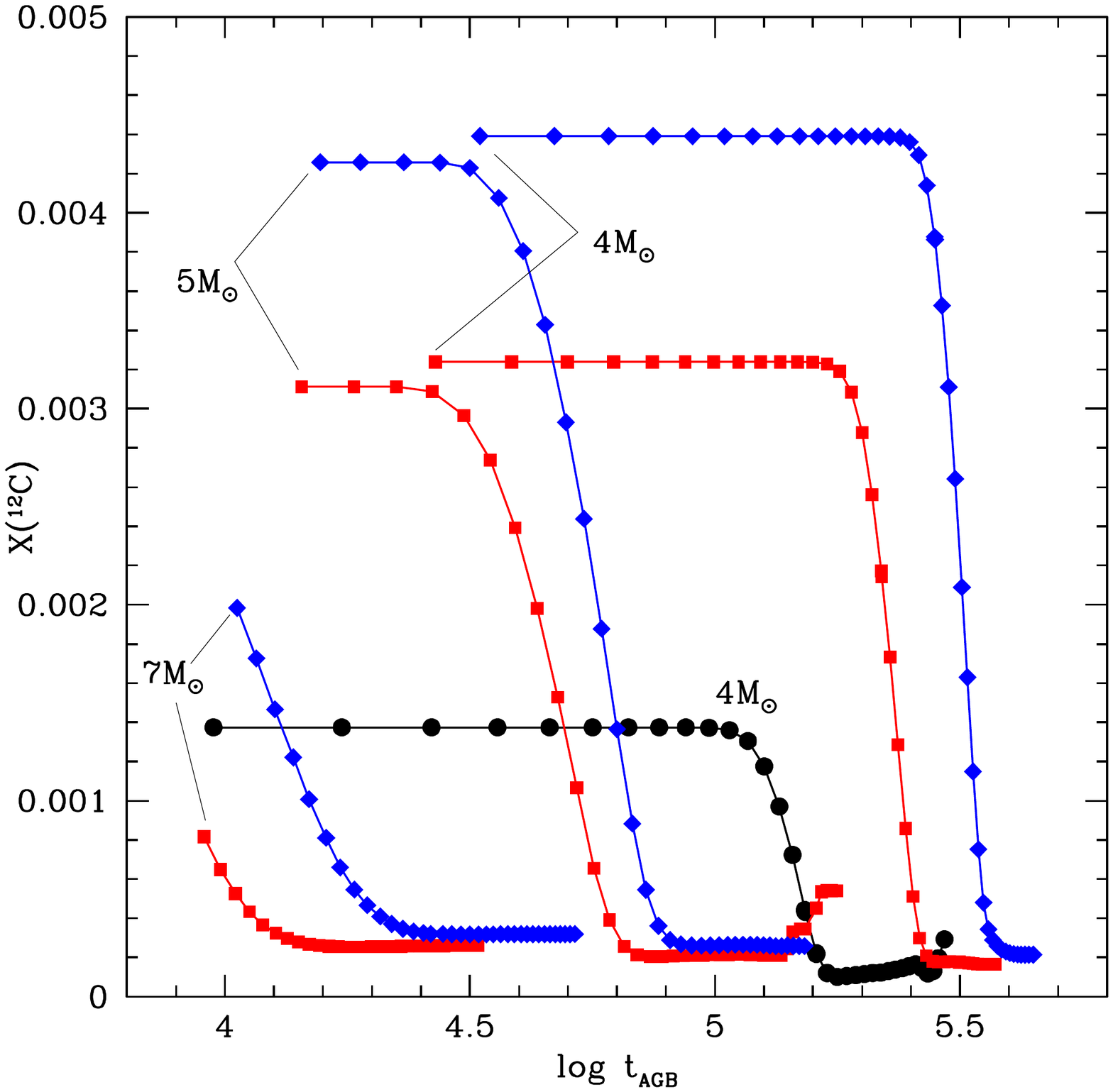}}
\end{minipage}
\begin{minipage}{0.48\textwidth}
\resizebox{1.\hsize}{!}{\includegraphics{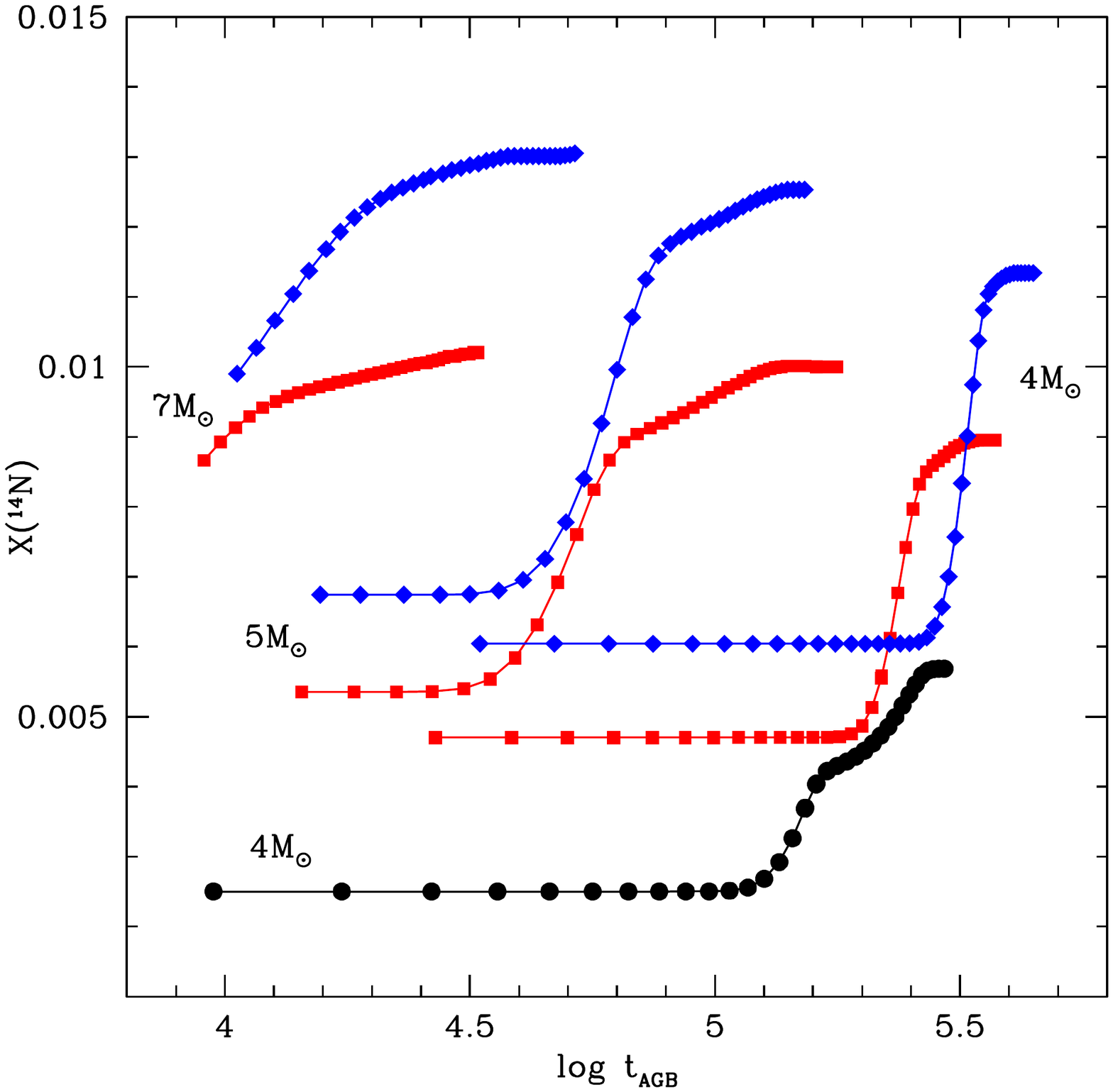}}
\end{minipage}
\vskip-80pt
\begin{minipage}{0.48\textwidth}
\resizebox{1.\hsize}{!}{\includegraphics{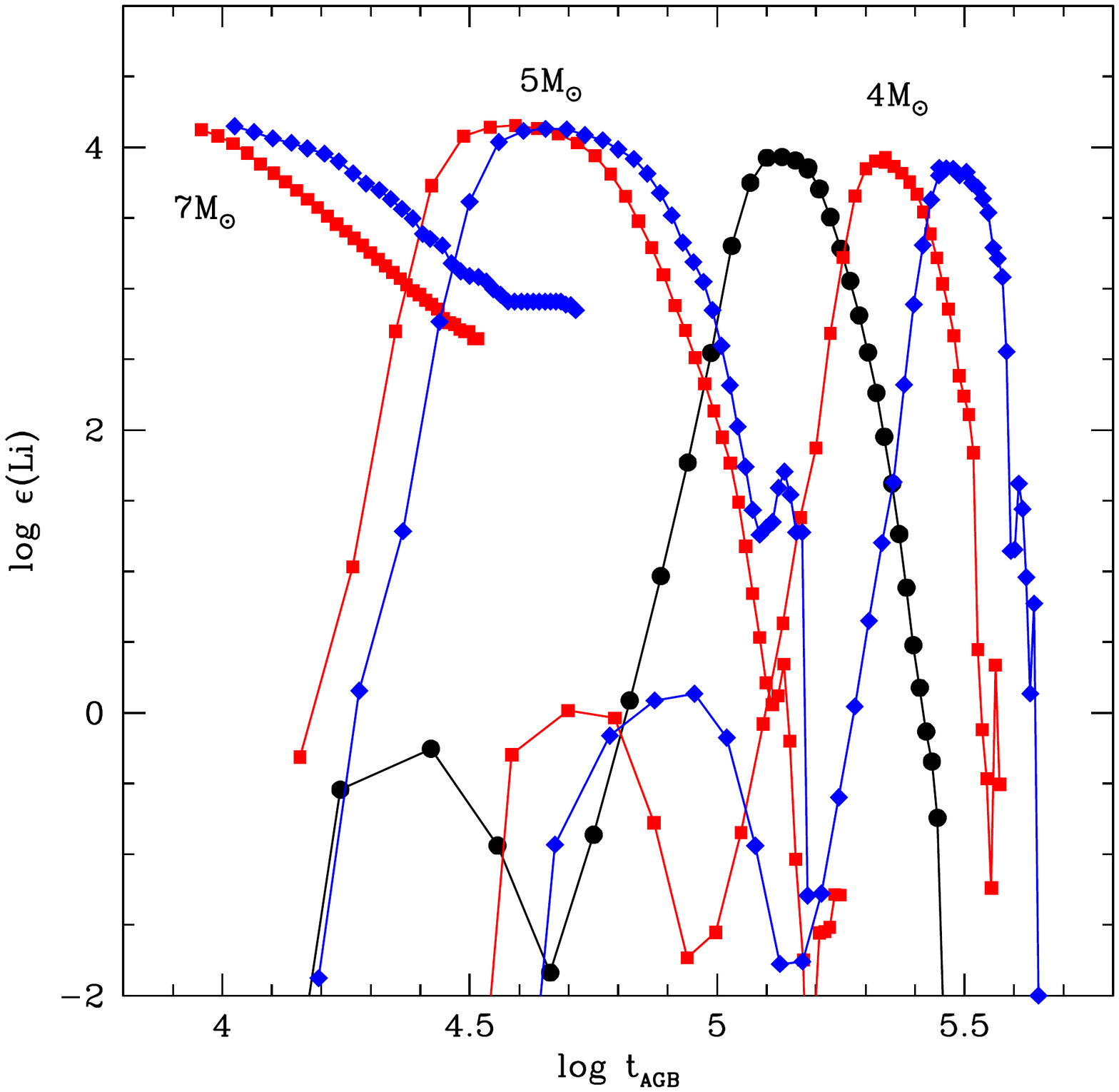}}
\end{minipage}
\begin{minipage}{0.48\textwidth}
\resizebox{1.\hsize}{!}{\includegraphics{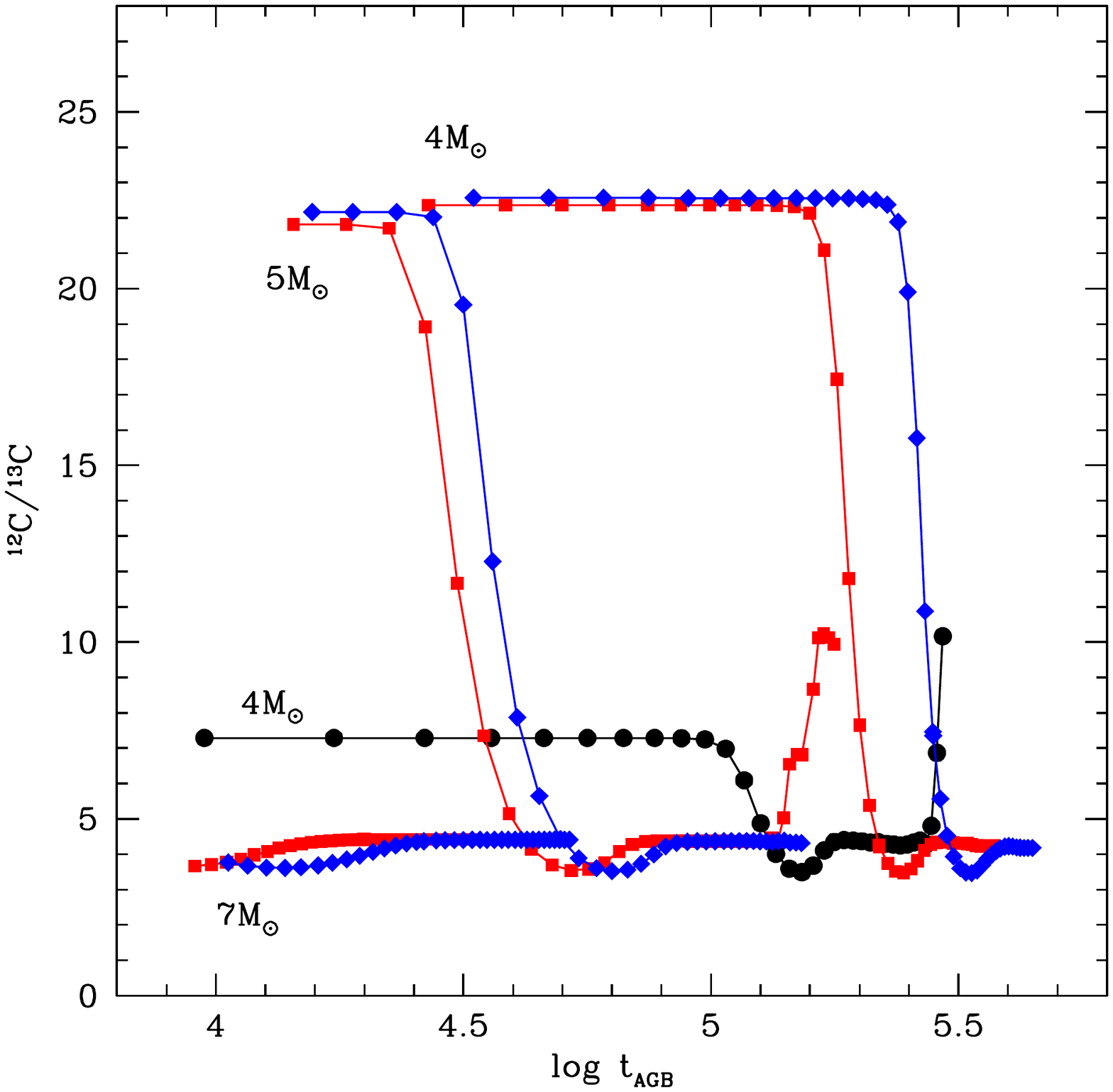}}
\end{minipage}
\vskip-60pt
\caption{The variations with time of the surface mass fraction of $^{12}$C
(top, left panel), $^{14}$N (top, right), lithium (bottom, left) and
the $^{12}$C/$^{13}$C ratio (bottom, right) of the same models shown in Fig.~\ref{f1phys}
(except for the $6~M_{\odot}$ model omitted here for sake of readibility).
The black dots refer to a $4~M_{\odot}$ star of solar metallicity,
published in \citet{ventura18}, while the other symbols are the same as
Fig.~\ref{f1phys}. }
\label{f1chem}
\end{figure*}

Both the average and the peak luminosity (the latter in the range 
$20-80 \times 10^3~L_{\odot}$) are strongly sensitive to the initial stellar mass.
The differences in the luminosity among stars of different mass reflect on the duration of 
the AGB phase, which varies from $\sim 5\times 10^4$ yr, for $8~M_{\odot}$ stars, to 
$\sim 3\times 10^5$ yr, for $4~M_{\odot}$ stars. The luminosities and temperatures reached 
by the $Z=0.03$ stars are generally higher than their $Z=0.04$ counterparts, the differences 
being below $\sim 15\%$ in all cases.

Fig.~\ref{f1chem} shows the variation of the surface chemical composition of the
stars reported in Fig.~\ref{f1phys}. The alteration of the surface chemistry is
mostly due to HBB, because the efficiency of TDU is low in this mass domain. On general
grounds, this behaviour is associated to the strength of TPs, which is weaker than in the 
lower-mass counterparts; a further effect, particularly relevant when convection is
described by means of the FST, is related to the strong HBB experienced, which 
accelerates the AGB evolution, in such a way that the external mantle is lost before 
deep TDU events take place. As reported in Table \ref{tabTDU}, the TDU efficiency is 
$\lambda < 0.2$ for the stars experiencing HBB. A similar behaviour was found in
the solar metallicity models discussed in \citet{ventura18}.

The ignition of HBB triggers the reduction of the surface carbon:  
the final surface $^{12}$C is depleted by a factor 15 (20) in $Z=0.03$ ($Z=0.04$) 
stars (see top, left panel of Fig.~\ref{f1chem}); the final $^{12}$C$/^{13}$C 
reaches the equilibrium value of $\sim 4$ in all cases. This is consistent with the
results shown in the bottom, right panel of Fig.~\ref{fagb}, where we see that 
the final C$/$O is below $\sim 0.1$ in the high-mass domain.
The initial mass affects the timing when the modification of the surface chemistry occurs:
while in the stars of higher mass carbon depletion starts since the very first 
thermal pulses, in the $4~M_{\odot}$ case this takes place only after the star has
experienced several TPs. This is consistent with the evolution of the $T_{\rm bce}$
of stars of different mass, shown in the right panel of Fig.~\ref{f1phys}.

In this metallicity domain the surface oxygen is only scarcely affected by HBB.
This is consistent with the results by \citet{ventura18}, who found 
that oxygen destruction is negligible in massive AGB stars of solar metallicity.
Therefore, the nucleosynthesis experienced at the base of the envelope is essentially  
pure CN cycling, which results in a direct relationship between the number of $^{12}$C 
nuclei destroyed and the number of $^{14}$N synthesized (see top, right panel of 
Fig.~\ref{f1chem}). The nitrogen abundance is higher in $Z=0.04$ stars compared to 
$Z=0.03$, due to the larger initial abundance of $^{12}$C. 

While HBB is potentially able to activate all the p-capture channels, until the
synthesis of silicon, we find that in the stars discussed here the Mg-Al chain
is not activated; this is related to the temperatures attained at the bottom of the
convective envelope (see Fig.~\ref{f1phys}), below $10^8$ K, the threshold required
to ignite the fist reaction of the chain, i.e. proton capture by $^{24}$Mg nuclei. This
is consistent with the analysis by \citet{flavia18}, focused on the temperatures at the
base of the external mantle required to activate the different nuclear channels, and 
the largest degree of nucleosynthesis activated in massive AGB stars of different
metallicity.

The temperatures at the base of the convective envelope
(Fig.~\ref{f1phys}) are above $40$ MK, which is the threshold required to 
activate the series of reactions that leads to the synthesis of lithium \citep{cameron}.
As shown in the bottom, left panel of Fig.~\ref{f1chem}, the stars evolve
as lithium-rich for a significant fraction of the AGB phase; this is 
different from the behaviour of their lower metallicity counterparts, where the survival
of lithium in the surface regions, particularly in the most massive stars, is limited to 
a very few TPs \citep{ventura00}. The long duration of the Li-rich phase is due to the
trend with metallicity of the temperatures at the bottom of the envelope, discussed above:
in metal-rich stars the T$_{\rm bce}$ are lower than in solar and sub-solar AGBs,
which allows a longer survival of $^3$He, the essential ingredient to lithium production
\citep{sackmann92}. While lithium production by massive AGB stars is larger in the
high metallicity domain, the overall lithium expelled by AGB stars is not sufficient to
explain the lithium vs. metallicity trend exhibited by Milky Way stars \citep{romano01},
which suggests that other sources must be responsible for the lithium enrichment of the
interstellar medium.

\begin{figure*}
\begin{minipage}{0.48\textwidth}
\resizebox{1.\hsize}{!}{\includegraphics{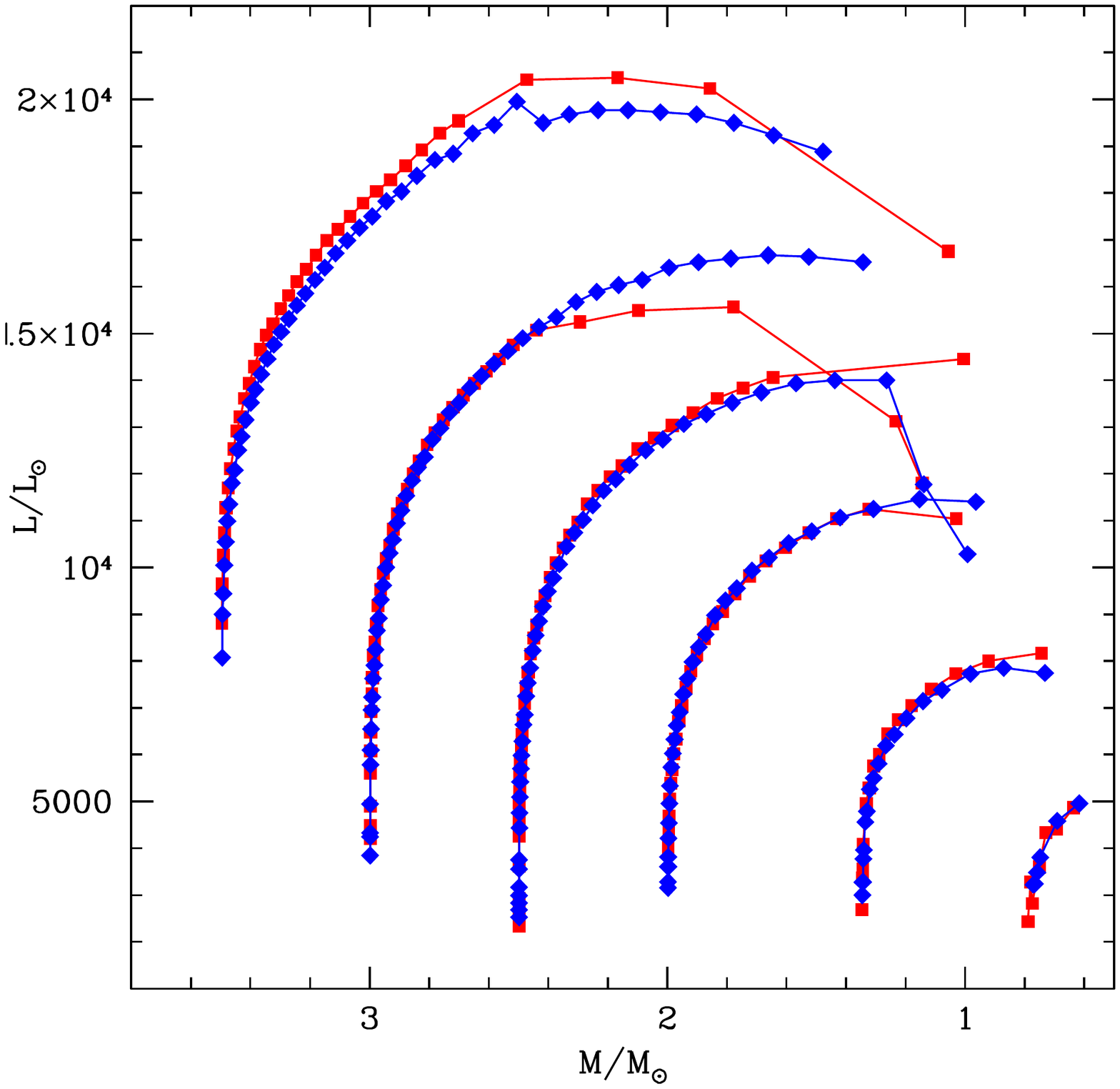}}
\end{minipage}
\begin{minipage}{0.48\textwidth}
\resizebox{1.\hsize}{!}{\includegraphics{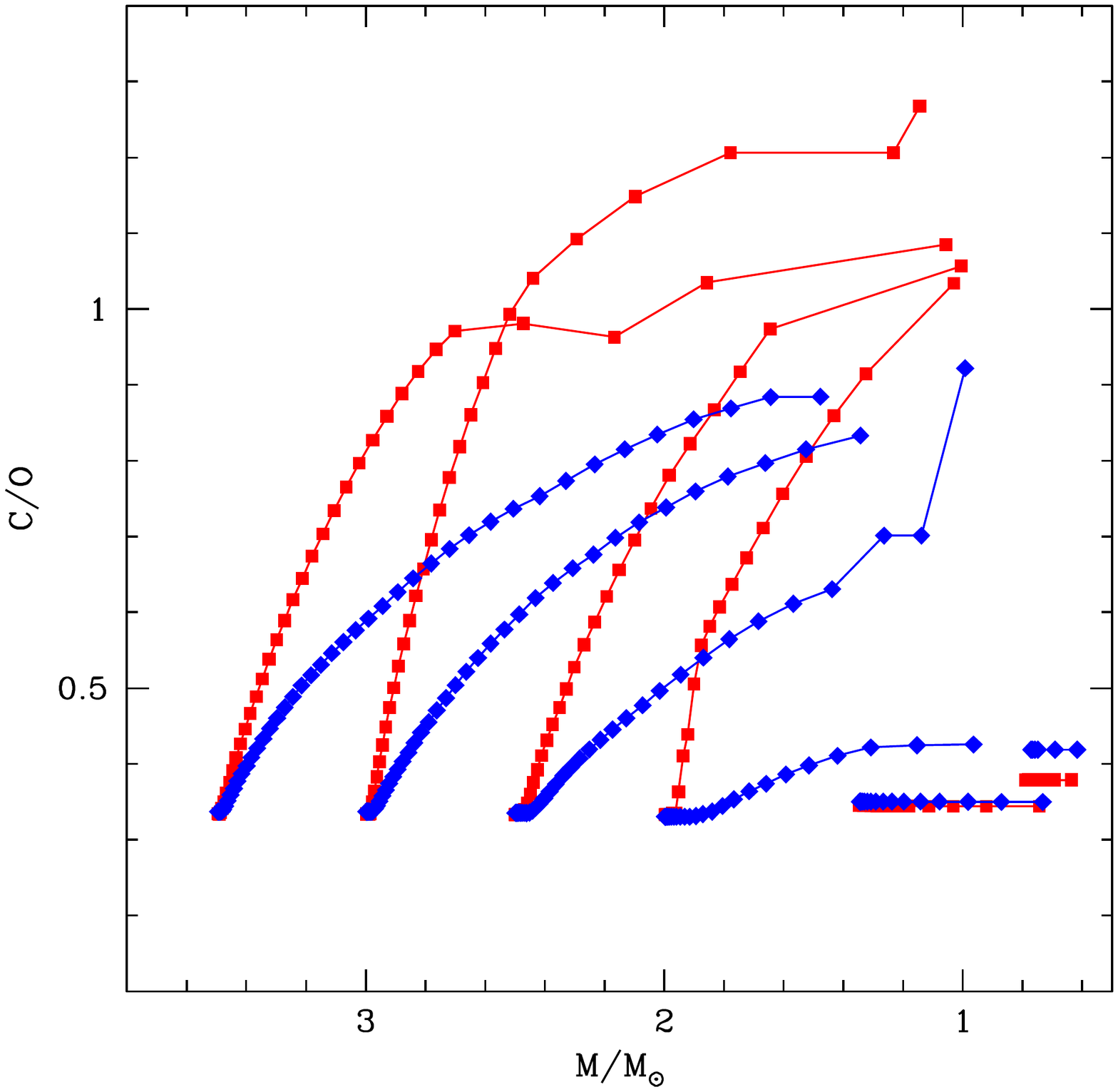}}
\end{minipage}
\vskip-60pt
\caption{The variation with the current mass of the star
of the luminosity (left panel) and surface C$/$O (right panel) of $Z=0.03$ (red squares) 
and $Z=0.04$ (blue diamonds) stars with 
initial mass in the range $1-3.5~M_{\odot}$.}
\label{flowm}
\end{figure*}

\subsection{Low mass AGB stars}
\label{lowm}
Stars of initial mass $M\leq 3~M_{\odot}$ do not experience HBB. The variation
of their surface chemical composition during the AGB life is only driven by TDU.
In this section we include stars of initial mass $3.5~M_{\odot}$, because they experience 
such weak HBB that the TDU plays the dominant role in the modification of their surface 
chemistry.

\begin{figure*}
\begin{minipage}{0.48\textwidth}
\resizebox{1.\hsize}{!}{\includegraphics{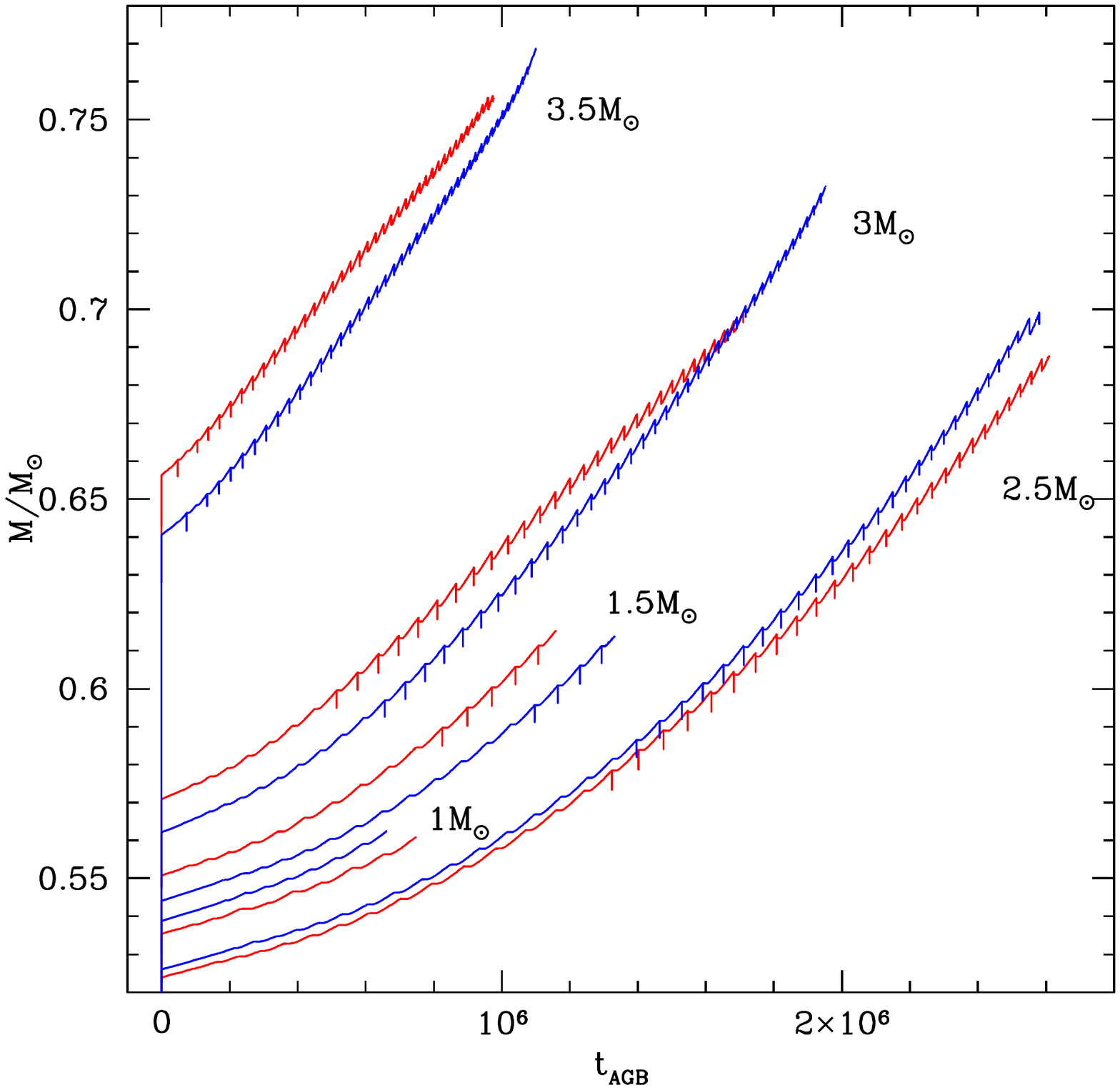}}
\end{minipage}
\begin{minipage}{0.48\textwidth}
\resizebox{1.\hsize}{!}{\includegraphics{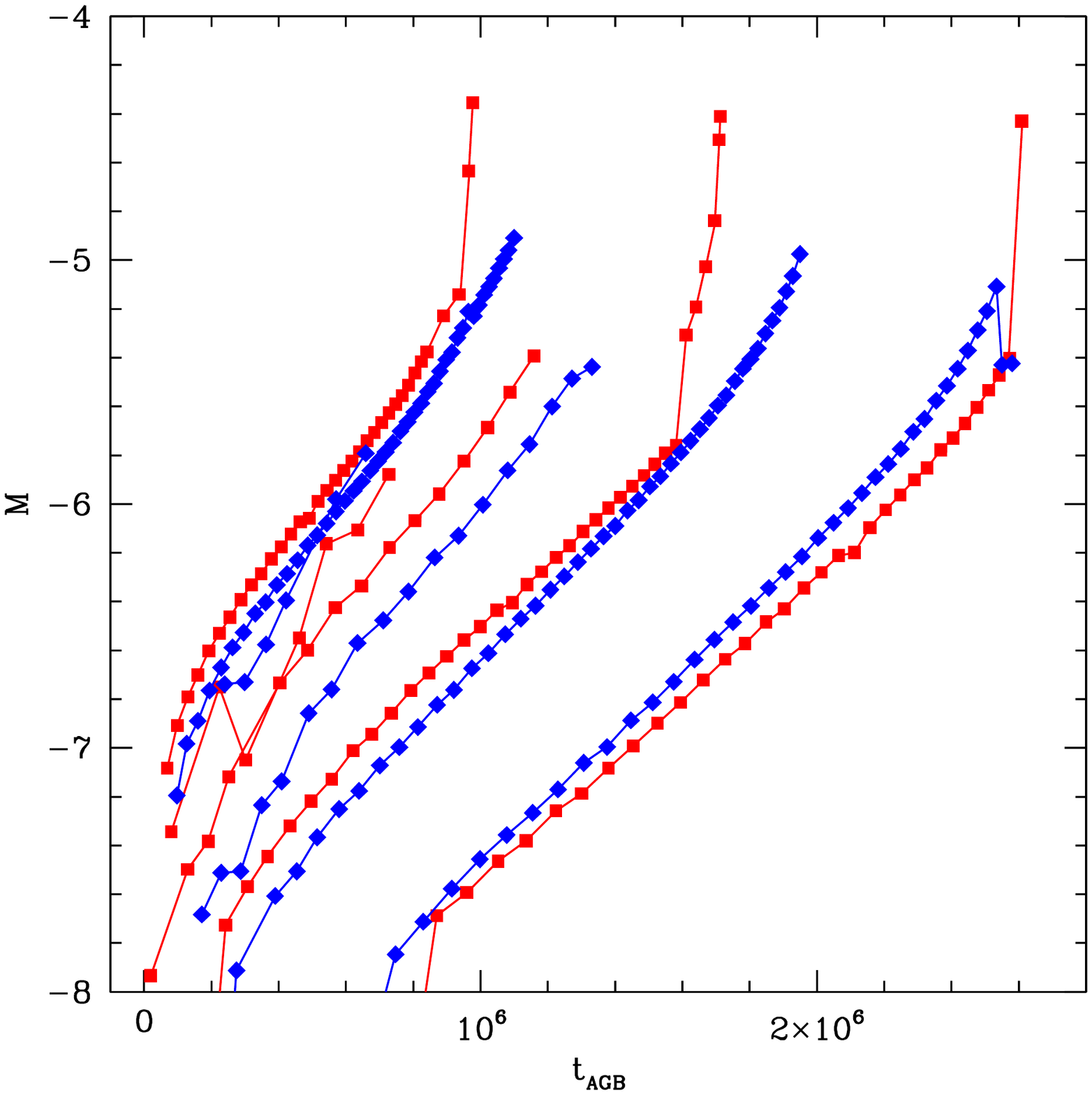}}
\end{minipage}
\vskip-60pt
\caption{The variation with time (counted since the beginning of the TP phase)
of the core mass (left panel) and rate of mass loss (right panel) of Z=0.03 and
Z=0.04 models of initial mass $1~M_{\odot}$, $1.5~M_{\odot}$, $2.5~M_{\odot}$,
$3~M_{\odot}$ and $3.5~M_{\odot}$.}
\label{flowmmdot}
\end{figure*}

Fig.~\ref{flowm} shows the variation of the luminosity and of the surface C$/$O ratio as 
a function of the current mass of the star. We report the evolution of all the stars with 
initial mass in the range $1-3.5~M_{\odot}$, with the exception of the $1.25~M_{\odot}$ 
model, omitted for readability. The variation with time of the core mass and the rate
of mass loss of AGB stars beloning to the low-mass domain are shown in Fig.~\ref{flowmmdot}.
The gradual increase in the luminosity is determined by 
the growth of the core mass. The maximum luminosity reached correlates 
with the initial mass of the star; also, the larger is the initial mass the higher the 
number of TPs experienced before the envelope is lost, and the larger is the final core 
mass. The efficiency of TDU increases as more and more TPs are experienced; 
consequently, the stars undergoing the deepest TDU events are those of initial mass
around $3~M_{\odot}$, for which $\lambda \sim 0.6$ during the final part of the AGB phase
(see Table \ref{tabTDU}). This can be seen in the left panel of Fig.~\ref{flowmmdot}.

Similarly to their higher mass counterparts, the external regions of the star
become cooler and cooler during the AGB, with the effective temperatures spanning
the same $T_{\rm eff}$ range (2500-3500 K) given in section \ref{highm}. The only
exceptions to this common behaviour are the $1~M_{\odot}$ models, whose effective
temperatures during the final AGB phase are $\sim 2800$ K, and the $3~M_{\odot}$
of metallicity $Z=0.03$, which after becoming carbon star readjust on a largely
expanded configuration, with $T_{\rm eff} \sim 2000$K.

The variations of the surface chemistry reflect the effect of the recurrent TDU events, 
which results in the gradual increase of the surface carbon and the consequent rise of the 
C$/$O ratio. The accumulation of the surface carbon increases with the initial mass of the 
star, because of the higher number of TDU episodes experienced. We find that at $Z=0.03$ 
only models of initial mass in the range $2.5-3.5~M_{\odot}$ become C-stars in the very 
final evolutionary phases. The present computations show that formation of C-stars is not 
expected for metallicities $Z>0.03$, except in the case of the $2.5~M_{\odot}$ star of 
$Z=0.04$.

\begin{figure*}
\begin{minipage}{0.48\textwidth}
\resizebox{1.\hsize}{!}{\includegraphics{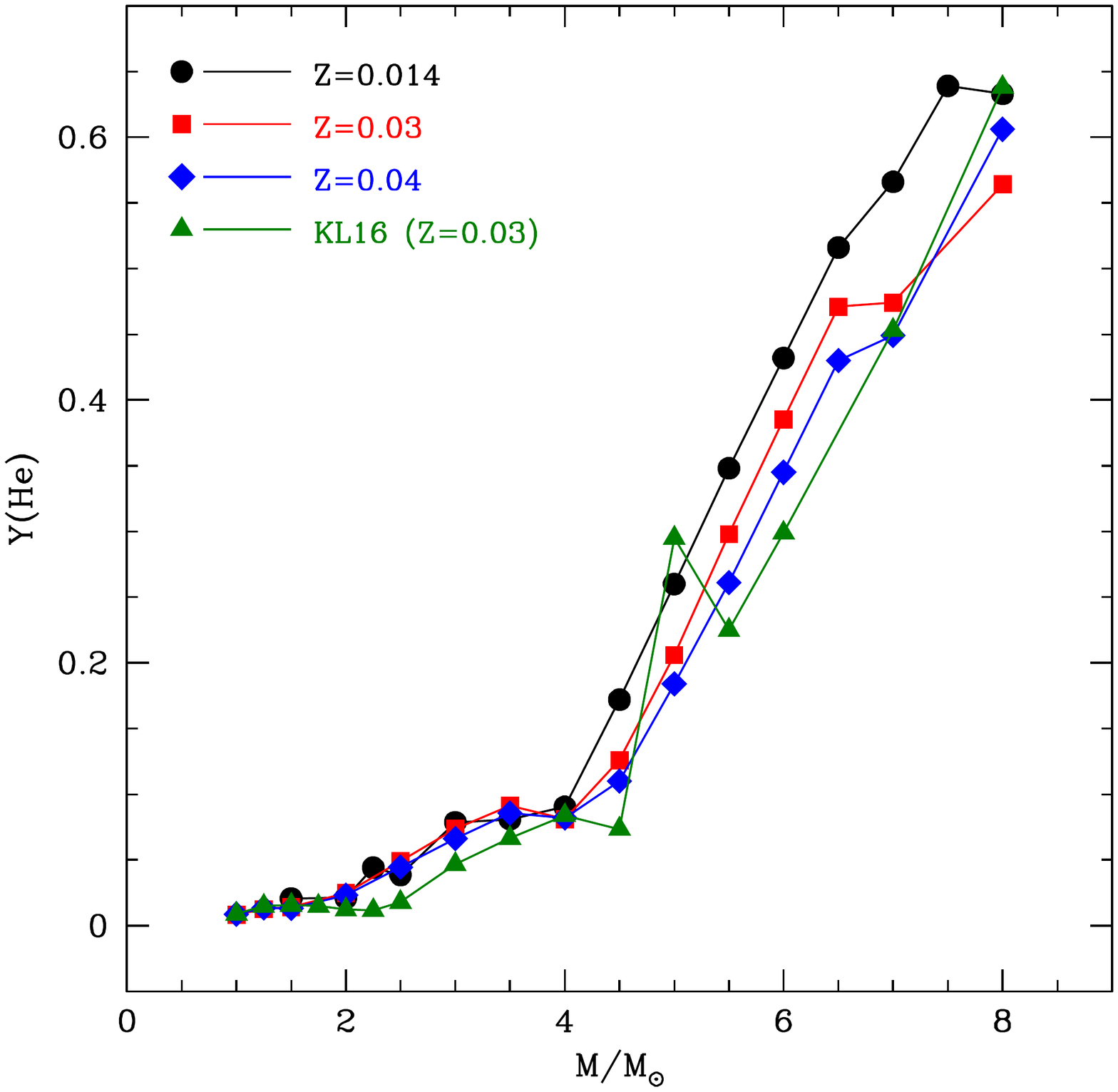}}
\end{minipage}
\begin{minipage}{0.48\textwidth}
\resizebox{1.\hsize}{!}{\includegraphics{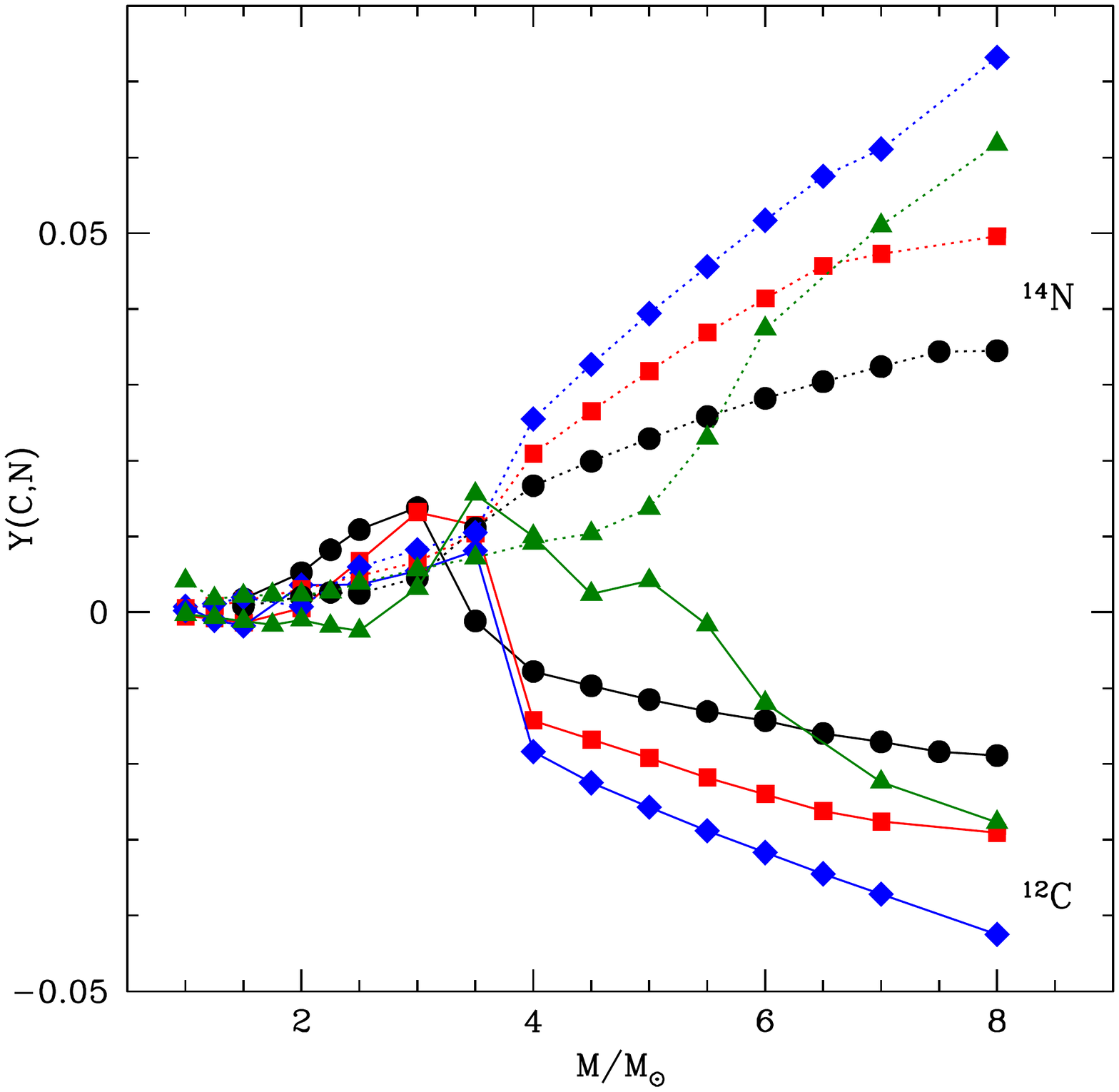}}
\end{minipage}
\vskip-80pt
\begin{minipage}{0.48\textwidth}
\resizebox{1.\hsize}{!}{\includegraphics{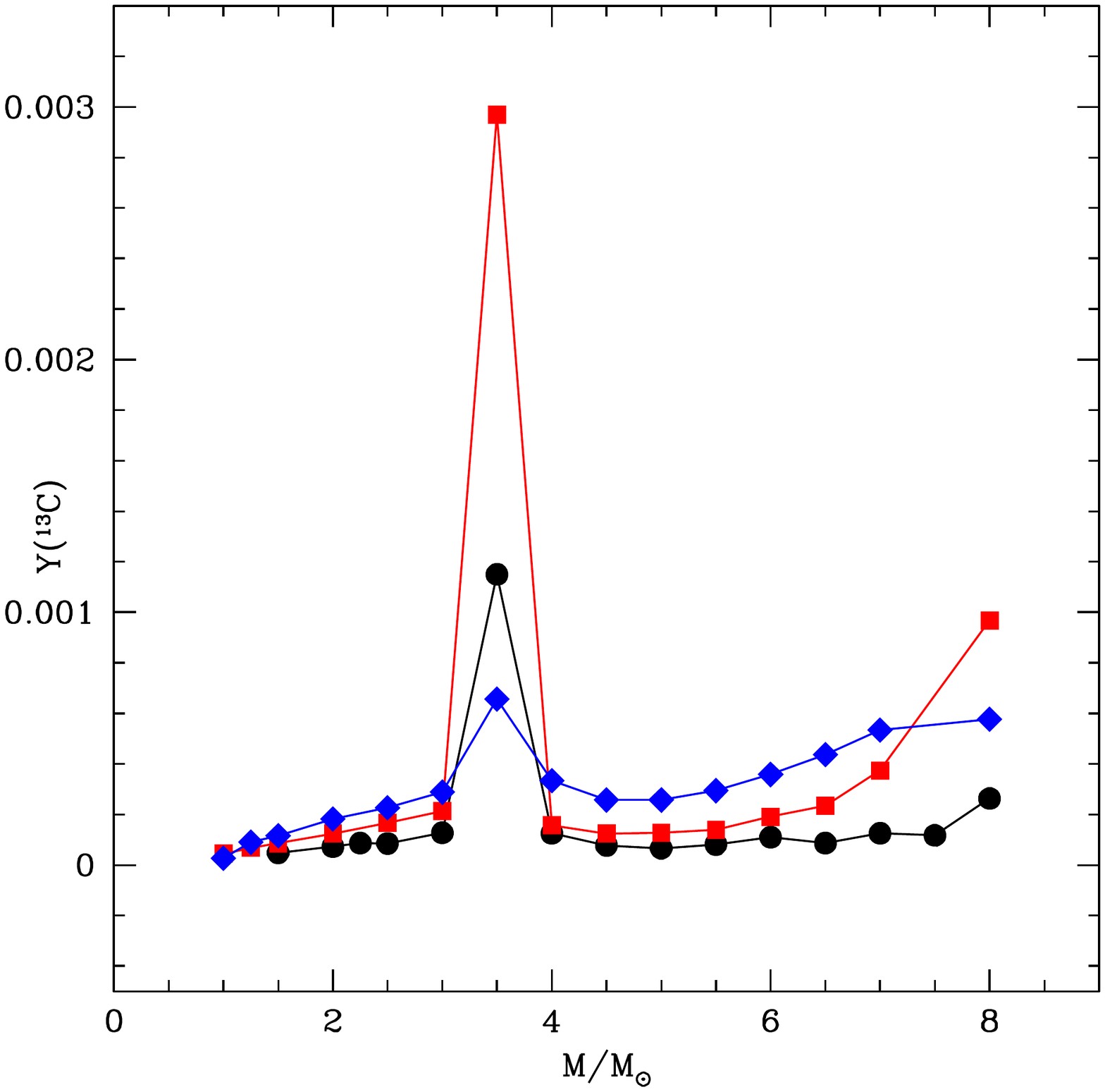}}
\end{minipage}
\begin{minipage}{0.48\textwidth}
\resizebox{1.\hsize}{!}{\includegraphics{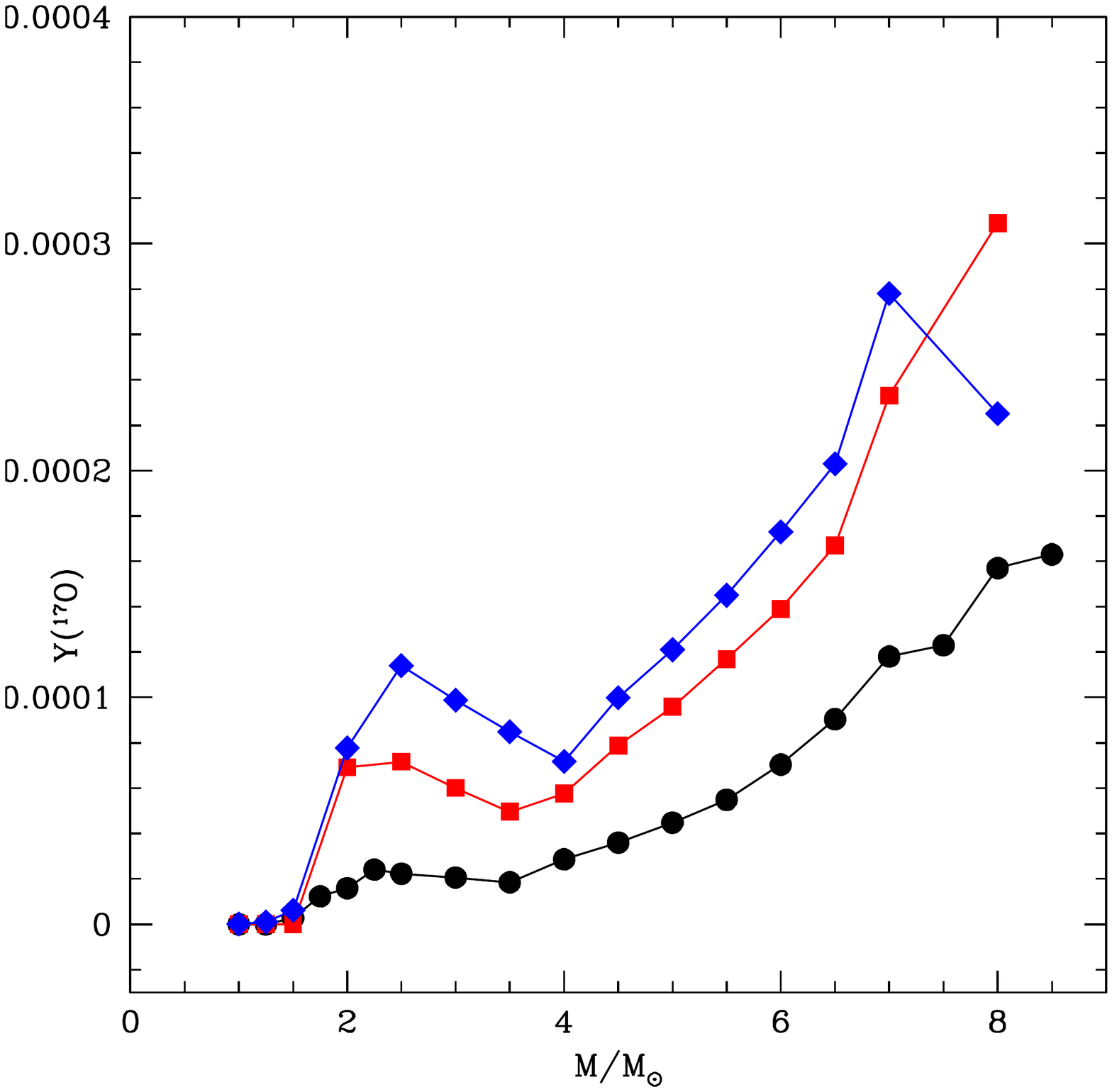}}
\end{minipage}
\vskip-60pt
\caption{The yields in solar masses, of helium (top, left panel) $^{12}$C and
$^{14}$N (top, right), $^{13}$C (bottom, left) and $^{17}$O (bottom, right) of our
$Z=0.03$ (red squares) and $Z=0.04$ (blue diamonds) AGB models 
and of solar-metallicity models by \citet{ventura18}. Green triangles 
indicate the helium, carbon and nitrogen yields published in \citet{karakas16}. 
}
\label{fyield}
\end{figure*}

\section{The stellar yields}
\label{yields}
Observations of chemical abundances and abundance ratios can be used to 
constrain the timescales of galaxy formation \citep{matteucci12}, as well as to pin down the 
shape of the stellar initial mass function (IMF) in galaxies where direct measurements are 
unfeasible (Romano et al. 2017, 2019; Zhang et al. 2018, and references therein). 
To interpret the observations, galactic chemical evolution models are needed, which require as input
stellar yields from stellar evolution calculations. 
As highlighted in the introduction, it is crucial to provide grids of stellar yields that cover the 
whole range of metallicities spanned by the observations. 

Here we follow the classical definition of the stellar yield, i.e., the yield of a given element $i$ is 
the net amount of newly-produced element that is ejected in the interstellar medium by a star 
during its life:

$$ 
Y_i=\int{[X_i-X_i^{init}] \dot M dt}
$$ 

The integral is calculated over the entire stellar lifetime, $X_i^{init}$ is the 
mass fraction of species $i$ at the beginning of the evolution, and $\dot M$ is the mass 
lost rate. If the element is destroyed in the stellar interior, then the yield is negative.

\begin{table*}
\setlength{\tabcolsep}{4pt}
\caption{The chemical yields of AGB stars of metallicity $Z=0.03$, $Z=0.04$ and (initial) mass in
the range $1-8~M_{\odot}$. The yields are given in solar masses and refer to the 
net produced quantity of each chemical species (see Section \ref{yields} for details).}             
\label{tabyield}      
\centering          
\begin{tabular}{l l l l l l l l l l l l}     
\hline
$Z=0.03$ & & & & & &  \\
\hline       
M$/$M$_{\odot}$ & H & He & $^{7}$Li & $^{12}$C & $^{13}$C & $^{14}$N & $^{16}$O &
$^{17}$O & $^{18}$O & $^{22}$Ne & $^{23}$Na  \\ 
\hline     
& & & & & Z=0.03 & & & & & \\  
\hline             
1.00 & -8.43(-3) & 8.34(-3) & -4-40(-9) & -5.28(-4) & 4.41(-5) & 5.71(-4) &  0.00(-0) & 9.46(-8) & -8.36(-7) & 0.00(+0) & 0.00(+0) \\
1.25 & -1.28(-2) & 1.27(-2) & -6-68(-9) & -8.01(-4) & 6.76(-5) & 8.65(-4) &  0.00(-0) & 9.48(-7) & -2.54(-6) & 0.00(+0) & 0.00(+0)  \\
1.50 & -1.45(-2) & 1.42(-2) & -8.85(-9) & -1.39(-4) & 8.61(-5) & 1.52(-3) & -1.01(-5) & 5.66(-6) & -5.04(-6) & 0.00(+0) & 0.00(+0)  \\
2.00 & -2.85(-2) & 2.48(-2) & -1.35(-8) &  5.33(-4) & 1.24(-4) & 3.02(-3) & -1.61(-4) & 7.08(-5) & -9.56(-6) &-6.61(-6) & 7.58(-6)  \\
2.50 & -6.13(-2) & 4.91(-2) & -1.81(-8) &  6.79(-3) & 1.67(-4) & 4.84(-3) & -9.41(-4) & 7.16(-5) & -1.45(-5) &-1.61(-5) & 2.34(-5)  \\
3.00 & -9.46(-2) & 7.38(-2) & -2.30(-8) &  1.32(-2) & 2.14(-4) & 6.63(-3) & -1.53(-3) & 6.00(-5) & -1.92(-5) &-2.61(-5) & 3.91(-5)  \\
3.50 & -1.17(-1) & 9.13(-2) & -2.14(-8) &  1.15(-2) & 2.97(-3) & 1.04(-2) & -1.74(-3) & 4.95(-5) & -5.92(-5) &-4.23(-5) & 5.40(-5)  \\
4.00 & -8.29(-2) & 8.11(-2) & -9.21(-9) & -1.43(-2) & 1.57(-4) & 2.09(-2) & -4.88(-3) & 5.77(-5) & -8.68(-5) &-5.41(-4) & 5.73(-4)  \\
4.50 & -1.28(-1) & 1.26(-1) & -7.17(-9) & -1.68(-2) & 1.24(-4) & 2.65(-2) & -7.49(-3) & 7.88(-5) & -1.03(-4) &-6.81(-4) & 7.18(-4)  \\
5.00 & -2.09(-1) & 2.06(-1) & -1.16(-8) & -1.92(-2) & 1.27(-4) & 3.18(-2) & -1.04(-2) & 9.58(-5) & -1.18(-4) &-8.71(-4) & 9.19(-4)  \\
5.50 & -3.00(-1) & 2.98(-1) & 8.10(-10) & -2.18(-2) & 1.39(-4) & 3.69(-2) & -1.31(-2) & 1.17(-4) & -1.32(-4) &-1.02(-3) & 1.08(-3)  \\
6.00 & -3.87(-1) & 3.85(-1) & 1.41(-8)  & -2.40(-2) & 1.91(-4) & 4.14(-2) & -1.54(-2) & 1.39(-4) & -1.46(-4) &-1.07(-3) & 1.13(-3)  \\
6.50 & -4.74(-1) & 4.71(-1) & 2.89(-8)  & -2.62(-2) & 2.35(-4) & 4.57(-2) & -1.76(-2) & 1.67(-4) & -1.60(-4) &-1.16(-3) & 1.21(-3)  \\
7.00 & -4.76(-1) & 4.74(-1) & 7.05(-8)  & -2.76(-2) & 3.74(-4) & 4.73(-2) & -1.77(-2) & 2.33(-4) & -1.71(-4) &-1.24(-3) & 1.30(-3)  \\
8.00 & -5.64(-1) & 5.64(-1) & 1.03(-7)  & -2.91(-2) & 9.67(-4) & 4.96(-2) & -1.90(-2) & 3.09(-4) & -1.92(-4) &-1.34(-3) & 1.38(-3)  \\
\hline
& & & & & Z=0.04 & & & & & \\    
\hline
1.00 & -8.55(-3) & 8.78(-3) & -4.42(-9) & -7.28(-4) & 6.42(-5) & 7.21(-4) &  0.00(-0) & 7.37(-8) & -7.28(-8) & 0.00(+0) & 0.00(+0) \\
1.25 & -1.35(-2) & 1.34(-2) & -6.72(-9) & -1.04(-3) & 8.98(-5) & 1.11(-3) &  0.00(-0) & 9.07(-7) & -3.22(-7) & 0.00(+0) & 0.00(+0)  \\
1.50 & -1.36(-2) & 1.33(-2) & -8.92(-9) & -1.78(-3) & 1.16(-4) & 1.95(-3) & -3.57(-6) & 1.05(-6) & -1.08(-6) & 0.00(+0) & 0.00(+0)  \\
2.00 & -3.63(-2) & 2.31(-2) & -1.33(-8) &  3.56(-3) & 1.82(-4) & 7.59(-4) & -4.40(-4) & 1.04(-4) & -2.25(-6) &-6.61(-6) & 7.58(-6)  \\
2.50 & -5.47(-2) & 4.46(-2) & -1.88(-8) &  3.62(-3) & 2.27(-4) & 6.00(-3) & -1.22(-3) & 1.14(-4) & -1.84(-5) &-1.61(-5) & 2.34(-5)  \\
3.00 & -7.98(-2) & 6.61(-2) & -2.26(-8) &  5.47(-3) & 2.88(-4) & 8.23(-3) & -2.38(-3) & 9.88(-5) & -2.37(-5) &-2.61(-5) & 3.91(-5)  \\
3.50 & -1.04(-1) & 8.56(-2) & -2.64(-8) &  8.09(-3) & 6.56(-4) & 1.05(-2) & -3.09(-3) & 8.48(-5) & -3.29(-5) &-4.23(-5) & 5.40(-5)  \\
4.00 & -8.44(-2) & 8.20(-2) & -5.61(-9) & -1.84(-2) & 3.34(-4) & 2.55(-2) & -4.96(-3) & 7.18(-5) & -1.13(-4) &-6.41(-4) & 6.73(-4)  \\
4.50 & -1.13(-1) & 1.10(-1) &  2.32(-9) & -2.25(-2) & 2.57(-4) & 3.27(-2) & -7.63(-3) & 9.98(-5) & -1.38(-4) &-7.98(-4) & 9.18(-4)  \\
5.00 & -1.87(-1) & 1.84(-1) &  5.08(-9) & -2.57(-2) & 2.77(-4) & 3.94(-2) & -1.10(-2) & 1.21(-4) & -1.57(-4) &-1.01(-3) & 1.11(-3)  \\
5.50 & -2.64(-1) & 2.61(-1) &  1.28(-8) & -2.88(-2) & 2.94(-4) & 4.56(-2) & -1.40(-2) & 1.45(-4) & -1.76(-4) &-1.12(-3) & 1.27(-3)  \\
6.00 & -3.49(-1) & 3.45(-1) &  2.28(-8) & -3.17(-2) & 3.58(-4) & 5.17(-2) & -1.72(-2) & 1.73(-4) & -1.95(-4) &-1.30(-3) & 1.43(-3)  \\
6.50 & -4.33(-1) & 4.30(-1) &  3.27(-8) & -3.45(-2) & 4.37(-4) & 5.75(-2) & -2.01(-2) & 2.03(-4) & -2.13(-4) &-1.40(-3) & 1.52(-3)  \\
7.00 & -4.52(-1) & 4.49(-1) &  1.17(-7) & -3.72(-2) & 5.34(-4) & 6.11(-2) & -2.09(-2) & 2.78(-4) & -2.31(-4) &-1.68(-3) & 1.80(-3)  \\
8.00 & -6.10(-1) & 6.06(-1) &  1.44(-7) & -4.25(-2) & 5.76(-4) & 7.32(-2) & -2.76(-2) & 3.09(-4) & -2.63(-4) &-1.84(-3) & 1.96(-3)  \\
\hline                  
\end{tabular}
\end{table*}

Fig.~\ref{fyield} shows the yields of helium and of the CNO elements of the high-metallicity
models discussed in the present work.
The helium yield increases with mass, ranging from almost null at the lowest masses, and
reaching values of the order of $\sim 0.6~M_{\odot}$ for $M \sim 8~M_{\odot}$. The change
in the slope of the Y$($He$)$ vs mass relationship is due to the onset of the second dredge-up,
occurring after the exhaustion of the core helium in stars of mass above $\sim 4~M_{\odot}$, 
which results in a significant increase in the surface helium. The trend with mass is
positive, because the higher is the mass of the star, the larger the extent of the inwards penetration of the convective envelope 
taking place during the second dredge-up \citep{ventura10}. Since the second dredge-up occurs
before the beginning of the TP-AGB phase, its results are more robust that those related to nucleosynthesis and mixing during the TP-AGB phase, 
whose description is affected by several uncertainties in the physical ingredients adopted \citep{karakas14b}.
The $^{12}$C yields are positive in the low-mass domain, owing to the effect of the TDU, 
which results in the increase of the surface carbon (see right panel of
Fig.\ref{flowm}). The largest yields of $^{12}$C, of the order of $0.015~M_{\odot}$, 
are produced by $\sim 3~M_{\odot}$ stars. Massive AGB stars destroy carbon via HBB (see 
top, left panel of Fig.~\ref{f1chem}), which is why they produce negative carbon yields. 
The $^{16}$O yields (not shown) are negative for $M\geq 4~M_{\odot}$, owing to the 
effects of HBB.
However, even in the most extreme cases, i.e., the $\sim 8~M_{\odot}$ stars, the yields are
not smaller than $\sim - 0.02~M_{\odot}$; this is because the HBB experienced by stars 
of the metallicities discussed here is not efficient enough to significantly deplete the
surface oxygen abundance. 

The nitrogen yields are always positive. In the low-mass domain this is 
due to the effects of the first dredge-up, which increases the surface
N abundance. No significant, further rise in the nitrogen content of the envelope is expected 
during the TP-AGB phase. The N yields of massive AGB stars reflect the effects of
HBB: independently on whether the sole CN or the full CNO cycle is activated, the outcome is
the synthesis of significant quantities of nitrogen. The N yields increase with the
mass of the star, reaching $\sim 0.06~M_{\odot}$ in the most massive cases.

The yields of the minor isotopes $^{13}$C and $^{17}$O are found to increase almost 
monotonically as a function of both stellar mass and metallicity. They increase with the 
initial mass and metallicity of the star, with the exception of a spike in $^{13}$C 
($^{17}$O) production occurring around $3.5 (2.5)~M_{\odot}$ 
(Fig.~\ref{fyield}, left- and right-handed lower panels, respectively). 
The prominent peak in the $^{13}$C yield is related to the combined effects of TDU and
HBB in $\sim 3.5~M_{\odot}$ stars, the former mechanism favouring the increase in the
surface $^{12}$C, later converted into $^{13}$C by proton captures.
It has been recently pointed out \citep{romano17, romano19, zhang18} that the 
CNO isotopic ratios may be used to probe the shape of the galaxy-wide IMF in high-redshift, 
massive dusty starbursts where direct measurements 
are unfeasible. These objects are thought to host a non-negligible fraction of super-solar 
metallicity stars (see, e.g., Johansson et al. 2012). Grids of super-solar metallicity 
yields as those presented in this study, therefore, will be extremely useful to 
spot IMF variations in extreme environments. 
Future implementation of our grids of super-solar metallicity stars 
into chemical evolution models for massive dusty starbursts 
will allow more robust 
predictions. 

\section{Comparison to other studies}
\label{comp}
Super-solar metallicity models were published by Karakas (2014a, hereinafter K14) 
and \citet{weiss09}. 

k14 models with metallicity $Z=0.03$ were calculated by means of the MONASH code.
The authors present an exhaustive discussion on the efficiency of the TDU in the
high-metallicity domain and how the initial mass and helium affect the AGB evolution.
In a subsequent paper, Karakas \& Lugaro (2016, hereainfter KL16) presented the gas
yields of K14 models, which we compare with the results presented in Section \ref{yields}.

Because the evolution of AGB stars is driven by the mass of the degenerate core we check 
for consistency between the K14 core masses and ours at the beginning of the
AGB phase. The core masses reported in Column 6 of Table 1 are similar to those by 
K14 for $M \leq 4~M_{\odot}$ stars. In the larger-mass domain our core masses are 
slightly higher than K14, with a $\sim 0.5~M_{\odot}$ shift in the initial mass versus 
core mass relationship.
 
The structure differences between our $Z=0.03$ models
and those published by K14 can be seen in the two panels of Fig.~\ref{fagb},
where the peak luminosities and the hottest temperatures reached during the AGB phase
are shown. A remarkable difference is that our massive AGB models 
are brighter and hotter than those of K14, with larger discrepancy for
higher initial masses. These differences tend to vanish in the
$M < 3~M_{\odot}$ domain.

Similar differences between ours and MONASH models were found for solar 
\citep{ventura18} and sub-solar metallicities \citep{ventura15}. The main reason 
for the dissimilarities is the treatment of turbulent convection: we calculate the temperature 
gradient within regions unstable to convection motions via the FST model (see 
Section \ref{agbinput}), whereas K14 uses the mixing length 
theory \citep{mlt}. As discussed in Section \ref{uncert}, the treatment of convection 
in the envelope deeply affects the physical behaviour of AGB stars, particularly 
under HBB conditions: \citet{ventura05a} showed that FST modelling 
favours a more efficient nucleosynthesis in the internal regions of the envelope than in 
the MLT case, which leads to higher luminosities and rates of mass loss. 

The difference with K14 are particularly evident in stars of mass  
corresponding to the minimum threshold required to activate HBB, which is 
$\sim 3.5~M_{\odot}$
in the present analysis, and $\sim 4.5~M_{\odot}$ in K14. This
discrepancy is entirely due to the treatment of convection; we rule out that 
dissimilarities in the core mass play a role here, because our values 
(Column 6 of Table 1) are the same within $0.02~M_{\odot}$ as those by 
K14 in the same mass range. 

The bottom, left panel of Fig.~\ref{fagb} shows the overall duration of the TP-AGB
phase. The results presented here and those by K14 are similar in the $M > 3~M_{\odot}$ 
domain, despite the large difference in the luminosity, which should make the duration
of the AGB phase of our models much shorter than that of K14. \citet{ventura18} found that  
at solar metallicity the time scale of the AGB phase calculated with the ATON code
is between 2 and 3 times shorter than the Monash models of the same
mass. The reasons for this is
in the treatment of mass loss. ATON uses the description by \citet{blocker95}, 
according to which $\dot M$ is strongly dependent on the luminosity. The
K14 computations instead are based on the \citet{vw93} treatment, where the rate of the mass
loss is set to increase with the period of the star. In AGB stars of lower metallicity than 
presented in this paper the \citet{blocker95} description leads to higher $\dot M$ because 
of the large luminosities typical of the stars experiencing HBB; in this case, the 
differences between
ATON and MONASH models is enhanced by the larger luminosities attained by ATON
models. For the higher metallicities presented in this paper the situation is different
because: a) HBB is weaker than at lower metallicity, thus the luminosities 
are smaller (compare the $Z=0.014$ and $Z=0.03$ lines in the top, left panel of
Fig.~\ref{fagb}); b) metal-rich stars evolve to larger radii, which
result in generally longer periods. Consequently, at $Z=0.03$, the
mass-loss rate of \citet{vw93} is more efficient than that of \citet{blocker95}. 
In massive AGB stars, these effects partly counterbalance those related
to the higher luminosities experienced by the ATON models calculated with the FST convection
description, making the duration of the whole TP-AGB phase of the present models 
similar to K14. 
The higher rates of mass loss predicted by the 
\citet{vw93} treatment is the reason why also in the $Z=0.03$ low-mass domain the duration 
of K14 models is shorter than in our models. 

The KL16 yields are compared to ours in Fig.~\ref{fyield}. The helium yields are similar,
with a $\sim 0.5~M_{\odot}$ shift related to the differences in the core masses 
outlined at the beginning of this section, and to the fact that most of the surface helium 
enrichment occurs during the second dredge-up, 
for which the results presented by different research groups are
in strong agreement \citep{ventura10}.
The carbon and 
nitrogen yields present significant differences, related to the different
physical behaviour discussed above. The carbon yields presented here are significantly
smaller than KL16 for $M>3~M_{\odot}$ stars, owing to the much more efficient HBB 
nucleosynthesis experienced. This is particularly evident in the $3.5-4.5~M_{\odot}$
mass range, which exhibits the largest differences in the temperatures at the bottom
of the envelope, as shown in the right panel of Fig.~\ref{fagb}. The $M>3~M_{\odot}$ 
$Z=0.03$ models experience a much stronger HBB than KL16, thus produce more nitrogen via
$^{12}$C proton capture reactions: this is the reason for the differences in the N
yields in the right panel of Fig.~\ref{fyield}.

The $Z=0.04$ models by \citet{weiss09} are compared to ours and to K14 in 
Fig.~\ref{fagb}. The comparison is limited to the duration of the AGB phase and to 
the final C$/$O, because neither the largest luminosity and temperature at the base 
of the envelope nor the chemical yields are given in \citet{weiss09}.

For stars of mass below $\sim 2.5~M_{\odot}$ the final carbon-to-oxygen ratio by 
\citet{weiss09} is practically the same as the results presented here and by K14, 
indicating that the efficiencies of TDU are similar in the three cases. For stars
of higher mass the results by \citet{weiss09} are somewhat intermediate between
the final C$/$O found in the present work and those by K14, suggesting that: a) 
the HBB experienced is weaker than in our case; b) the TDU efficiency is smaller
than in K14. While the weaker HBB is explained by the difference in convection
modelling, point (b) above is related to the description of convective borders
adopted (as discussed in Section \ref{uncert}, K14 results are based on the 
algorithm proposed by \citet{lattanzio86}, whereas \citet{weiss09} impose an 
exponential decay of velocities from the border of all the convective regions formed) 

The results shown in the bottom, left panel of Fig.~\ref{fagb} indicate that the 
AGB evolution predicted by \citet{weiss09} is faster than in the present study and
in K14, for almost all the masses considered. This is due to the treatment of
mass loss, because the fitting formula by \citet{jacco05} leads to mass loss
rates $2-7$ times higher than the predictions based on the \citet{blocker95}
and \citet{vw93} formulae in the super-solar metallicity domain.

\section{Dust production}
\label{dust}

\begin{figure*}
\begin{minipage}{0.48\textwidth}
\resizebox{1.\hsize}{!}{\includegraphics{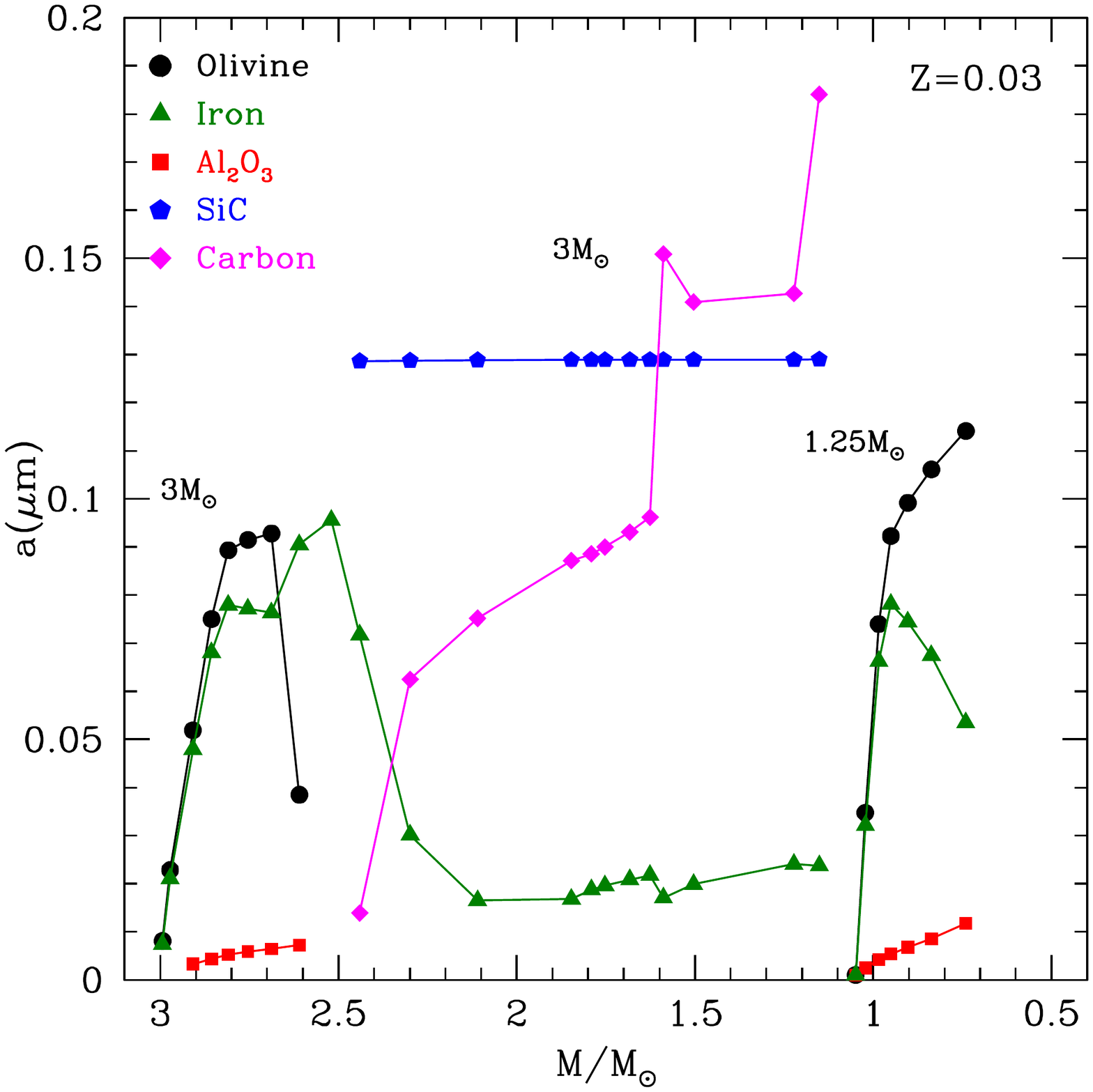}}
\end{minipage}
\begin{minipage}{0.48\textwidth}
\resizebox{1.\hsize}{!}{\includegraphics{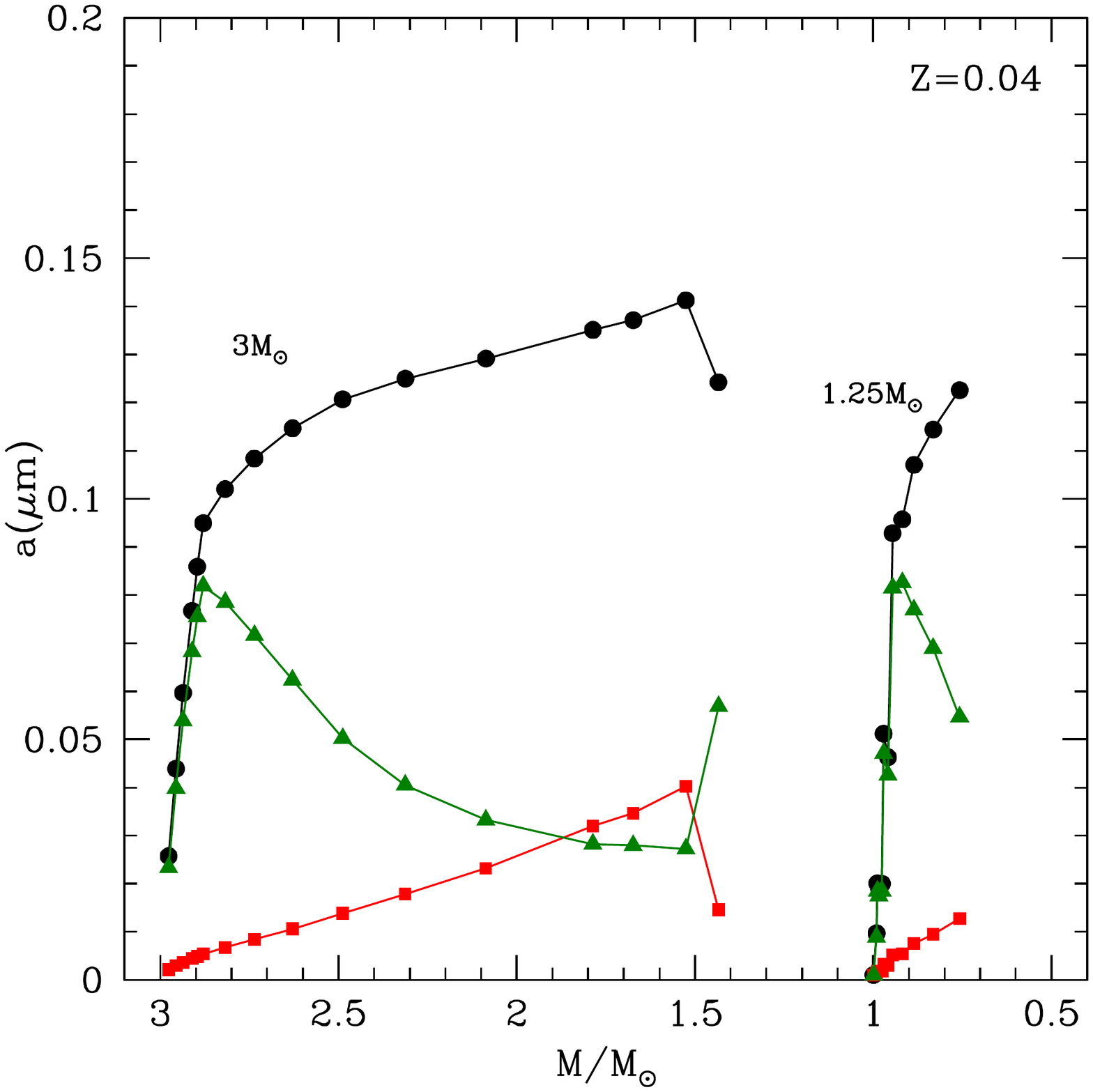}}
\end{minipage}
\vskip-80pt
\begin{minipage}{0.48\textwidth}
\resizebox{1.\hsize}{!}{\includegraphics{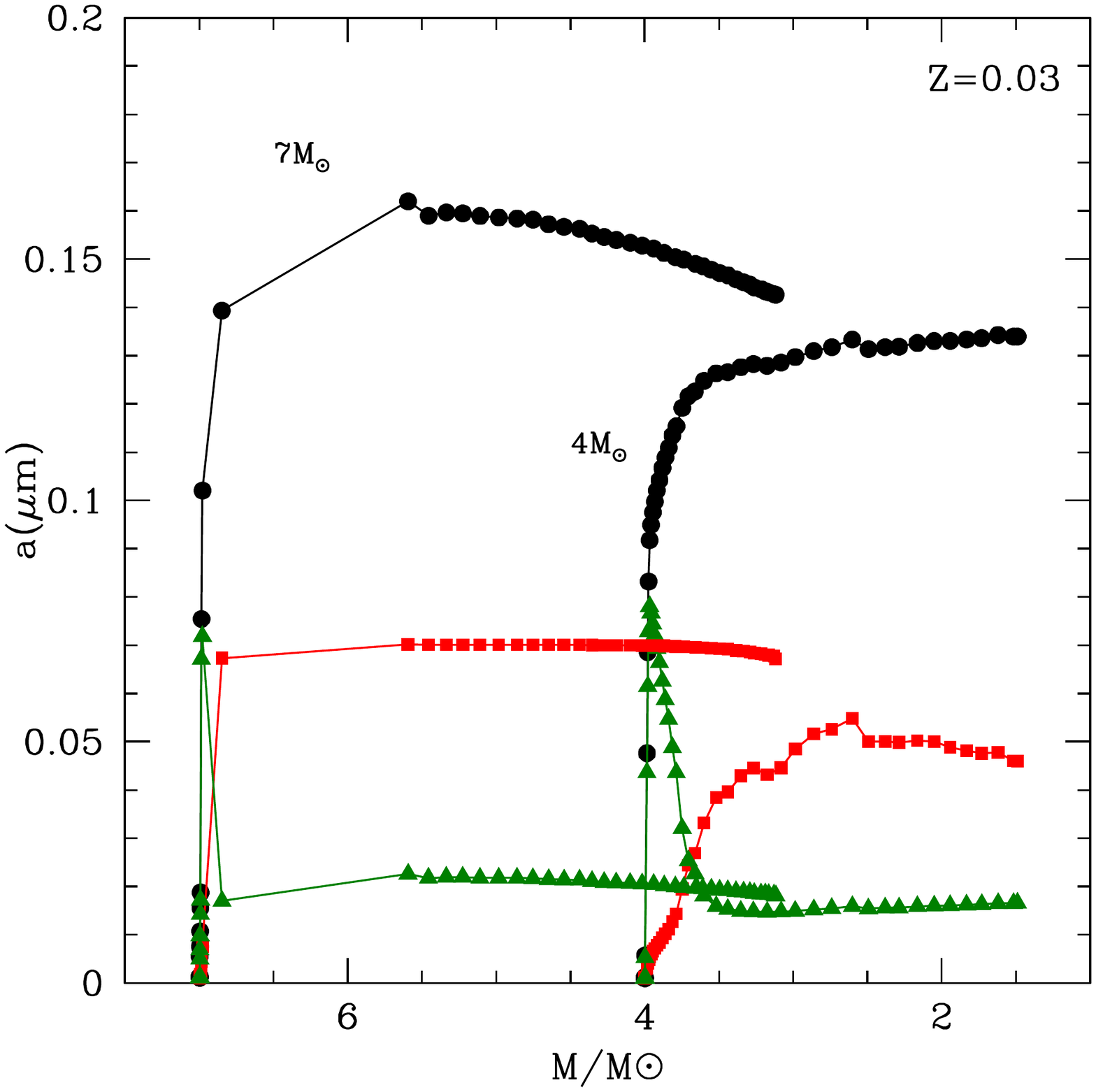}}
\end{minipage}
\begin{minipage}{0.48\textwidth}
\resizebox{1.\hsize}{!}{\includegraphics{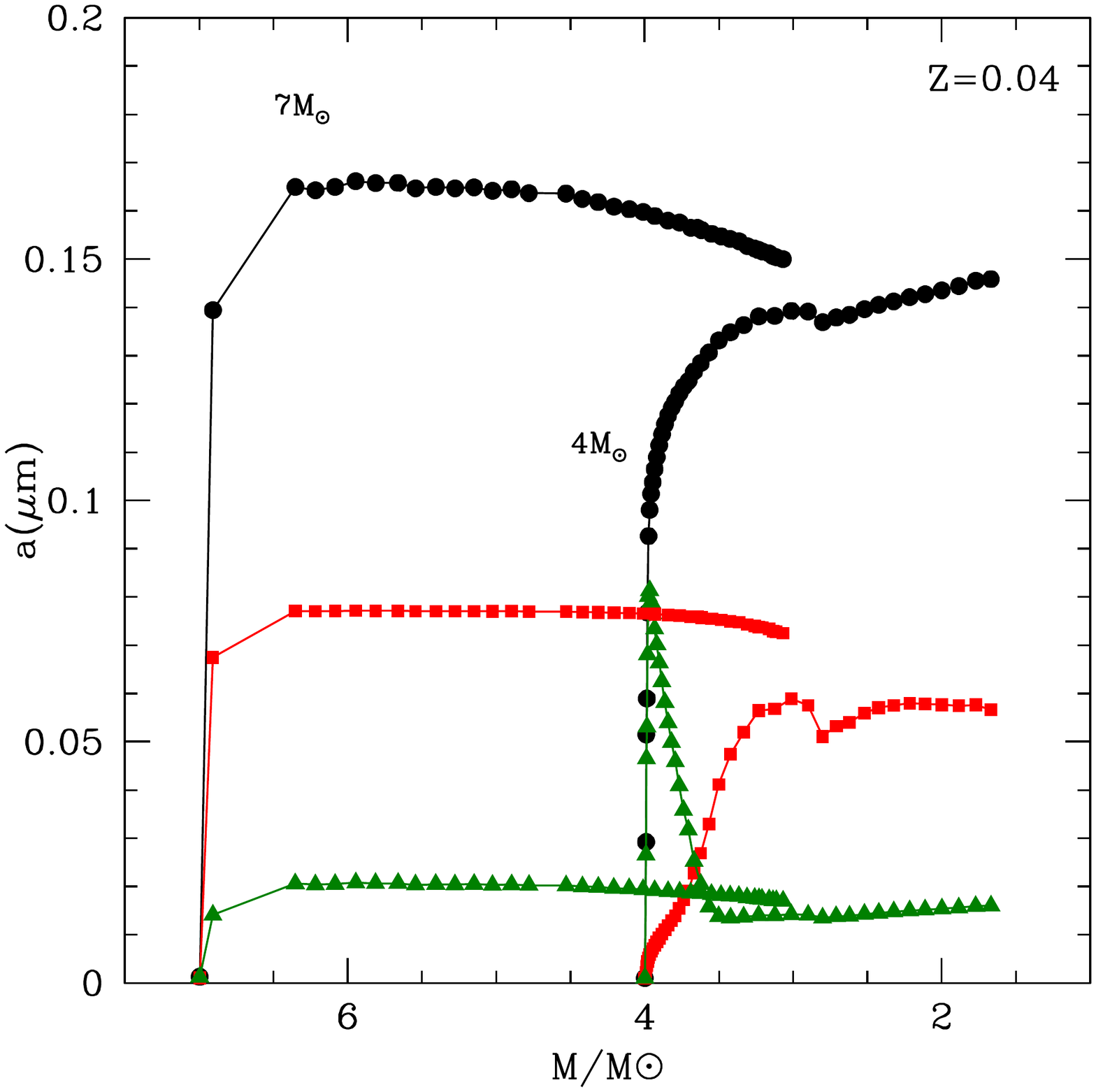}}
\end{minipage}
\vskip-60pt
\caption{The size (radius) of the different dust species (listed in the legend) formed in the wind of
AGB stars as a function of the current stellar mass. 
The left and right
panels refer to the $Z=0.03$ and $Z=0.04$ metallicities, respectively.
For sake readability we only show the 
$1.25~M_{\odot}$ and $3~M_{\odot}$ cases for the low-mass domain (top panels), and the $4~M_{\odot}$ 
and $7~M_{\odot}$ models, for the HBB domain (bottom panels).
}
\label{fsize}
\end{figure*}

The amount and mineralogy of the dust formed in the circumstellar envelope of AGB stars
are determined by the surface chemical composition and by global physical
parameters, in particular by a) the rate of the mass loss, which determines the
density of the wind, and in turn the number of gaseous molecules available to condensation;
b) the luminosity of the star, which affects the radiation pressure acting on the newly
formed dust grains; and c) the effective temperature, which sets the location of the condensation
zone \citep{fg06}.
Among the various species, those most thermodynamically stable are alumina dust 
and SiC, for oxygen-rich AGBs and carbon stars, respectively. The condensation zone of 
these compounds is more internal than that of silicates and solid carbon, which  
are characterized by higher extinction coefficients \citep{fg06}. 

In the schematization adopted here the dust grains of the different species form in the 
condensation zone and grow from nano-size dimensions until they reach an asymptotic size
when the gas densities drop below the level at which the rate of growth becomes
much smaller than the velocity with which the wind moves outwards. It is to this
asymptotic dimension that we refer to in the following discussion and in Fig.~\ref{fsize} 
showing the size of the grains of the considered dust species, formed in
the wind of stars of different mass. 

Before starting the discussion of the results obtained in the present investigation,
we believe important to underline that the modelling of dust formation used here, mostly
based on the schematization proposed by the Heidelberg group, is a simplified description
of a much more complex situation, where the combination of shocks and pulsation effects
carry gas molecules into more external and cooler regions of the circumstellar envelope,
where dust formation may occur in conditions substantially different than those found
in the static wind proposed here. The most serious limitation in the present description
is that mass loss is assumed apriori, whereas it should be the outcome of a self-consistent
treatment of the wind dynamics, in agreement with the studies by \citet{lars08} and the
most recent investigations by \citet{bladh19a, bladh19b}. The results in terms of dust
production presented here must be taken with some caution, particularly for what attains
the production of carbonaceous species. Indeed the present findings are based on the
rate of mass loss calculated by means of the \citet{wachter02, wachter08} works, which,
as discussed in section \ref{uncert}, neglects any role of the carbon excess: this assumption
may lead to an overestimation of the $\dot M$'s found here \citep{bladh19b}, considering 
that in the few models that reach the C-star stage the carbon excess is small, owing to the 
large oxygen content of the surface regions of the star. It goes without saying that the 
growth of dust particles, particularly of the carbon grains, might be overestimated as
well. 

Table \ref{tabdust} reports the mass of the total 
and the different species of dust produced by our models of different mass during the AGB 
life. Given the significant differences in the evolution properties of AGB stars of 
different masses, we divide the description below in three mass ranges.

\subsection{Production of silicates in low-mass AGB stars ($< 2 M_{\odot}$)}
\label{m125}
The $1.25~M_{\odot}$ models shown in Fig.~\ref{flowm} represent the low-mass AGB stars 
with initial mass below $2~M_{\odot}$ that
experience only a few ($\sim 10$) TPs before to loose the entire envelope. Considering
the small number of TDU events, these stars do not reach the C-star phase and evolve
as O-rich stars for the whole AGB (right panel
of Fig.~\ref{flowm}).
Based on the previous discussion, the dust formed in their envelope is 
composed of silicates, alumina dust, and solid iron, with no traces of carbonaceous particles.

The dominant contribution to the dust budget is provided by silicates, particularly
by olivine grains, whose size increases steadily during the AGB phase until it reaches
dimensions of the order of $0.12 ~ \mu$m towards the final evolutionary stages, when
the percentage of silicon locked into dust particles approaches $\sim 20\%$.
Negligible quantities of alumina dust are formed, owing to the
low rates of mass loss experienced below $\sim 10^{-6}~M_{\odot}/$yr.
We see in Fig.~\ref{fsize} that the size of the solid iron particles formed is
anticorrelated with olivine; this is because the iron dust formation layer is more
external than that of silicates, thus a significant production of the latter results in 
a significant wind acceleration, which favours the drop in the gas density and the decrease of 
the rate of growth of iron grains. During the first TPs almost $25\%$ of gaseous iron is 
condensed into dust grains of size slightly below $0.1 ~ \mu$m; 
the size of iron particles drops to $\sim 0.05~\mu$m during the 
final AGB stages, when silicates form in 
significant quantities, 
with the fraction of iron locked into dust being below $10\%$. 

The metallicity has some effects on the dust formed by low-mass AGBs, the 
$Z=0.04$, $1.25~M_{\odot}$ star produces more dust than its $Z=0.03$ counterpart, 
owing to the higher quantities of silicon, aluminium and iron in the surface layers,
and to the higher rates of mass loss experienced; however, as shown in Fig.~\ref{fsize},
the differences in the size of the particles formed are within $10\%$.

\subsection{Dust production by stars close to the mass threshold to activate HBB 
($2~M_{\odot} < M < 3.5~M_{\odot}$)}
Stars of initial mass close to $\sim 3~M_{\odot}$ experience several TPs
and TDU episodes, with the consequent increase in the surface carbon. These stars
are represented by the $3~M_{\odot}$ models in Fig.~\ref{fsize}. A distinction based on
metallicity is mandatory here, because, as discussed in Section \ref{lowm}, the $Z=0.03$ 
stars in this mass domain reach the C-star stage, whereas most of the $Z=0.04$ stars evolve
as O-rich (see right panel of Fig.~\ref{flowm}).

We start from the $Z=0.04$ case, which is similar to the lower-mass stars discussed 
in Section \ref{m125}. The dust formed in the wind has a 
dominant contribution from silicates and the amounts of iron dust and silicates are
anti-correlated. The size of the silicates grains formed is  
$\simeq 0.15~\mu$m, with $\sim 25\%$ of silicon locked into dust. The production of silicates is
more efficient than in the $1.25~M_{\odot}$ case, because the rate of the mass loss 
reaches higher values, up to $\sim 10^{-5}~M_{\odot}/$yr in the final evolutionary phases.
An additional difference, also due the higher rate of mass loss, is that the formation 
of alumina dust is not negligible in this case: during the final phases the fraction of aliminium 
condensed into alumina dust is $\sim 15\%$ and the size of Al$_2$O$_3$ grains is 
$\sim 0.05\mu$m (see the red squares in the top, right panel of Fig.~\ref{fsize}, which 
represent the size of Al$_2$O$_3$ grains). According to the present modelling the 
C-star stage is reached only by stars of initial mass $\sim 2~M_{\odot}$, during the
very final AGB phases: however, the carbon excess to oxygen is very small, thus the
formation of carbonaceous dust particles is negligible.

The behaviour of the $3~M_{\odot}$ star of metallicity $Z=0.03$ is qualitatively different,
because it becomes a C-star during the final $10\%$ (in time) of the AGB evolution.
Note that although the duration of the phase during which the star is O-rich is much
longer than the C-star phase, $80\%$ of the mass is lost after the star becomes a C-star
(see right panel of Fig.~\ref{flowm}): the dust produced by these stars is therefore 
dominated by carbonaceous particles and solid iron.
The amount of silicates produced is smaller than in the higher metallicity $3~M_{\odot}$ 
star and the formation of alumina dust is negligible: this is because the production of 
these species is limited to the first part of the AGB phase, when the rate of mass loss 
is below a few $\sim 10^{-6}~M_{\odot}/$yr. 

During the C-star phase the dust mineralogy is dominated 
by SiC and solid iron, due to the low surface C excess, with C/O ratio 
below $\sim 1.2$ for a significant fraction of the C-rich phase.
The SiC grains reach dimensions $\sim 0.13~\mu$m, whereas the size of solid carbon particles is 
below $\sim 0.1~\mu$m. The results reported in the top, left panel of Fig.~\ref{fsize} 
indicate that the size of SiC grains keeps constant during the whole AGB phase: this is 
a saturation effect due to the fact that all the residual silicon not already locked 
into SiS molecules, i.e., $\sim 55\%$ of the original silicon abundance, is condensed into dust. 
Conversely, only $\sim 5\%$ of carbon is locked into solid particles. The 
consumption of gaseous silicon by formation of SiC dust was explored in 
AGB stars of sub-solar metallicity by \citet{ventura14b}.

These conditions prove extremely favourable to the formation of large amounts of solid
iron; we see in Fig.~\ref{fsize} that during the phases just before and immediately
after the star becomes C-star, solid iron is the dominant species, the grain size reaching
$\sim 0.1 ~ \mu$m, with almost half of gaseous iron condensed into dust.
The formation of solid iron grains is partly due to the large
availability of iron in the surface regions of high-metallicity stars. Furthermore,
compared to lower metallicity stars, the low C excess 
affects the efficiency of the growth of solid carbon grains and favours the
formation of solid iron grains, which is otherwise inhibited by the fast acceleration of the wind
triggered by the large extinction coefficients of solid carbon particles.

During the latest evolutionary phases, after more carbon is accumulated 
in the surface regions via the recurrent TDU events, and the surface 
C$/$O exceeds $1.2$, the size of carbon grains formed reaches $\sim 0.18~\mu$m, 
with $10-15\%$ of carbon condensed into dust. The amount of carbon dust exceeds SiC 
only during these late phases.

The extinction properties of solid carbon grains favour the acceleration of
the wind, which in the $3~M_{\odot}$ star shown in the top, left panel of
Fig.~\ref{fsize}, reaches velocities slightly below 30 Km/s. These results indicate
that the winds of carbon stars in the super-solar metallicity domain are slower
than in the lower metallicity case, owing to the smaller values of the carbon
excess in the surface regions. In the $Z=0.03$ stars of initial mass 
$2.5~M_{\odot}$ and $3.5~M_{\odot}$ we find smaller velocities during the C-star
phase, in the range 20-30 Km/s. These results must be taken with some caution,
because the analysis by \citet{wachter02, wachter08}, used in the present 
investigation to derive the mass loss rates, holds for values of the carbon
excess above 8.2, whereas in these stars, as reported in Table \ref{tabTDU}, we
find $C-0 < 8.1$.

\citet{ventura12} warned that the model predictions for the production of silicates
during part of the O-rich phase of low-mass stars are not robust. This is due to the 
fact that the wind is not accelerated, which renders the results sensitive to the 
assumed velocity with which the wind enters the condensation zone. We stress here that 
in all the cases discussed here so far of dust production in low-mass AGB stars and the
formation of silicates during the O-rich AGB phases of the $3~M_{\odot}$ star of 
metallicity $Z=0.03$, the wind is accelerated after the formation of dust, until
reaching velocities of the order of 10 Km/s. This result is due to the large amounts 
of silicon and aluminium available in the envelope of super-solar metallicity stars, and 
renders the present findings more robust than for stars of lower metallicity, as they are 
independent on the assumptions regarding the initial velocity \citep{ventura12}.

\subsection{HBB and dust formation ($M > 3.5 M_{\odot}$)}
One of the most significant differences between the evolution of low-mass and massive
AGB stars is the overall energy release: comparing the results 
in the left panel of Fig.~\ref{f1phys} and in the left panel of Fig.~\ref{flowm}
it is clear that while the luminosities of low-mass AGBs are in the range $5\times 10^3 -
2\times 10^4~L_{\odot}$, the stars experiencing HBB evolve at higher luminosities
$2\times 10^4~L_{\odot} < L < 8\times 10^4~L_{\odot}$. This difference affects 
the rate of mass loss, when considering 
the high sensitivity of the mass loss description by \citet{blocker95} to the luminosity.

In fact, massive AGB stars experience rates of mass loss 
in the range  $10^{-5}-10^{-4}~M_{\odot}/$yr, significantly larger than those experienced 
by their lower mass counterparts. These large rates of mass loss favour
high wind densities, which trigger the production of large quantities of dust 
\citep{ventura12}. The bottom panels of Fig.~\ref{fsize} 
show that the dimension of olivine, alumina dust, and solid iron grains are generally
higher than in the low-mass domain, shown in the top panels of the same figure.
The trends with mass, mostly due to the behaviour of luminosity and mass loss rate, are
clear in Fig.~\ref{fsize}: the size of olivine grains slightly increases from 
$\sim 0.14 ~\mu$m, for the $4~M_{\odot}$ star, to $\sim 0.16~\mu$m, in the $7~M_{\odot}$ 
case. The fraction of silicon condensed into dust is in the range $30-40\%$. 
The rates of mass loss experienced by massive AGB stars are sufficiently large to favour 
copious production of alumina dust; the Al$_2$O$_3$ grains formed reach dimensions 
ranging from $\sim 0.05 ~\mu$m ($4~M_{\odot}$) to $\sim 0.08 ~\mu$m ($7~M_{\odot}$). 
The percentage of aluminium locked into Al$_2$O$_3$ is generally above $50\%$. In the 
models of higher mass we see in Fig.~\ref{fsize} that the size of alumina dust
particles, $\sim 0.08~\mu$m, are unchanged during the AGB phase; this is due to the 
saturation effect, first described in massive AGB stars of solar chemical composition by
\citet{flavia14a}, with almost all the gaseous aluminium available locked into dust grains.
The formation of silicates enhances the effects of radiation pressure, acting on
dust grains. For the stars experiencing HBB we find that the final wind velocities 
are slightly in excess of 20 Km/s. This results is very similar to the velocities
found by \citet{ventura18} for solar metallicity stars: the latter models produce lower
amounts of silicates (see Fig.~\ref{fdust}) than their higher-metallicity counterparts,
but this is compensated by the larger luminosities (see Fig.~\ref{fagb}), which increase 
the effects of radiation pressure on the acceleration of the wind.

We note some metallicity effects, where the higher-metallicity stars form dust particles
of slightly larger size, though the differences are within $5\%$.

\begin{table*}
\caption{The total dust mass and the masses of the individual species produced by our 
$Z=0.03$ and $Z=0.04$ models during the AGB phase.
}             
\label{tabdust}      
\centering          
\begin{tabular}{c c c c c c c}     
\hline
$M/M_{\odot}$ & $M_{\rm dust}$ & $M_{\rm sil}$ & $M_{\rm Al_2O_3}$ & $M_{\rm iron}$ &
M$_{\rm C}$ & M$_{\rm SiC}$ \\ 
\hline                    
& & & $Z=0.03$ & & &  \\
\hline       
1.00 & 2.00(-4) & 1.44(-4) & 1.15(-7) & 5.34(-5) & - & - \\
1.25 & 4.38(-4) & 3.45(-4) & 5.45(-7) & 8.85(-5) & - & - \\
1.50 & 8.42(-4) & 6.99(-4) & 2.40(-5) & 1.13(-4) & - & - \\
2.00 & 1.37(-3) & 1.15(-3) & 4.12(-6) & 2.08(-4) & - & - \\
2.50 & 1.91(-3) & 3.57(-4) & 4.34(-7) & 8.72(-4) & 3.33(-5) & 6.34(-4) \\
3.00 & 2.82(-3) & 1.16(-4) & 3.88(-8) & 4.14(-4) & 1.33(-3) & 9.44(-4) \\
3.50 & 3.17(-3) & 3.32(-4) & 8.63(-7) & 2.51(-3) & 7.76(-5) & 2.36(-4) \\
4.00 & 4.34(-3) & 4.09(-3) & 1.61(-4) & 8.59(-5) & - & - \\
5.00 & 6.13(-3) & 5.69(-3) & 3.98(-4) & 3.71(-5) & - & - \\
5.50 & 7.38(-3) & 6.78(-3) & 5.60(-4) & 3.63(-5) & - & - \\
6.00 & 8.59(-3) & 7.86(-3) & 6.92(-4) & 3.93(-5) & - & - \\
6.50 & 9.56(-3) & 8.63(-3) & 8.90(-4) & 4.37(-5) & - & - \\
7.00 & 1.22(-2) & 1.11(-2) & 1.07(-3) & 5.85(-5) & - & - \\
8.00 & 8.15(-3) & 3.92(-3) & 7.70(-7) & 3.85(-3) & - & - \\
\hline
& & & $Z=0.04$ & & &  \\
\hline       
1.00 & 2.35(-4) & 1.80(-4) & 1.76(-7) & 5.25(-5) & - & - \\
1.25 & 4.67(-4) & 3.96(-4) & 9.64(-7) & 6.70(-5) & - & - \\
1.50 & 1.03(-3) & 9.23(-4) & 4.62(-6) & 9.75(-5) & - & - \\
2.00 & 1.86(-3) & 1.69(-3) & 1.12(-5) & 1.56(-4) & - & - \\
2.50 & 2.28(-3) & 1.86(-3) & 1.30(-5) & 3.96(-4) & - & - \\
3.00 & 3.06(-3) & 2.76(-3) & 1.95(-5) & 2.75(-4) & - & - \\
3.50 & 3.75(-3) & 2.95(-3) & 6.08(-5) & 7.29(-4) & - & - \\
4.00 & 6.88(-3) & 6.38(-3) & 3.98(-4) & 1.01(-4) & - & - \\
4.50 & 6.87(-3) & 6.30(-3) & 5.33(-4) & 3.89(-5) & - & - \\
5.00 & 8.08(-3) & 7.33(-3) & 7.05(-4) & 3.80(-5) & - & - \\
5.50 & 9.17(-3) & 8.30(-3) & 8.36(-4) & 3.60(-5) & - & - \\
6.00 & 1.10(-2) & 9.87(-3) & 1.09(-3) & 4.10(-5) & - & - \\
6.50 & 1.15(-2) & 1.03(-2) & 1.15(-3) & 3.92(-5) & - & - \\
7.00 & 1.37(-2) & 1.23(-2) & 1.36(-3) & 5.04(-5) & - & - \\
\hline                  
\end{tabular}
\end{table*}

\subsection{The dust production rate}
To understand the feedback from AGB stars to the cycle of matter 
of the host system, we discuss the rate at which AGB stars 
eject dust in the interstellar medium, i.e., the dust production rate 
(DPR). The determination of the DPR is crucial 
not only to understand how dust is produced in single galaxies 
\citep{srinivasan09, srinivasan16, raffa14, flavia16, flavia18, flavia19}, 
but also, on more general grounds, to assess whether the dust produced 
by stars is able to reproduce interstellar dust abundances \citep{svetlana08}.

\begin{figure*}
\begin{minipage}{0.48\textwidth}
\resizebox{1.\hsize}{!}{\includegraphics{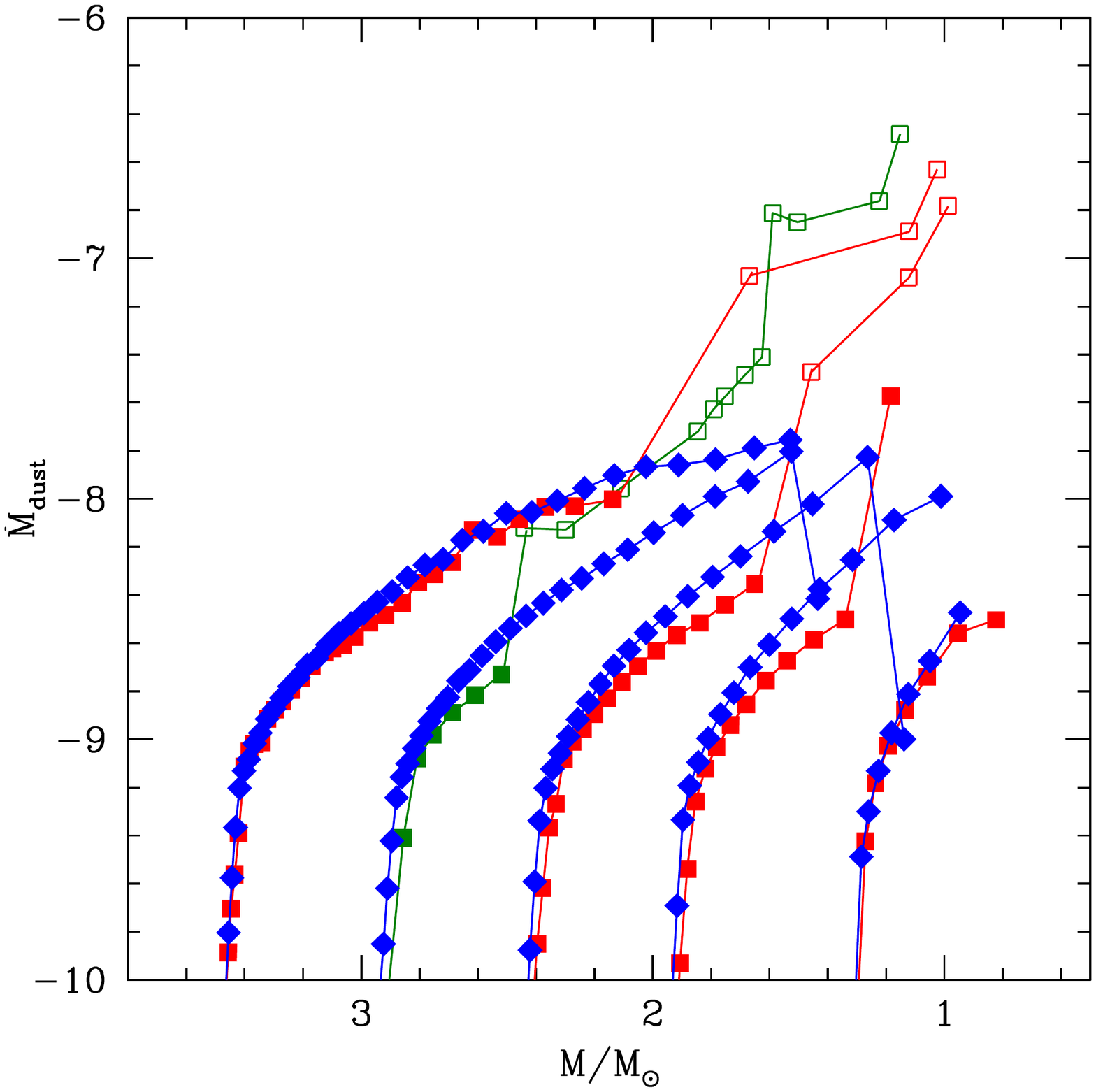}}
\end{minipage}
\begin{minipage}{0.48\textwidth}
\resizebox{1.\hsize}{!}{\includegraphics{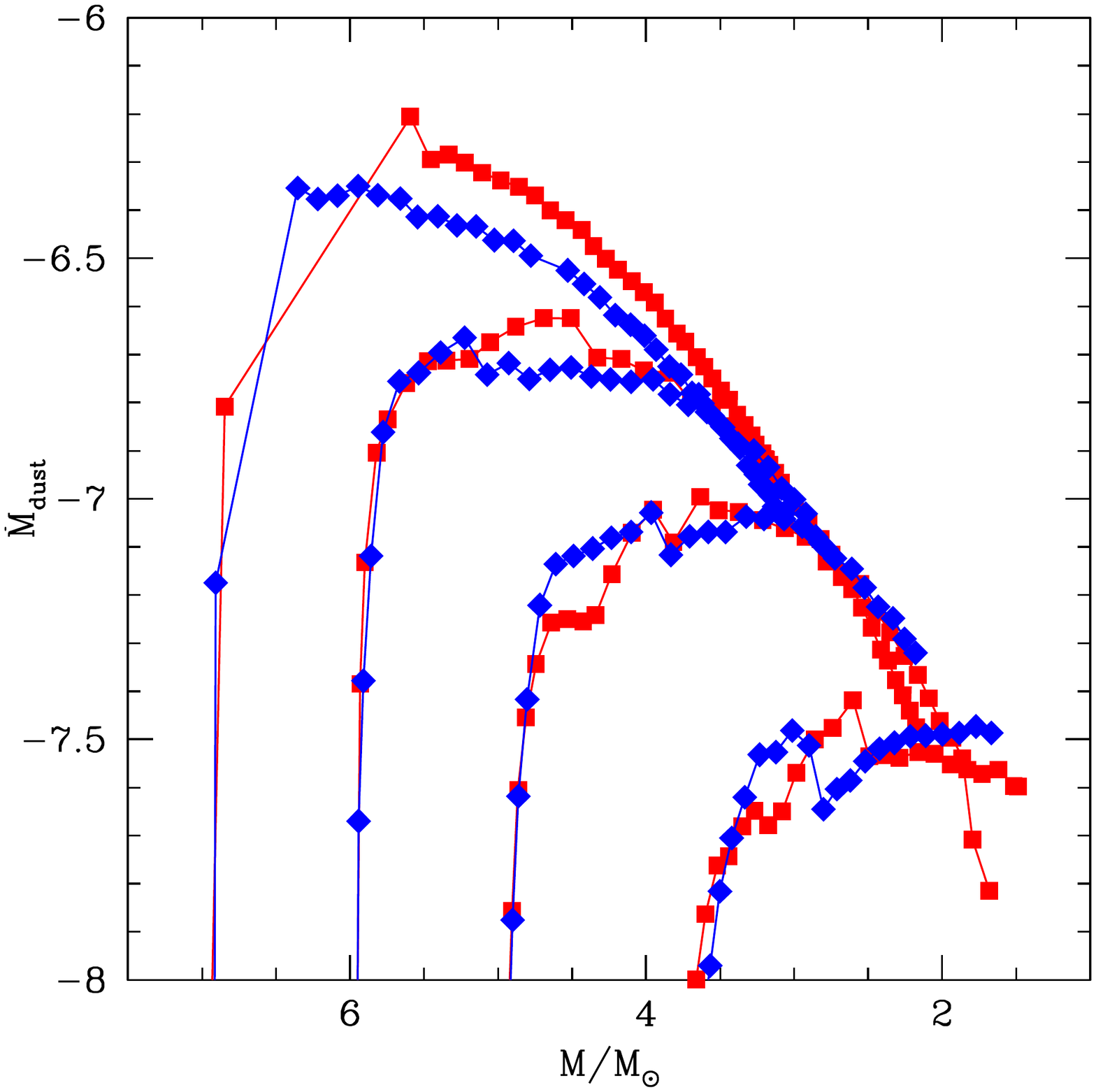}}
\end{minipage}
\vskip-60pt
\caption{The variation of the dust production rate with the stellar mass
for low-mass stars (left panel) and massive AGBs (right panel). Red squares and blue
diamonds refer to $Z=0.03$ and $Z=0.04$, respectively. Full points indicate stages 
during which the star is O-rich, while open points refer to carbon stars. The track 
referring to the $2.5~M_{\odot}$ stars of metallicity $Z=0.03$ is green, to better 
distinguish it from the $2~M_{\odot}$ and $3~M_{\odot}$ models. 
}
\label{fmdot}
\end{figure*}

In Fig.\ref{fmdot} we show the variation of the DPR of the same models 
shown in Fig. \ref{f1phys} and \ref{flowm} during the AGB life. The lines 
represent the overall DPR at each evolutionary stage obtained by summing up the 
contributions from all the individual dust species.

The DPR of low-mass stars increases during the AGB phase, consistently with the
earlier discussion and the results shown in the top panels of Fig.~\ref{fsize}. The DPR
are generally in the range $10^{-8}-10^{-7} ~ M_{\odot}/$yr and the main contribution
is provided by silicates. The sole exceptions to this general behaviour are given by
$2.5-3.5~M_{\odot}$ stars of metallicity $Z=0.03$: these
stars become carbon stars and during the C-star phase they form dust with rates up to
$\sim 5\times 10^{-7} ~ M_{\odot}/$yr in the final evolutionary stages.

Massive AGB stars (right panel of Fig.\ref{fmdot}) behave 
differently: the trend of DPR follows the variation of the luminosity because the dust 
production is strictly connected with the efficiency of HBB, as it is the overall energy 
flux. The DPR is higher during the initial AGB phases, when the luminosities and the 
temperatures at the base of the convective envelope are higher (see Fig.~\ref{f1phys}), 
then decreases towards the end of the AGB evolution. 
A trend of DPR with mass is clearly visible in Fig.~\ref{fmdot}: the largest DPR
grows from a few $10^{-8} ~ M_{\odot}/$yr, in the stars of mass $4~M_{\odot}$,
up to $\sim 5\times 10^{-7} ~ M_{\odot}/$yr in the $7~M_{\odot}$ case.

\subsection{The overall dust mass by super-solar metallicity AGB stars}

\begin{figure}
 \resizebox{\hsize}{!}{\includegraphics{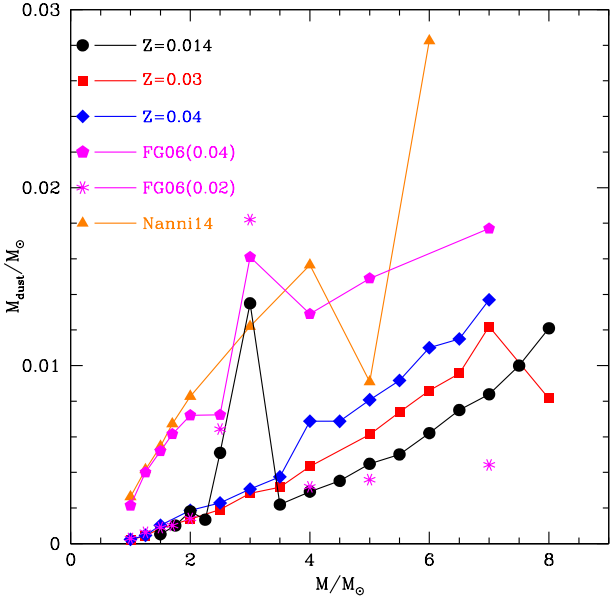}}
\vskip-70pt
 \caption{The mass of dust produced during the whole AGB phase of our models of
 different mass and metallicities. We also show, for comparison,  the solar metallicity 
 models by \citet{ventura18} and the results from \citet{fg06} and \citet{nanni14}.
 }
 \label{fdust}
\end{figure}

The amounts of dust produced by our models of different mass during the AGB life 
are summarised in Fig.\ref{fdust}, where we also show for comparison the dust masses for 
solar metallicity models of \citet{ventura18}, the results at $Z=0.02$ and $Z=0.04$ of 
\citet{fg06} and the $Z=0.04$ models by \citet{nanni14}.

The mass of dust increases with the initial mass of the star, spanning the
range $0.0001-0.012~M_{\odot}$ for $Z=0.03$ and $0.0001-0.014~M_{\odot}$ for $Z=0.04$.
Most of the dust is under the form silicates, with the exception of the $Z=0.03$ stars
with initial mass $2.5-3.5~M_{\odot}$, which produce mostly carbonaceous dust; this
is consistent with the earlier study by \citet{nanni14}.
On the average, the amount of dust produced is higher than in solar metallicity AGB
stars. A striking difference between stars of different metallicity is the behaviour 
of the $2-3~M_{\odot}$ stars of solar metallicity, which produce large amount of carbon dust,
of the order $0.01-0.015~M_{\odot}$, comparable with those of the most massive AGB stars.
This important difference is due to the higher carbon excess reached by solar metallicity
stars with respect to the $Z=0.03$ counterparts studied here.

Our $Z=0.04$ results are on average a factor $\sim 2$ lower than those of \citet{fg06}.
This difference is due to the adopted description of the mass loss. While we use the 
treatment by \citet{blocker95}, \citet{fg06}, like K14, adopted the mass loss versus period 
relationship by \citet{vw93}. As discussed in Section \ref{comp}, in metal-rich AGB
models the mass loss rates determined by applying the \citet{vw93} are generally higher
than \citet{blocker95}, which leads to higher rates of dust production in the wind.

The difference in the total dust produced during the AGB life is particularly relevant in 
the low-mass domain, where the dust masses by \citet{fg06} are $2-3$ times higher than
those calculated here. The difference is smaller for the stars experiencing HBB, as 
the present models experience stronger HBB conditions and evolve at larger luminosities 
compared to \citet{fg06}.

The dust masses found by \citet{nanni14}, indicated with orange triangles in
Fig.~\ref{fdust}, are similar to those by \citet{fg06} in the low- and intermediate-mass 
domain. For stars of mass $\sim 6~M_{\odot}$ the amount of dust by \citet{nanni14} are 
substantially higher; this is connected with the formation reaction for silicates used 
by \citet{nanni14}, discussed in \citet{nanni13}, which makes the formation of silicates 
to take place at temperatures of the order of 1400 K, closer to the stellar surface than 
in the present and in \citet{fg06} models.

\section{Conclusions}
\label{end}
We studied the evolution through the AGB phase of $1-8~M_{\odot}$ stars of super-solar 
metallicities, $Z=0.03$ and $Z=0.04$.
We find that stars of mass above $\sim 3~M_{\odot}$ experience HBB at the
base of the convective mantle. These objects evolve at luminosities 
$20-80\times 10^3~L_{\odot}$ during the AGB phase, with time scales ranging from $\sim 1$ Myr,
for $M\sim 4~M_{\odot}$ stars, to $\sim 5\times 10^4$ yr, for $M\sim 8~M_{\odot}$. The
surface chemical composition is driven by proton captures in the internal regions
of the convective envelope, with the destruction of the surface $^{12}$C and the synthesis
of large quantities of $^{14}$N. Significant amounts of lithium and sodium are produced,
whereas the destruction of the surface $^{16}$O is modest.

Low-mass stars not exposed to HBB experience several TDU events, which gradually increase
the surface carbon. The C-star stage is reached only by $2.5-3.5~M_{\odot}$ stars of
metallicity $Z=0.03$, with the surface C$/$O reaching a maximum of $\sim 1.2$. It is not
expected that C-stars are formed at $Z=0.04$. 

Dust production at super-solar metallicity is dominated by the formation of
silicates particles, mostly olivine, with a $\sim 10\%$ contribution of alumina dust
and solid iron. The only exception is represented by $2.5-3.5~M_{\odot}$ stars of
metallicity $Z=0.03$, whose dust production is dominated by the carbonaceous species,
silicon carbide and solid carbon, formed during the final AGB phases. 
The total mass of dust formed depends on the initial mass of the star, ranging from 
$\sim 0.001~M_{\odot}$, for stars of mass $1~M_{\odot}$, to 
$\sim 0.012~M_{\odot}$, for $M=8~M_{\odot}$. 

Regarding the gas pollution, our new models extend the available grids of stellar yields 
to  
super-solar metallicity. Nearly 70\% of solar $^{14}$N is synthesised in 
AGB stars, according to recent chemical evolution model results \citep{romano19}. This 
number is expected to be even higher in the inner disc of the Galaxy. Since the $^{14}$N 
yield continues to increase with stellar metallicity (see Fig~\ref{fyield}), with the highest 
yields found for 
the most metal-rich stars, stellar yields from super-solar metallicity 
models are required in chemical evolution studies to properly assess the $^{14}$N enrichment 
in the most evolved environments. Another example is $^{12}$C, whose yields are also dependent 
on metallicity. Carbon is observed in massive local ellipticals, as well as in their 
dust-obscured progenitors at high redshifts, and can be used to constrain the timescales 
of formation of these systems (e.g. Johansson et al. 2012). A meaningful quantification of 
the timescale for the most massive ellipticals require the adoption of stellar yields 
computed for super-solar metallicity stars. Finally, as mentioned in Section 4, the adoption 
of super-solar metallicity yields of $^{13}$C and $^{17}$O derived from self-consistent low- and 
intermediate-mass star models is mandatory for improving the modeling of dusty starburst 
galaxies at high redshifts and, consequently, the derivation of their galaxy-wide stellar 
IMF (see Romano et al. 2017; Zhang et al. 2018).

Future work also involves comparison of the nucleosynthetic predictions from the 
ATON AGB models to the isotopic composition of meteoritic stardust grains. For 
ATON such comparison has the advantage, with respect to those performed using 
other codes, of also being able to predict the expected types, amounts, and 
sizes of the dust produced by the potential AGB parent stars of the meteoritic 
grains. Such comprehensive, comparative investigation should include both 
carbonaceous and oxide dust, and both the light elements and those heavier than 
iron, which are produced in AGB stars by neutron captures. For the latter, an 
initial application of this innovative approach of combining AGB nucleosynthesis 
and dust formation predictions has been presented in relation to large 
($\sim \mu$m-sized) SiC grains by Lugaro et al. (2020, submitted), who concluded 
that these large grains originated from AGB stars of super-solar metallicity and 
that the initial number of dust seeds should decrease with increasing 
metallicity. However, such work combined neutron-capture nucleosynthesis 
predictions from the MONASH models and dust formation predictions from the 
ATON models. As the development of tools that allow predictions for 
neutron-capture nucleosynthesis using ATON are almost completed (Yague et al., 
in preparation), in the near future a more self-consistent approach will be 
possible.

\begin{acknowledgements}
      P.V. and D.R. benefited from the International Space Science Institute (ISSI, 
      Bern, CH, and ISSI-BJ, Beijing, CN) thanks to the funding of the team 
      {\emph "Chemical abundances in the ISM: the litmus test of stellar IMF 
      variations in galaxies across cosmic time"}. M.L. acknowledges the financial support 
      of the Hungarian National Research, Development and Innovation Office (NKFI), grant 
      KH\_18 130405 and the Lend\"ulet grant (LP17-2014) of the Hungarian Academy of 
      Sciences.
\end{acknowledgements}

%
%

\end{document}